\documentclass[a4paper,useAMS,usenatbib,onecolumn]{mnrasfile}
\usepackage{amssymb}
\usepackage{tabularx}
\usepackage{hyperref}
\usepackage{amsmath}
\usepackage{graphicx}
\usepackage{enumerate}
\usepackage{subfigure}
\usepackage[T1]{fontenc}
\usepackage{aecompl}
\usepackage{verbatim}
\usepackage[english]{babel}
\usepackage{pdfpages}

\newcommand{\df}{\delta}

\title[Covariance matrix of the matter power spectrum]{Perturbative approach to covariance matrix of the matter power spectrum}

 \author[I. Mohammed, U. Seljak, Z. Vlah] {Irshad
 Mohammed,\thanks{mohammed@fnal.gov}$^1$, Uro\v s Seljak$^{2,3}$  and
 Zvonimir Vlah$^{4,5}$\\ \\
 $^1${Fermilab Center for Particle Astrophysics, Fermi National Accelerator Laboratory, Batavia, IL 60510-0500, USA} \\
 $^2${Physics, Astronomy Department, University of California, Berkeley, CA 94720, USA}\\
 $^3${Lawrence Berkeley National Laboratory, Berkeley, CA 94720, USA}\\
 $^4${Stanford Institute for Theoretical Physics and Department of Physics, Stanford University, Stanford, CA 94306, USA}\\
 $^5${Kavli Institute for Particle Astrophysics and Cosmology, SLAC and Stanford University, Menlo Park, CA 94025, USA}
 }

\begin{document}
\maketitle


\begin{abstract}

We evaluate the covariance matrix of the matter power spectrum using perturbation theory up to dominant terms at 1-loop order and compare it to numerical simulations. We decompose the covariance matrix into the disconnected (Gaussian) part, trispectrum from the modes outside the survey (beat coupling or super-sample variance), and trispectrum from the modes inside the survey, and show how the different components contribute to the overall covariance matrix. We find the agreement with the simulations is at a 10\% level up to $k \sim 1 h {\rm Mpc^{-1}}$. We show that all the connected components are dominated by the large-scale modes ($k<0.1 h {\rm Mpc^{-1}}$), regardless of the value of the wavevectors $k,\, k'$ of the covariance matrix, suggesting that one must be careful in applying the jackknife or bootstrap methods to the covariance matrix. We perform an eigenmode decomposition of the connected part of the covariance matrix, showing that at higher $k$ it is dominated by a single eigenmode. The full covariance matrix can be approximated as the disconnected part only, with the connected part being treated as an external nuisance parameter with a known scale dependence, and a known prior on its variance for a given survey volume. 
Finally, we provide a prescription for how to evaluate the covariance matrix from small box simulations without the need to simulate large volumes.  

\end{abstract}



\section{Introduction}\label{sec:intro}

The distribution of matter in the Universe contains a wealth of information about the energy content of the Universe, its properties, and evolution. Initial distribution is thought to be a Gaussian random field, but as a result of the gravitational instability the tiny fluctuations in the initial matter distribution evolves nonlinearly and produces non-Gaussian correlations. The simplest statistic for data analysis is the two point correlation function, or its Fourier transform the power spectrum. This contains significant cosmological information since it is sensitive to many parameters, and much of the information comes from deeply in the nonlinear regime. The two-point analysis is the prime focus of many cosmological observables like weak lensing (WL) or galaxy clustering, which are the key observable probes in the current and future generation surveys like Dark Energy Survey (DES\footnote{http://www.darkenergysurvey.org}), Dark Energy Spectroscopic Instrument (DESI\footnote{http://desi.lbl.gov}), Large Synoptic Survey Telescope (LSST\footnote{https://www.lsst.org}), Euclid\footnote{http://www.euclid-ec.org} etc.

Our ability to extract useful constraints on the cosmological parameters from these surveys depend on our ability to model the statistical properties of the distribution of matter in the Universe, particularly the matter power spectrum, in the non-linear regime ($k > 0.2 h^{-1}{\rm Mpc}$). In recent years, there have been many efforts in providing accurate estimates of the matter power spectrum for various cosmological models and redshifts using perturbation theory \citep{2002PhR...367....1B},  the halo model \citep{2002PhR...372....1C}, simulations \citep{2010ApJ...715..104H,2009ApJ...705..156H,2010ApJ...713.1322L}, and semi-analytic models \citep{2014MNRAS.445.3382M,2015PhRvD..91l3516S}. The current precision from simulations is at 1\%, and can easily be improved further with simulations if necessary. There are additional complications such as the baryonic effects \citep{2011MNRAS.415.3649V, 2011MNRAS.417.2020S, 2013MNRAS.434..148S, 2014arXiv1410.6826M}, which redistribute the matter within the halo centers and change the power spectrum at high $k$, and it is likely that WL will not be able to extract much reliable information at high $k$ (or $l$). We will not model these processes here and focus on  the dark matter part only.

While the matter power spectrum predictions, in the absence of baryonic effects, are under control, for a complete analysis one also needs its covariance matrix. The covariance matrix of the matter power spectrum is important in order to perform any statistical inference analysis on the cosmological data. Therefore, future surveys would require an accurate estimation of the covariance matrix of the matter power spectrum in order to perform the cosmological parameters estimation. An accurate quantification of the covariance matrix is crucial in order to derive constraints on the cosmological parameters using observables modeled on the matter power spectrum. Failing to do so will mislead the interpretation of the data.

In the linear regime, computing the covariance matrix is reasonably straight-forward, even if it can be computationally expensive for surveys with complicated masks. Due to the non-linear evolution of the matter density field, the Fourier modes become correlated, and this generates a non-Gaussian covariance matrix \citep{1999ApJ...527....1S, 2001ApJ...554...67H,2001ApJ...554...56C}, which we also call the connected part. It affects both diagonal and non-diagonal elements of the covariance matrix. Its characterization has been developed in terms of the physically motivated halo model \citep{2013PhRvD..87l3504T,2014MNRAS.445.3382M}, or numerical simulations \citep{2009ApJ...700..479T,2011ApJ...734...76S,2012MNRAS.423.2288H,2013PhRvD..88f3537D,2014PhRvD..89h3519L, 2015MNRAS.446.1756B,2015arXiv151205383B}. 

Another approach is using perturbation theory (PT), which is the focus of the work here. Our approach is to identify the perturbative terms that dominate the covariance matrix, and PT can be used to understand better the structure of the covariance matrix. This paper is organized as follows: In section \ref{sec:newmodel}, we review the approach and describe the perturbation theory up to partial 
1-loop calculations for the full covariance matrix of the matter power spectrum. In section \ref{sec:sims}, we outline a comparison of our model with two sets of cosmological simulations, with and without SSC term. In section \ref{sec:decomposition}, we decompose the covariance matrix based on simulations and the analytic model into a vector that fully describes the full covariance matrix, with and without SSC term, using a principle component analysis. In section \ref{sec:fisher}, we study the degeneracies between covariance matrix model and other cosmological parameters and perform a Fisher information  matrix analysis. We also address the question of how  to evaluate covariance matrix from small box simulations. Finally we conclude with a  discussion in section \ref{sec:discussion}. During the completion of our work an independent analysis including all PT terms up to 1-loop has been presented here \cite{2015arXiv151207630B}, with  results comparable to ours.


\section{Covariance matrix in perturbation theory}\label{sec:newmodel}

The simplest of the statistics is the two-point correlation function of the matter density perturbations $\delta$, or its Fourier transform matter power spectrum $P(k)$,

\begin{equation}
        \langle \tilde{\delta}(\mathbf{k}) \tilde{\delta}^{\star}(\mathbf{k}') \rangle = (2\pi)^3 \delta_D(\mathbf{k-k'}) P(k)
\end{equation}
\\
where, $\delta_D$ is the Dirac delta function, $k$ is the magnitude of the wavevector $\mathbf{k}$ and $\tilde{\delta}(\mathbf{k})$ is the Fourier transform of the matter density perturbations $\delta$, defined as,

\begin{equation}
            \tilde{\delta}(\mathbf{k}) = \int \delta(\mathbf{x}) e^{i \mathbf{k.x}} d^3{\mathbf x}
            \label{eqn:deltak}
\end{equation}
\\
For a given survey volume $V$, one can compute $\tilde{\delta}(\mathbf{k})$ using equation \ref{eqn:deltak}. Then the power spectrum can be estimated by dividing the survey volume into shells centered at $k_i$ as,

\begin{equation}
        P(k_i) = V_f\int_{k_i} \dfrac{d^3\mathbf{k}}{V_s(k_i)} \tilde{\delta}(\mathbf{k})\tilde{\delta}(\mathbf{-k})
\end{equation}
\\
where, $V_s(k_i) = 4\pi k_i^2 \Delta k$ is the volume of the $i-$th shell, $\Delta k$ is the bin width, and $V_f=(2\pi)^3/V$ is the volume of the fundamental cell in Fourier space.

The matter power spectrum is a two-point function, it's statistical properties defining its uncertainties, precision to which it can be measured, and correlations at various scales can be quantified in terms of a covariance matrix $\mathbf{Cov}(k_i, k_j)$, which is a four-point function and defined as,

\begin{equation}
        \mathbf{Cov}(k_i, k_j) \equiv \langle P({\bf k_i}) P({\bf k_j}) \rangle -
                    \langle P({\bf k_i}) \rangle
                    \langle P({\bf k_j}) \rangle,
    \label{eqn:covdefinition}
\end{equation}
\\
where, the angle brackets represent an ensemble average. We interchangeably uses the notation $\mathbf{Cov}(k_i,k_j)$ and $\mathbf{Cov}_{ij}$. Similarly, we also interchangeably use $P(k_i)$ or $P_i$ for the matter power spectrum.

A full covariance matrix can be decomposed into,

\begin{equation}
        \mathbf{Cov}_{ij}^{\rm Full} = \mathbf{Cov}_{ij}^{\rm G} +
        \mathbf{Cov}_{ij}^{\rm NG} .
\end{equation}

The first contribution is the disconnected contribution, also known as the Gaussian contribution,
and is always present due to the random phases of the modes of the density field. As the matter power spectrum is computed by averaging over those Fourier modes, a smaller number of modes gives larger statistical uncertainty over the mean power spectrum. Therefore, this contribution is inversely proportional to the number of modes. As we have only one survey volume to observe, the largest scale modes are very few, and hence the covariance matrix is dominated by this part for low $k$. When noise can be ignored it is known as the sampling variance. It dominates along the diagonal elements of the covariance matrix: for a finite volume of the survey the modes are mixed, but if we bin the modes with binning $\Delta k>2\pi/R$, where $R^3=V$ is the survey volume, then the window is diagonal. The Gaussian contribution can be estimated as,

\begin{equation}
    \mathbf{Cov}_{ij}^{\rm G} = \dfrac{2}{N_k} \delta_{ij} P(k_i) P(k_j),
\end{equation}
\\
where, $N_k = V_s / V_f$ is the total number of $k$ modes in the corresponding shell, and $\delta_{ij}$ is the Kronecker delta function which is unity for $i=j$, and zero otherwise. It scales inversely with the fourth power of the growthfactor for its redshift evolution.

The second contribution is the connected part, or the non-Gaussian part, and can be expressed in terms of the trispectrum as

\begin{equation}
    \mathbf{Cov}_{ij}^{\rm NG} = \dfrac{\bar{T}(\bf{k_1,-k_1,k_2,-k_2})}{V}
    \label{ngcon}
\end{equation}
\\
where, $\bar{T}$ is the bin averaged trispectrum, the fourth-order connected moment of the density perturbation, given by,

\begin{equation}
    \bar{T}(\bf{k_1,-k_1,k_2,-k_2}) = \int_{V_{k_i}}  \int_{V_{k_j}}
    \dfrac{d^3 \bf{k_1}}{V_{k_i}} \dfrac{d^3 \bf{k_2}}{V_{k_j}} T(\bf{k_1,-k_1,k_2,-k_2})
    \label{eqn:tt}
\end{equation}
\\
where the two integral on the right are across the $i-$th and $j-$th shell.

The Gaussian part of the covariance depends on the binning scheme in the $k$ shells, whereas the non-Gaussian part is fairly independent of any binning scheme, except for a small dependence for low $k$ (see section \ref{sec:trispec} for details). However, both terms scale inversely with the survey volume.

We split the non-Gaussian contribution into the parts that comes from modes outside the survey (SSC) and inside the survey, and  we further split the latter into tree level and 1-loop terms, 

\begin{equation}
    \mathbf{Cov}_{ij}^{\rm NG} =
            \mathbf{Cov}_{ij}^{\rm SSC}+
            \mathbf{Cov}_{ij}^{\rm Tree-level} +
            \mathbf{Cov}_{ij}^{\rm 1-loop}
\end{equation}
\\
where, the terms on the right are the beat coupling or super-sample covariance (SSC), tree level, and 1-loop contributions, respectively. The first two terms are both tree-level contributions, and their split is motivated by the different calculational approach to the two terms. To this, we add a specific 1-loop term because it significantly improves the results. In principle, we should include all 1-loop terms as well as add counter-terms to 1-loop, but we will not explore these extensions here (see \cite{2015arXiv151207630B, 2016arXiv160401770B}). 


\subsection{Beat coupling covariance}

The first NG contribution is the beat coupling effect, also known as the Super Sample Covariance (SSC). It is caused by the coupling between all $k$ modes that are larger than the survey, and the modes inside the survey. It is caused by the survey window effects mixing the modes and does not show up in periodic box simulations. Its leading effect can be derived as a tree level trispectrum in the squeezed limit. This contribution was first pointed out by \cite{2006MNRAS.371.1188H}, and further studied by \cite{2006PhRvD..74b3522S,2009ApJ...701..945S, 2009ApJ...700..479T,2013MNRAS.429..344K,2007NJPh....9..446T, 2009MNRAS.395.2065T,2012JCAP...04..019D}. Because this contribution comes from the mode that is constant across the survey, it can be viewed as a local curvature term within the survey, which can further be mimicked by a change in the background density $\delta_b$. Therefore, this term can be modeled completely by the response of the matter power spectrum to the change in background density \citep{2013PhRvD..87l3504T, 2014PhRvD..89h3519L,2014PhRvD..90j3530L}. Further, this response can be accurately calculated using the separate universe simulations, where the same effect can be mimicked by a change in cosmological parameters \citep{2005ApJ...634..728S,2011ApJS..194...46G,2011JCAP...10..031B,2015MNRAS.448L..11W,2015arXiv151101465B}. This term is the dominant contribution in the quasi-linear regime, and can be modeled using the response of the mean background density of the Universe to the matter power spectrum,

\begin{equation}
    \mathbf{Cov}_{ij}^{\rm SSC} = \sigma_b^2 \dfrac{\partial P(k_i)}{\partial \delta_b} \dfrac{\partial P(k_j)}{\partial \delta_b}
    \label{eqn:ssc}
\end{equation}
\\
where, $\partial P(k_i)/\partial \delta_b$ is the response of the matter power spectrum to the change in the background density $\delta_b$, and $\sigma^2_b$ is the variance of the mean density field in the survey window, defined as

\begin{equation}
    \sigma_b^2 \equiv \langle \delta_b^2\rangle = \dfrac{1}{V^2} \int
        \dfrac{d^3\mathbf{q}}{(2\pi)^3} \tilde{W}(\mathbf{q})^2 P(q),
\label{SSCsig}
\end{equation}
\\
where, $\tilde{W}(\mathbf{q})$ is the survey window function (with units of volume). The responses depend on whether we divide by the local density (SSC-local), in which case the low $k$ limit is

\begin{equation}
\dfrac{d\ln P(k)}{d\delta_b}_{\rm SSC-local} ={5 \over 21}-{1 \over 3}{d \ln P(k) \over d \ln k},
\end{equation}
\\
or we do not divide by the local density (SSC-global), which gives low $k$ limit of

\begin{equation}
\dfrac{d\ln P(k)}{d\delta_b}_{\rm SSC-global} ={47 \over 21}-{1 \over 3}{d \ln P(k) \over d \ln k}.
\end{equation} 

We will use numerical derivative obtained from separate universe simulations \cite{2014PhRvD..89h3519L}. In general SSC term is well understood and can easily be computed by running two separate universe  simulations.


\subsection{Tree level covariance}\label{sec:trispec}

\begin{figure}
    \centering
    \includegraphics[width=0.95\textwidth]{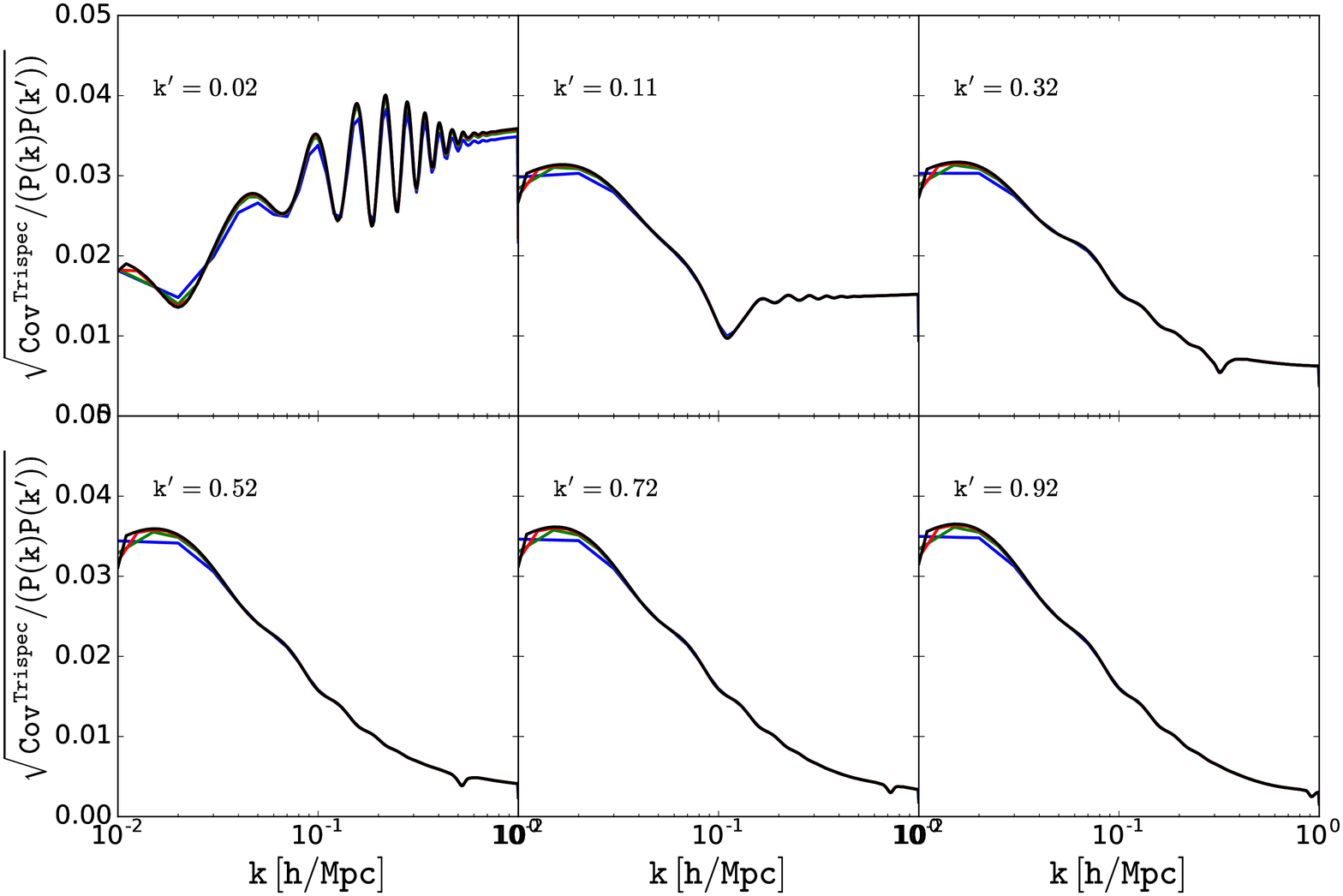}
    \caption{The trispectrum contribution to the covariance for different binning schemes in different colors.
            Blue, green, red and black corresponds to a linear bin width $(\Delta k)$ of 0.01, 0.025, 0.05 and 0.001 respectively.}
    \label{fig:trispectrum}
\end{figure}

In the tree-level perturbation theory, and ignoring the window function effects that lead to SSC term above, the full trispectrum can be written as,

\begin{equation}
    T(\mathbf{k_1,k_2,k_3,k_4}) = 4\left[ F_2(\mathbf{k_{12}},-\mathbf{k_1})
            F_2(\mathbf{k_{34}},\mathbf{k_4}) P(k_1)P(k_{34})P(k_3) + \rm{cyc.}\right]
            + 6\left[  F_3(\mathbf{k_1,k_2,k_3}) P(k_1)P(k_2)P(k_3) + \rm{cyc.}
            \right]
\end{equation}
\\
where $\mathbf{k_{12} = k_1+k_2}$, and cyc denotes the cyclic permutations of the arguments. The kernels $F_2$ and $F_3$ are the solutions to the equations of motion of gravitational instability to 2nd and 3rd order respectively. Although this is the full expression for the trispectrum, its particular configurations contribute to the non-Gaussian covariance, namely parallelogram, as in equation \ref{eqn:tt}. Therefore, the parallelogram configuration of the trispectrum that contributes to the non-Gaussian covariance is given by \citep{1999ApJ...527....1S},

\begin{align}
\bar T(k_i, k_j) = \int_{k_i} &\frac{d^3 k_1}{V_s(k_i)}\int_{k_j}\frac{d^3 k_2}{V_s(k_j)} ~ \bigg(
12 F_3^{(s)}(\mathbf k_1, -\mathbf k_1, \mathbf k_2) P_{\rm L}(\mathbf k_1)^2 P_{\rm L}(\mathbf k_2) \nonumber\\
&+8 F^{(s)}_{2}(\mathbf k_1 - \vec k_2, \mathbf k_2)^2 P_{\rm L}(\mathbf k_2)^2 P_{\rm L}(\mathbf k_2 - \mathbf k_1 ) \nonumber\\
&+8 F^{(s)}_{2}(\mathbf k_1 - \vec k_2,\mathbf k_2)F^{(s)}_{2}(\mathbf k_2 - \mathbf k_1,\mathbf k_1)
P_{\rm L}(\mathbf k_2) P_{\rm L}(\mathbf k_1) P_{\rm L}(\mathbf k_2 - \mathbf k_1 ) \nonumber\\
&+ \{ \mathbf k_1\leftrightarrow \mathbf k_2 \} \bigg),
\end{align}
\\
where, $F^{(s)}_2$ and $F^{(s)}_3$ are the standard Eulerian PT kernels (see e.g. \cite{2002PhR...367....1B}). Diagonal part $\bar T(k_i, k_i)$ can now be explicitly written as

\begin{align}
\frac{\bar T(k_i, k_i)}{ \left[P_{\rm L}(k_i)\right]^3} = -\frac{44}{63}
+ \frac{1}{49}\int_{-1}^1 d\mu \left( 3+10\mu \right)^2 \frac{P_{\rm L}\left( k \sqrt{2(1-\mu)} \right)}{P_{\rm L}(k_i)}
\end{align}

Using the approximation suggested in \cite{1999ApJ...527....1S} one gets $\bar T(k_i, k_i) / \left[P_{\rm L}(k_i)\right]^3\simeq 232/441$, although somewhat better approximation (at low $k$) can be obtained $\bar T(k_i, k_i) / \left[P_{\rm L}(k_i)\right]^3\simeq 454/441 - 62/343 ~n_s\simeq 2620/3087$.

Figure \ref{fig:trispectrum} shows the trispectrum contribution to the covariance for few different binning schemes. Binning makes essentially no effect, as expected. This would appear to be in conflict with \cite{2008MNRAS.389.1675T}, where they studied binning dependence of the ratio between nonlinear and linear mode, finding strong dependence. However, this is caused by division with the linear mode, which induces Gaussian fluctuations in the ratio.


\subsection{One loop covariance}

In this section, we describe the calculations of the one loop non-Gaussian covariance contribution.  A full calculation has been recently presented in \cite{2015arXiv151207630B, 2016arXiv160401770B}. Here we will instead do a simplified calculation, where we first derive a functional derivative of high $k$ power to the low $q$ power. In the next step, we compute the fluctuations of low $q$ power due to the finite volume effects, i.e. the sampling variance fluctuations. Finally, we obtain the covariance by combining the two.

One loop PT on the power spectrum has been intensively studied in previous works such as \cite{1981MNRAS.197..931J,1983MNRAS.203..345V,1984MNRAS.209..139J,1990MNRAS.243..171C,1991PhRvL..66..264S,1992PhRvD..46..585M,1994ApJ...431..495J,1994MNRAS.270..183B,1996ApJ...467....1L,1996ApJ...473..620S,2002PhR...367....1B}. Here we briefly review these results. In perturbation theory framework, one can describe the full matter power spectrum as,

\begin{equation}
    P(k,z) = P^{(0)}(k,z) + P^{(1)}(k,z) + ... + P^{(n)}(k,z)
\end{equation} 
\\
where, the superscript $n$ denotes the $n-$loop contribution. $P^{(0)}(k,z)$, or the zero loop contribution, is just the linear power spectrum, such that $P^{(0)}(k,z) = P_L(k)D^2(z)$. The one loop contribution, $P^{(1)}(k,z)$, consists of two terms and can be written as,

\begin{align}
    P^{(1)}(k,z) & = P_{13}(k,z) + P_{22}(k,z) \nonumber\\
                & = 6\int F_3^{(s)}(\mathbf{k, \tilde{q},-\tilde{q}})P_L(k,z)P_L(\tilde{q},z) d^3\tilde{q}
                    + 2\int \left( F_2^{(s)}(\mathbf{k-\tilde{q},\tilde{q}})\right)^2
                        P_L(|\mathbf{k-\tilde{q}}|,z)P_L(\tilde{q},z) d^3 \tilde{q}
\end{align}

Here $P_{ij}$ denotes the amplitude given by a connected diagram representing the contribution from $\langle \delta_i \delta_j \rangle_c$ to the power spectrum. We have assumed Gaussian initial conditions, for which $P_{ij}$ vanishes if $i+j$ is odd. The two contributions, $P_{13}$ and $P_{22}$, have different structure. $P_{13}$ is in general negative and describes the decorrelation of the propagator, while $P_{22}$ is positive definite and describes the effects of mode coupling between modes with wavevectors $\mathbf{k-q}$ and $\mathbf{q}$.

The functional derivatives of these function with respect to the linear power spectrum yields,

\begin{align}
N \frac{\delta P_{13}( \mathbf k)}{\delta P_\text{L}( \mathbf q)} &=
3 P_\text{L}(\mathbf k) \int d^3 \tilde{q}~ F_3^\text{(s)}(\mathbf{k},\tilde{ \mathbf q },-\tilde{ \mathbf q }) \delta^D(\mathbf q-\tilde{ \mathbf q})
+ 3 ~\delta^D (\mathbf k - \mathbf q) \int \frac{d^3 \tilde{q}}{(2\pi)^3} ~ F_3^\text{(s)}(\mathbf{k},\tilde{ \mathbf q },-\tilde{ \mathbf q })
P_\text{L}(\tilde{ \mathbf q})  \nonumber\\
& = 3 F_3^\text{(s)}(\mathbf{k},\mathbf{q},-\mathbf{q}) P_\text{L}(\mathbf k) + 3 P_{13}(\mathbf k)/P_{\rm L}(\mathbf k) \delta^D (\mathbf k - \mathbf q)  \nonumber\\
N \frac{\delta P_{22}(\mathbf k)}{\delta P_\text{L}(\mathbf q)} &=
4 \int d^3 \tilde{q} \left[ F_2^\text{(s)}(\mathbf{k}-\tilde{ \mathbf q },\tilde{ \mathbf q })\right]^2
P_\text{L}(\mathbf{k}-\mathbf{ \vec q }) \delta_D (\mathbf q-\tilde{\mathbf q}) \nonumber\\
& = 4 \left( F_2^\text{(s)}(\mathbf{k}-\mathbf{{q}},\mathbf{{q}})\right)^2
P_\text{L}(\mathbf{k}-\mathbf{q}).
\label{eq:fncder_p22p13}
\end{align}

Where $N$ is the normalization prefactor to be chosen below. 
Dropping the delta function part of the result we have for the total one-loop result

\begin{align}
 N  \frac{\df P_{\rm 1-loop} ( \mathbf k)}{\df P_\text{L}( \mathbf q)}
& =  N  \frac{\delta P_{22}(\mathbf k)}{\delta P_\text{L}(\mathbf q)} 
+ 2 N \frac{\delta P_{13}( \mathbf k)}{\delta P_\text{L}( \mathbf q)} \nonumber\\
& = 4 \left( F_2^\text{(s)}(\mathbf{k}-\mathbf{{q}},\mathbf{{q}})\right)^2 P_\text{L}(|\mathbf{k}-\mathbf{q}|)
+  6 F_3^\text{(s)}(\mathbf{k},\mathbf{q},-\mathbf{q})P_\text{L}(k).
\end{align}

Preforming the angle averaging $\langle X(\mathbf q) \rangle_\Omega = \int \frac{d \Omega_{\vec q}}{4\pi} X(\vec q)$, we get

\begin{align}
2 N \left\langle \frac{\df P_{\rm 1-loop} ( \mathbf k)}{\df P_\text{L}( \mathbf q)}  \right\rangle_\Omega
& = 
2 \int d\mu~\left( F_2^\text{(s)}(\mathbf{k}-\mathbf{{q}},\mathbf{{q}})\right)^2 P_\text{L}(|\mathbf{k}-\mathbf{q}|)
+ 3 \int d \mu ~ F_3^\text{(s)}(\mathbf{k},\mathbf{q},-\mathbf{q})P_\text{L}(k).
\label{eq:oneLfd}
\end{align}

\begin{figure}
\centering
\includegraphics[width=0.8\linewidth]{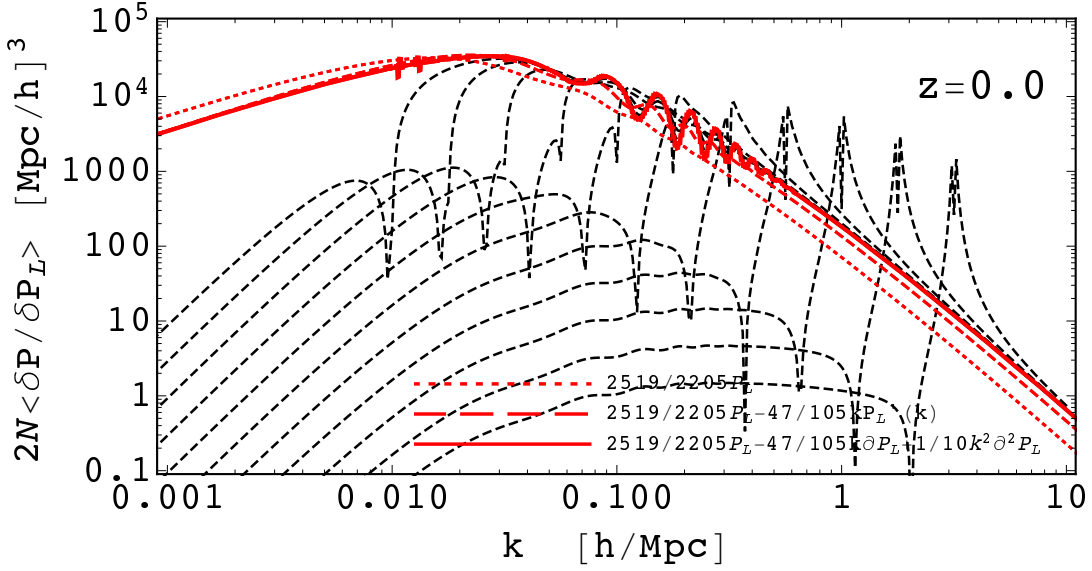}
\caption{One loop functional derivatives $2N \langle \df P_{\rm 1-loop} (k)/\df P_\text{L}(q) \rangle_\Omega$ (black dashed lines) given in Eq.\eqref{eq:oneLfd} at redshift $z=0.0$.
Values of $q$ are from left to right are $\{${0.01, 0.018, 0.032, 0.056, 0.10, 0.18, 0.32, 0.56, 1.0, 1.8, 3.2}$\}\left[h^{-1}{\rm Mpc}\right]$.
Red lines correspond to contributions in the limiting cases $k \gg q$ in Eq. \eqref{eq:oneLfdlimir}.}
\label{fig:oneLfd}
\end{figure}

It is instructive to look at the limiting cases. From above it readily follows \citep{2012PhRvD..85j3523S}

\begin{align}
2 N \left\langle \frac{\df P_{\rm 1-loop} ( \mathbf k)}{\df P_\text{L}( \mathbf q)}  \right\rangle_\Omega
\sim \frac{2519}{2205} P_\text{L}( \mathbf k)
- \frac{47}{105} k P_\text{L}'( \mathbf k) + \frac{1}{10} k^2 P_\text{L}''( \mathbf k) ,~~{\rm as}~~k \gg q,
\label{eq:oneLfdlimir}
\end{align}

and similarly

\begin{align}
2 N \left\langle \frac{\df P_{\rm 1-loop} ( \mathbf k)}{\df P_\text{L}( \mathbf q)}  \right\rangle_\Omega
\sim \frac{9 k^4}{49 q^4} P_\text{L}( \mathbf q)
- \left( \frac{61 k^2}{315 q^2} - \frac{4 k^4}{105 q^4} \right) P_\text{L}( \mathbf k) ,~~{\rm as}~~k \ll q.
\end{align}
The result of the fill one loop functional derivative given in Eq.\eqref{eq:oneLfd} and the corresponding 
$k \gg q$ expansion is shown in Figure \ref{fig:oneLfd}.

We can introduce the normalized functional derivative, $\mathbf{V(q,k)}$, by dividing the expression above by the power per mode $\Delta^2(q)=4\pi q^3P(q)/(2\pi)^3$ \cite{2014arXiv1411.2970N},

\begin{equation}
    \mathbf{V}(q,k) = {P_L(q) \over \Delta^2(q)}\left\langle \frac{\df P_{\rm 1-loop} ( \mathbf k)}{\df P_\text{L}( \mathbf q)}  \right\rangle_\Omega .
\end{equation}

In the low $q$ limit, i.e., $q\ll k$, the functional derivative becomes independent of $q$ 

\begin{equation}
    \lim_{q/k\to 0} \mathbf{V}(q,k) = W(k) P(k)
\end{equation}
\\
where,

\begin{equation}
    \mathbf{W}(k) = \dfrac{2519}{2205} -
                    \dfrac{47}{105} \dfrac{d\ln P(k)}{d\ln k} +
                    \dfrac{1}{10} \dfrac{d^2\ln P(k)}{d\ln k^2}.
    \label{eqn:w}
\end{equation}

The expressions above give the mean response at a given $k$ to the power at a given $q$. At low $q$, the power at a given $q$ will fluctuate due to the finite volume. The Gaussian part of the covariance scales as $2P^2/N_k$ and number of modes $N_k$ scales as $N_k=Vk^3 \Delta \ln k /(2\pi)^3$. The full expression for one loop covariance contribution can now be written as,

\begin{equation}
        \mathbf{Cov}_{ij}^{\rm 1-loop} = \left(\dfrac{1}{V\pi^2}
                                    \int P^2_{\rm Lin}(q) q^2
                                    \mathbf{V}(q,k_i) \mathbf{V}(q,k_j)
                                    dq\right).
                                    \label{eqn:cov1loop}
\end{equation}
\\

Equation \ref{eqn:cov1loop} is an integral of the term $P^2_{\rm Lin}(q)q^2$ and the functional derivatives. We can define

\begin{equation}
S=\left( \dfrac{1}{V\pi^2} \int P^2_{\rm Lin}(q) q^2dq \right)
\label{S}
\end{equation}

In the low $q$ limit this separates into

\begin{equation}
                \mathbf{Cov}_{ij}^{\rm 1-loop} =
                                                                        S\mathbf{W}(k_i)P(k_i)
                                                                        \mathbf{W}(k_j)P(k_j).
                                                                        \label{eqn:cov1loopq}
\end{equation}

\begin{figure}
    \centering
    \includegraphics[width=0.8\textwidth]{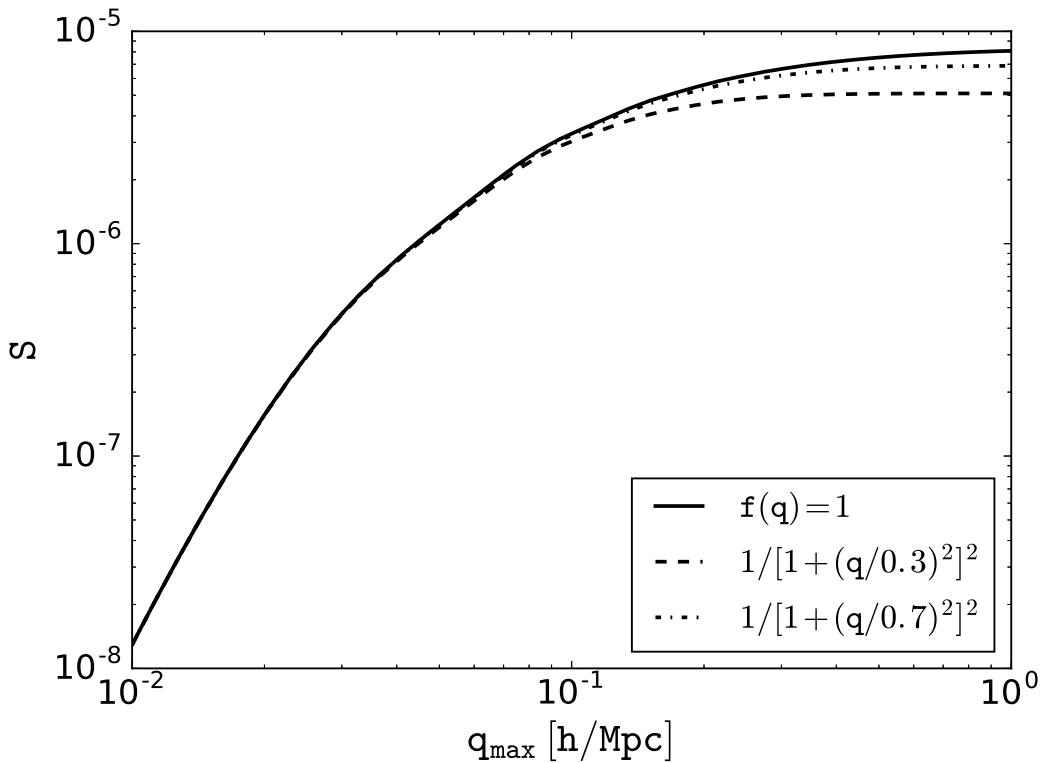}
    \caption{The 1-loop variance $S$ as a function of $q_{max}$ for a volume of $1 (h^{-1} {\rm Gpc})^3$. The integral converges
    at $q_{max} \sim 0.3 h {\rm Mpc}^{-1}$. We also show the effect of damping for $z=0$ ($q_{nl}=0.3 h {\rm Mpc^{-1}}$) and for 
$z=1$ ($q_{nl}=0.7 h {\rm Mpc^{-1}}$).}
    \label{fig:p2k2}
\end{figure}

Figure \ref{fig:p2k2} shows the integral of $S$ to $q_{\rm max}$, showing that the scales at which this contribution is important are mostly linear ($q<q_{nl}\sim 0.2 h {\rm Mpc^{-1}}$). 
This suggests the nonlinear corrections are likely to be small, especially at higher redshifts.

We will also explore simple extensions beyond the 1-loop. Since the integral in equation \ref{S} extends into the nonlinear regime its high $q$ contribution may not be reliable. We expect the response to high $q$ modes to be suppressed in functional derivatives, so their effect is reduced 
\cite{2014arXiv1411.2970N}. The transition is governed by the nonlinear scale $q_{nl}$ where $\Delta^2(q)=\int q^2dqP(q)W_R^2(q)/2\pi^2 = 1.35^2$, 
where $W_R$ is a gaussian window $W_R=\exp[-(0.46q/q_{nl})^2/2]$, so the expectation is that $q_{nl} \sim 0.3 h {\rm Mpc^{-1}}$ at $z=0$. 
Following \cite{2014arXiv1411.2970N} we will assume a Lorentzian damping form and explore the 1-loop expression

\begin{equation}
                \mathbf{Cov}_{ij}^{\rm NL} = \left(
                                                \dfrac{1}{V\pi^2} 
                                                \int {P^2_{\rm Lin}(q)\over (1+ (q/q_{nl})^2)^2} q^3
                                                                        \mathbf{V}(q,k_i) \mathbf{V}(q,k_j)
                                                                        d\ln q 
                                             \right).
                                                                        \label{eqn:cov1loopd}
\end{equation}
\\
Figure \ref{fig:p2k2} also shows the integral of $S$ along with the damping for $z=0$. We see that damping slightly reduces the value of $S$ at $z=0$. 
The damping effect is even smaller for higher redshifts. 


\section{Comparison with simulations}\label{sec:sims}

\begin{table}
    \begin{center}
    \begin{tabular}{ | l | c c c  c  c  c  c  c | }
      \hline
      Parameter & Number of simulations & Boxsize ($h^{-1}$Mpc) & $\Omega_m$ & $\sigma_8$ & $h$ & $n_s$ & $\Omega_b$ & $w$\\
      \hline
      L14 & 3584 & 500 & 0.286 & 0.82 & 0.7 & 0.96 & 0.047 & -1.0\\
      B15 & 12288 & 650 & 0.257 & 0.801 & 0.72 & 0.963 & 0.044 & -1.0 \\
      \hline
    \end{tabular}
    \end{center}
    \caption{Specifications for the two sets of the cosmological simulations used for comparison.}
    \label{tbl:cosmo}
\end{table}

We used two sets of cosmological simulations, from \cite[][B15 hereafter]{2015MNRAS.446.1756B} and \cite[][L14 hereafter]{2014PhRvD..89h3519L}, to compare with our model. For both simulations, the cosmology is the flat $\Lambda$CDM with particular cosmological parameters values listed in table \ref{tbl:cosmo}. The simulation suite from L14 also contains an explicit calculation of the super-sample covariance term measured using a corresponding separate universe simulations. B15 have a much larger volume, with a total effective volume greater than 3000 $h^{-3}{\rm Gpc}^3$. L14 gives 3584 matter power spectra at redshift zero, which we use to generate a full covariance matrix. B15 provides their covariance matrix at four different redshifts: 0.0, 0.5, 1.0, 2.0.

\begin{figure}
    \centering
    \includegraphics[width=0.47\textwidth]{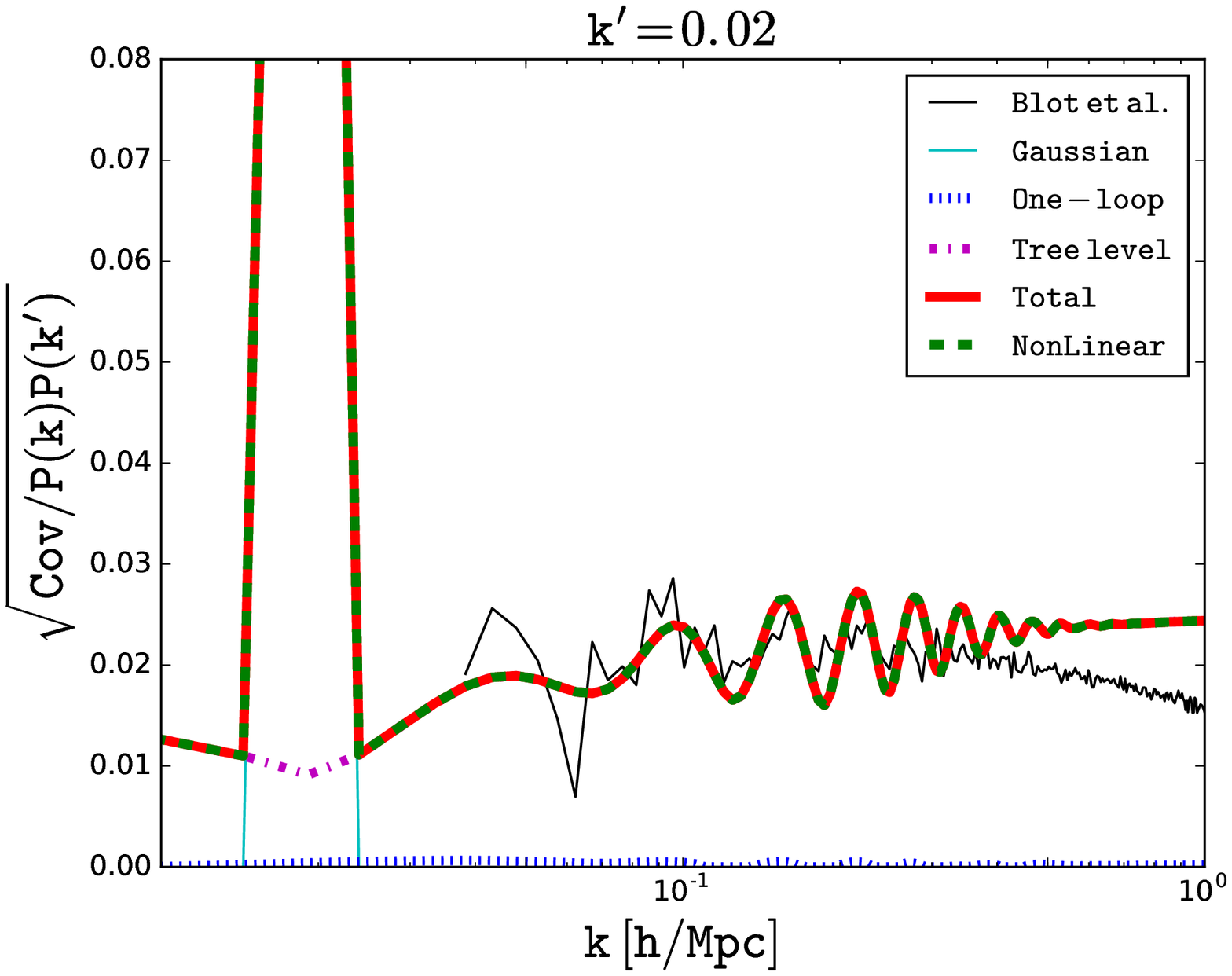}
    \includegraphics[width=0.47\textwidth]{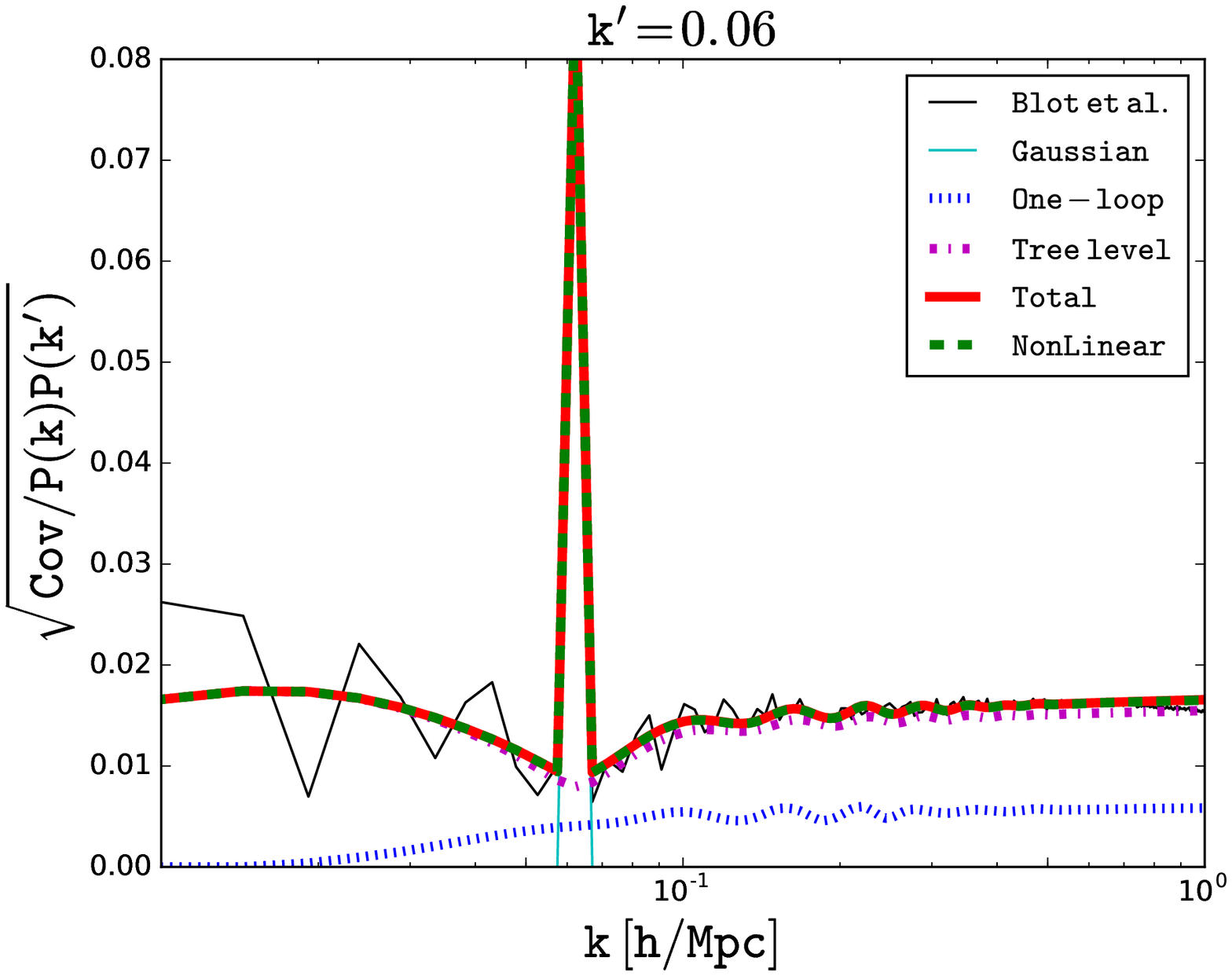}
    \includegraphics[width=0.47\textwidth]{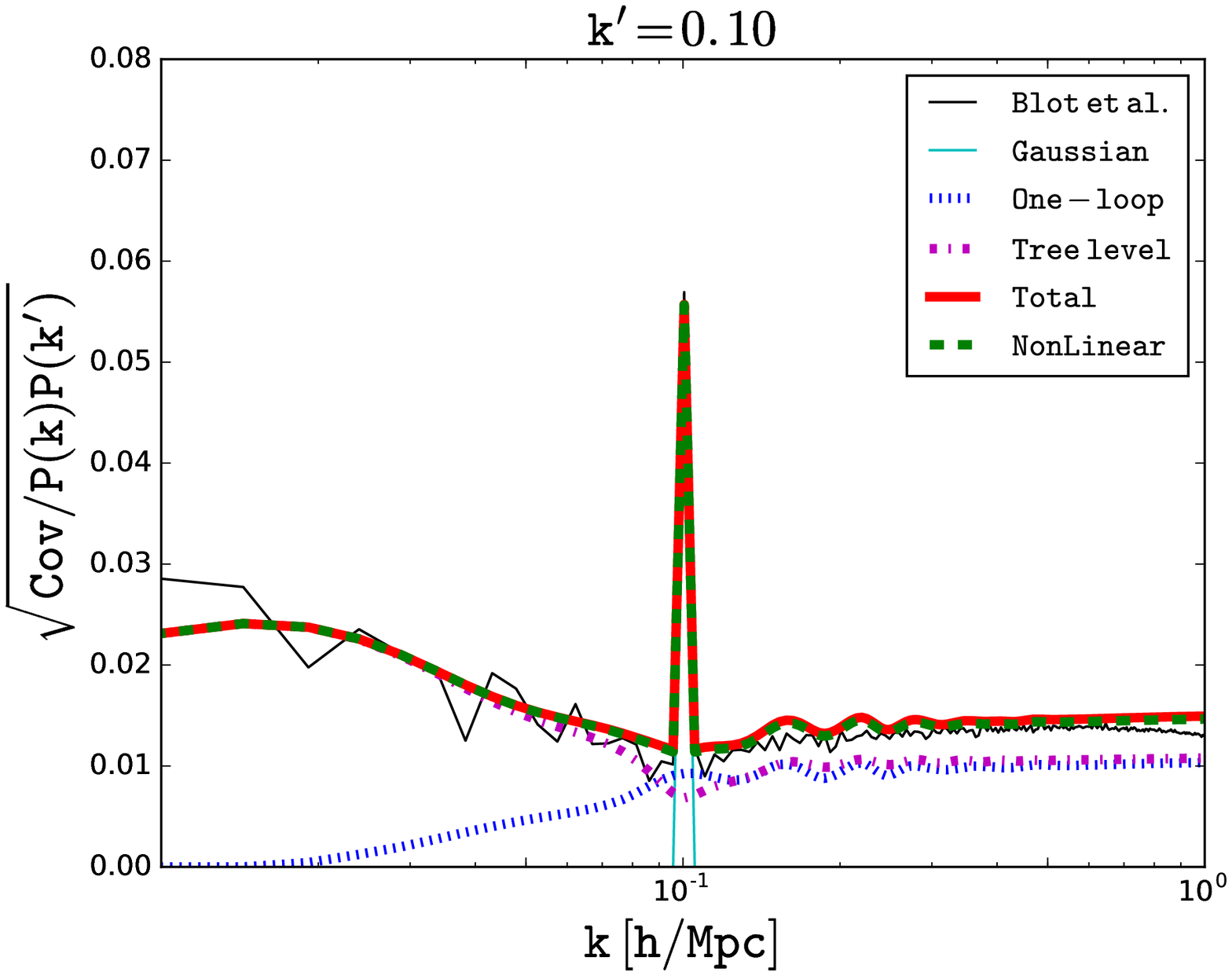}
    \includegraphics[width=0.47\textwidth]{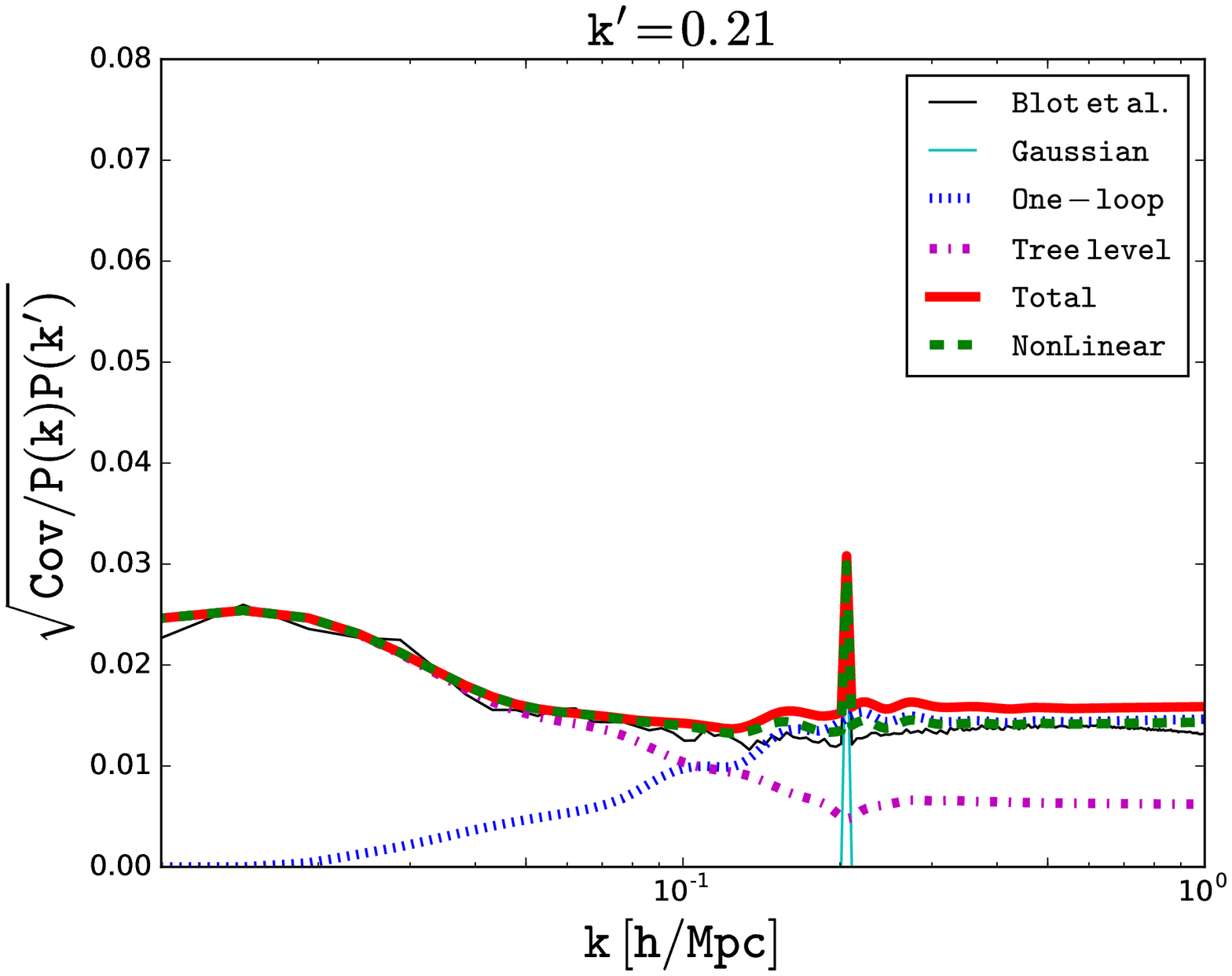}
    \includegraphics[width=0.47\textwidth]{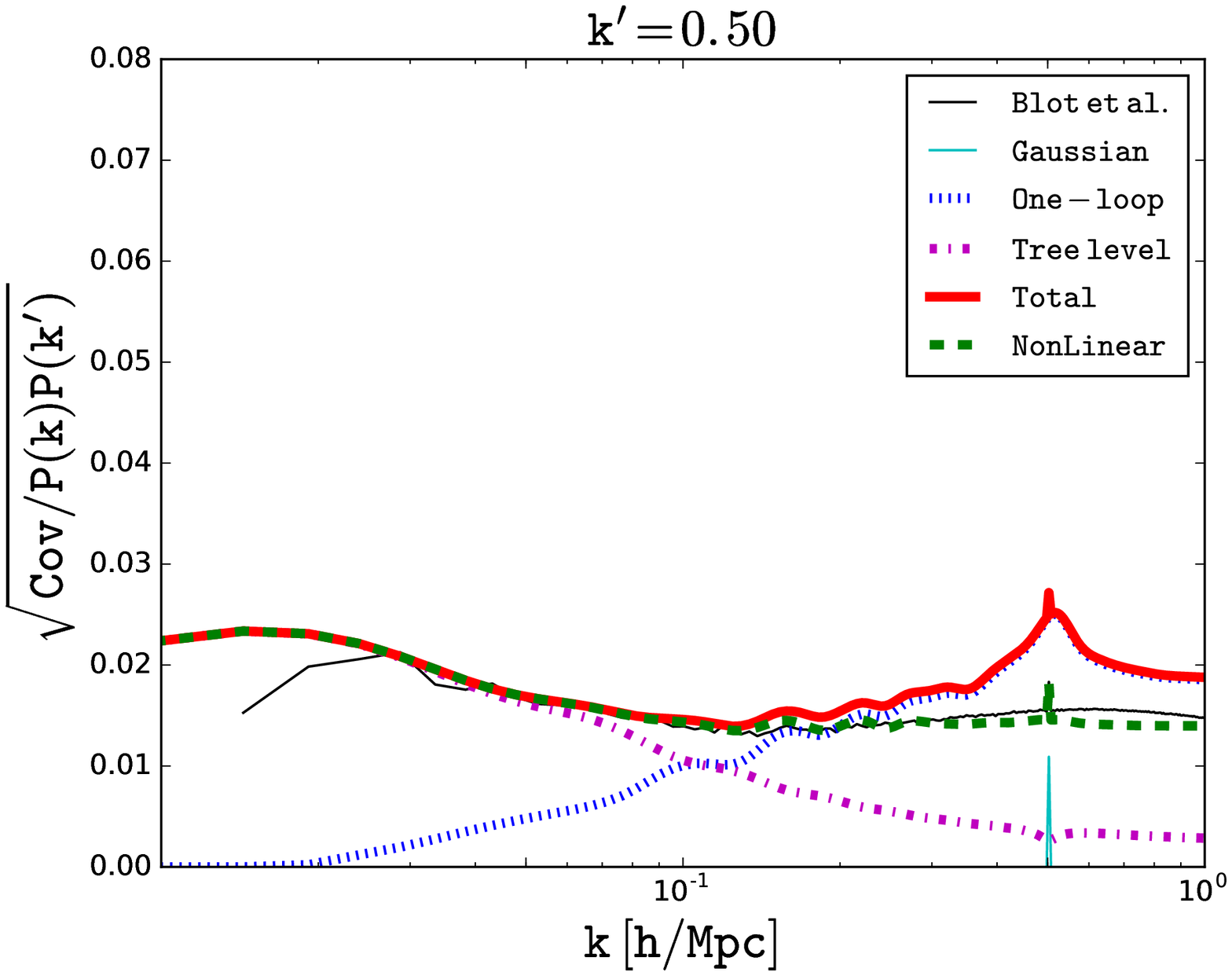}
    \includegraphics[width=0.47\textwidth]{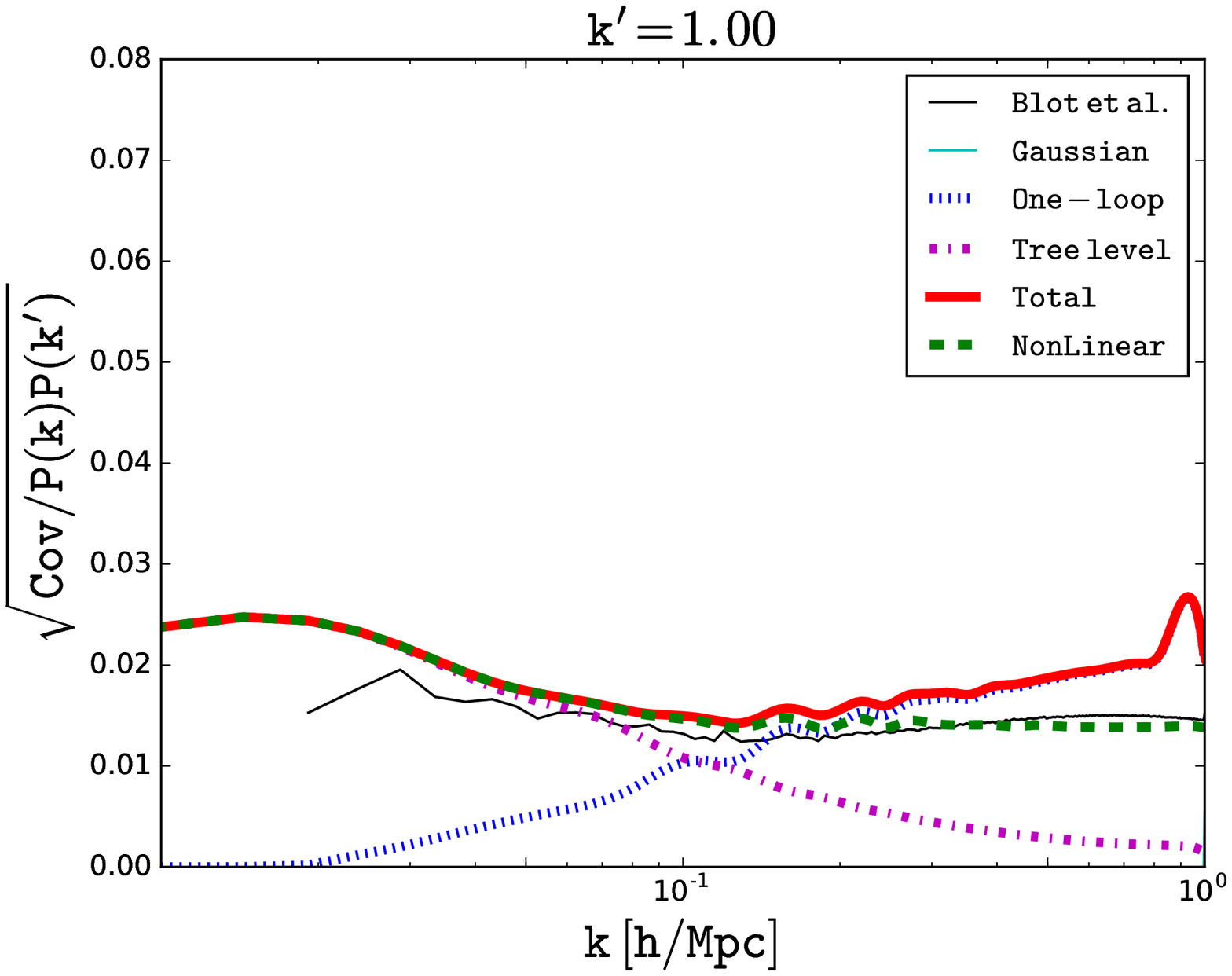}
    \caption{Comparing analytic model with B15 covariance using 1-loop exact functional derivatives at redshift 0.0.}
    \label{fig:blot_1loop00}
\end{figure}

\begin{figure}
    \centering
    \includegraphics[width=0.47\textwidth]{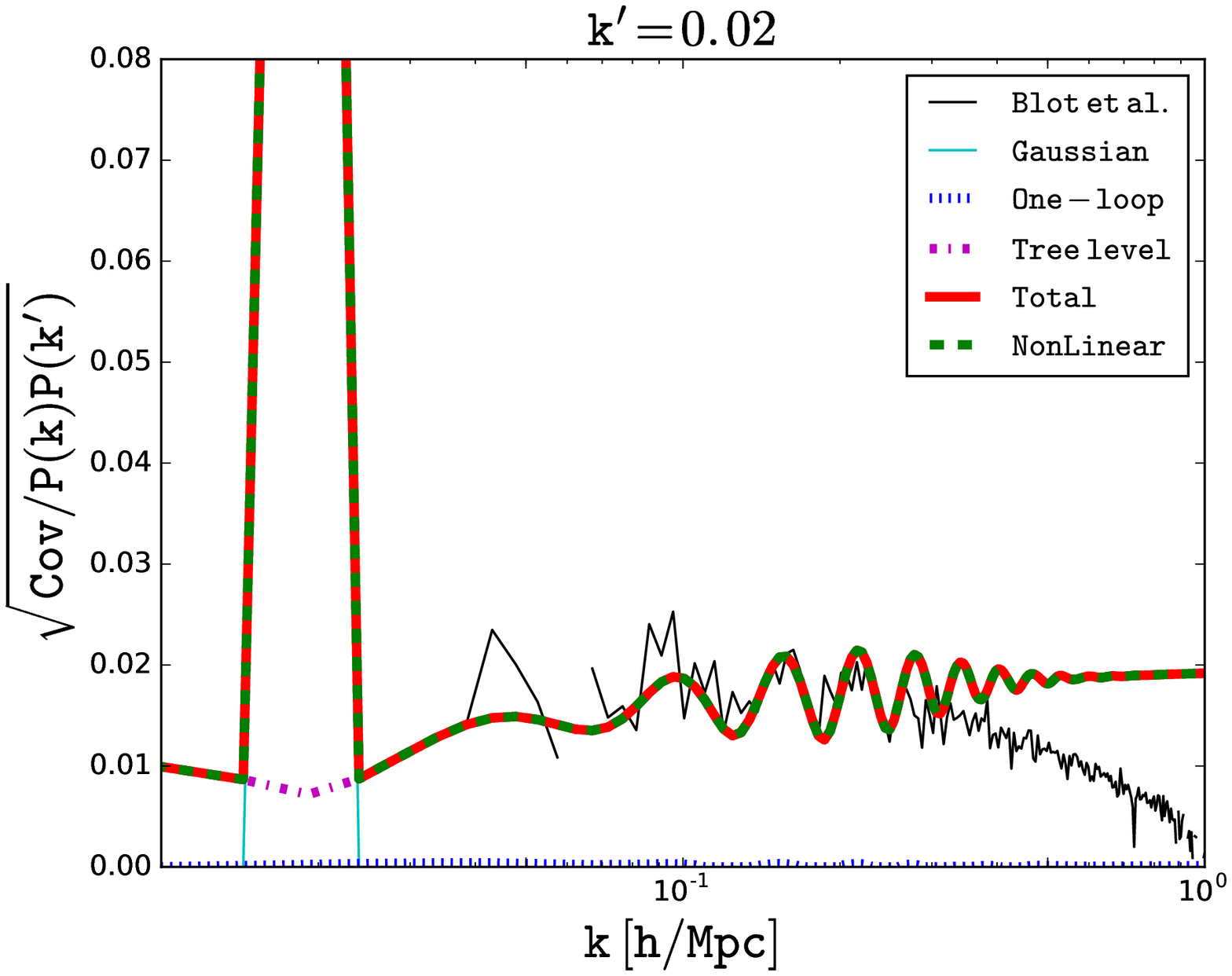}
    \includegraphics[width=0.47\textwidth]{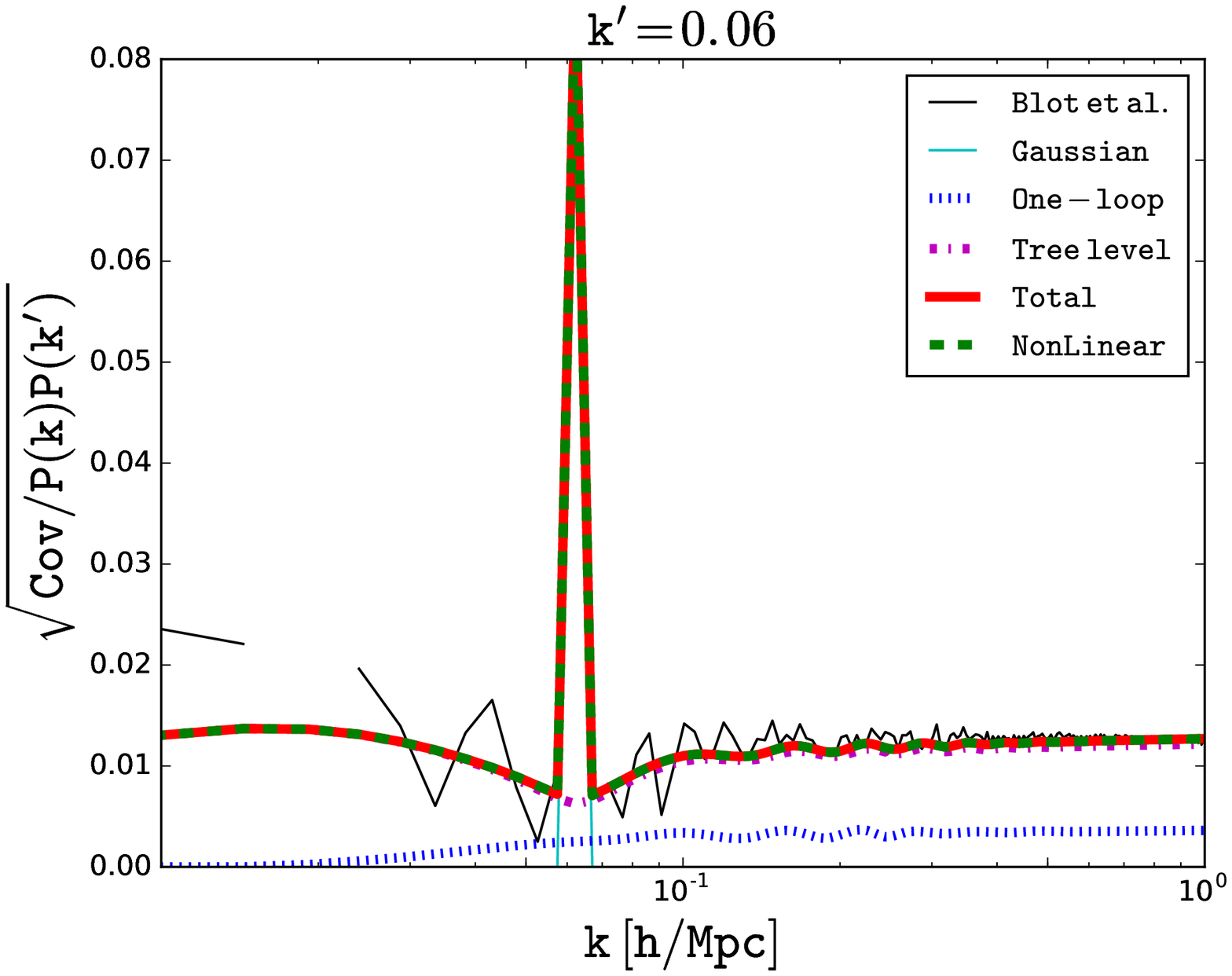}
    \includegraphics[width=0.47\textwidth]{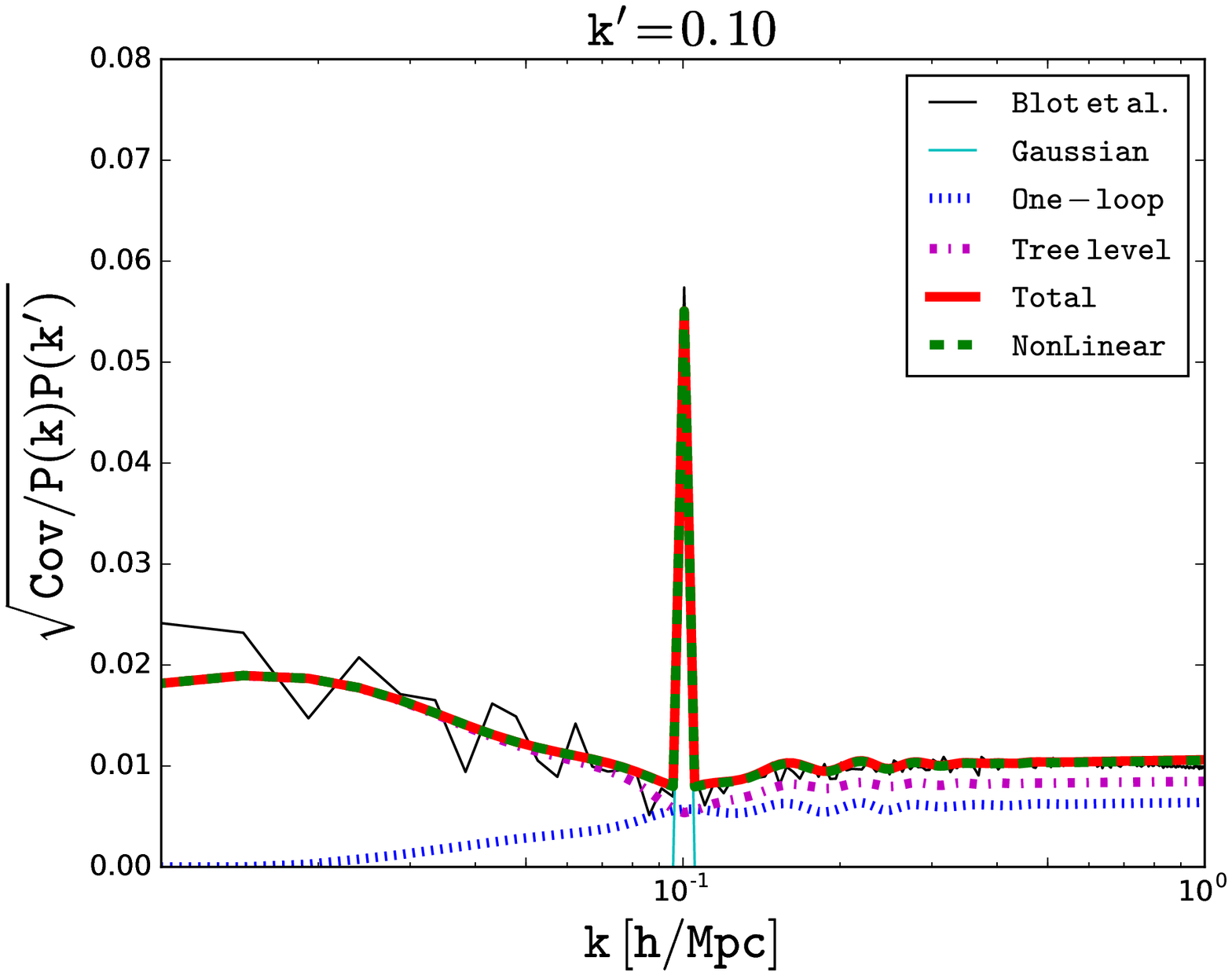}
    \includegraphics[width=0.47\textwidth]{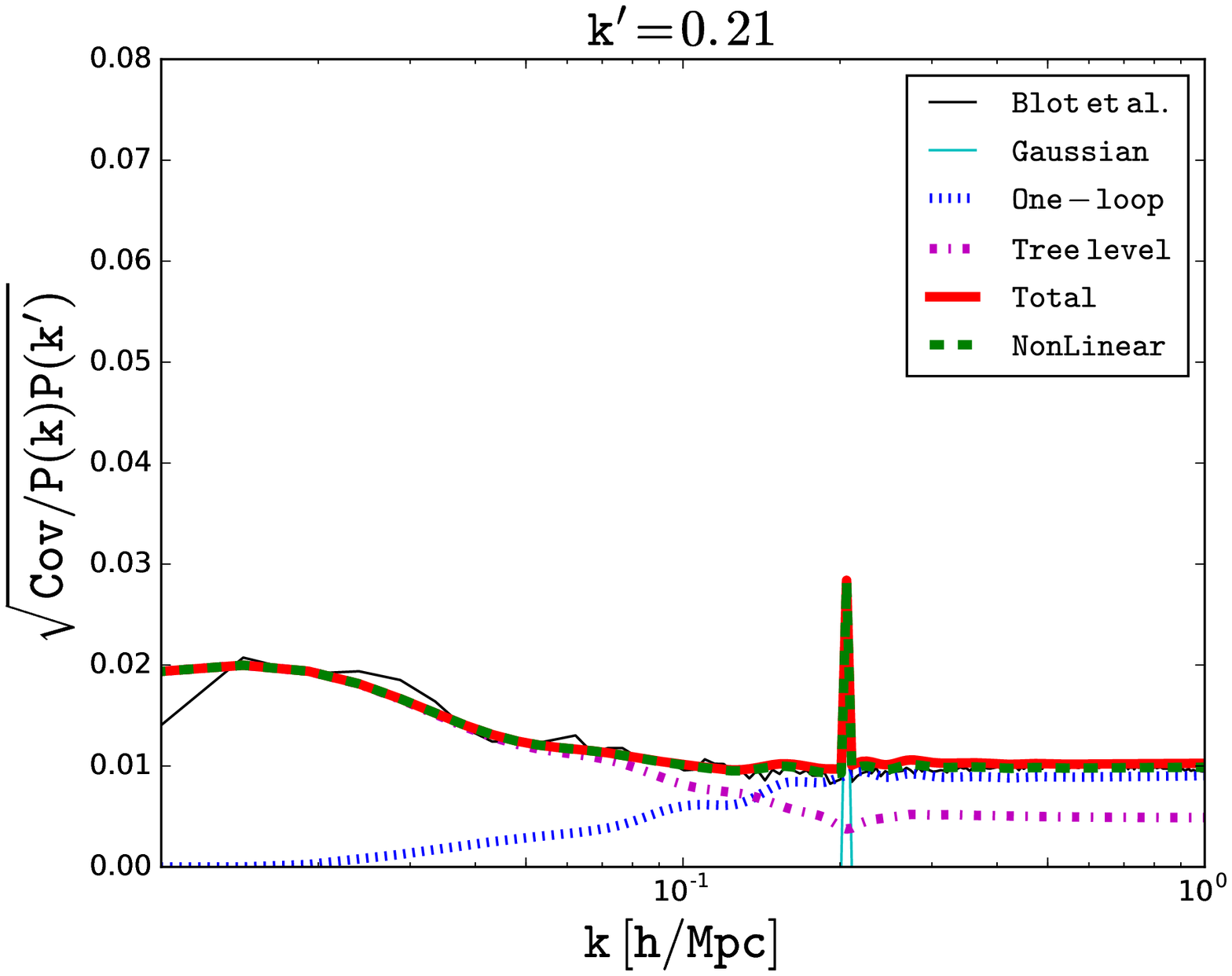}
    \includegraphics[width=0.47\textwidth]{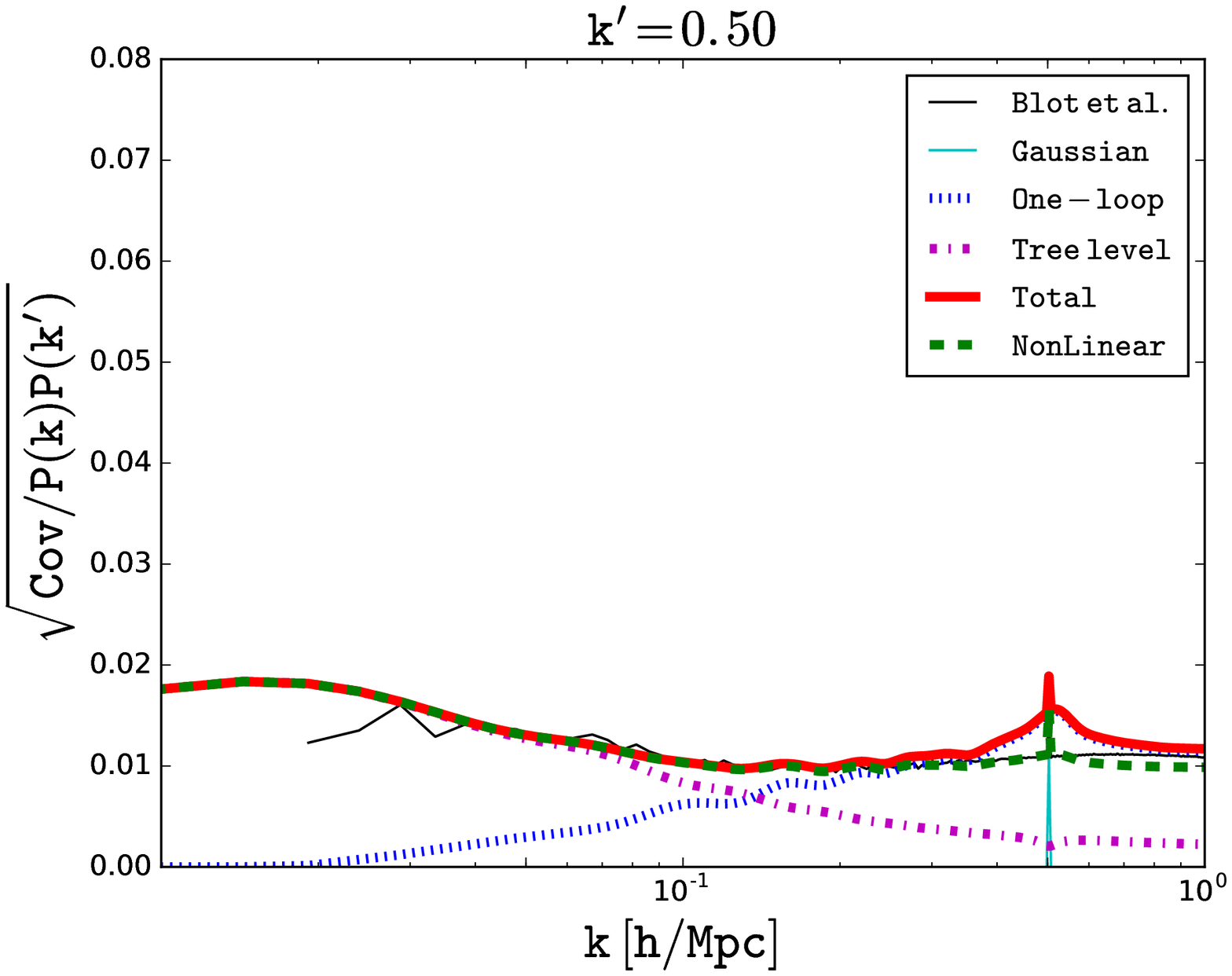}
    \includegraphics[width=0.47\textwidth]{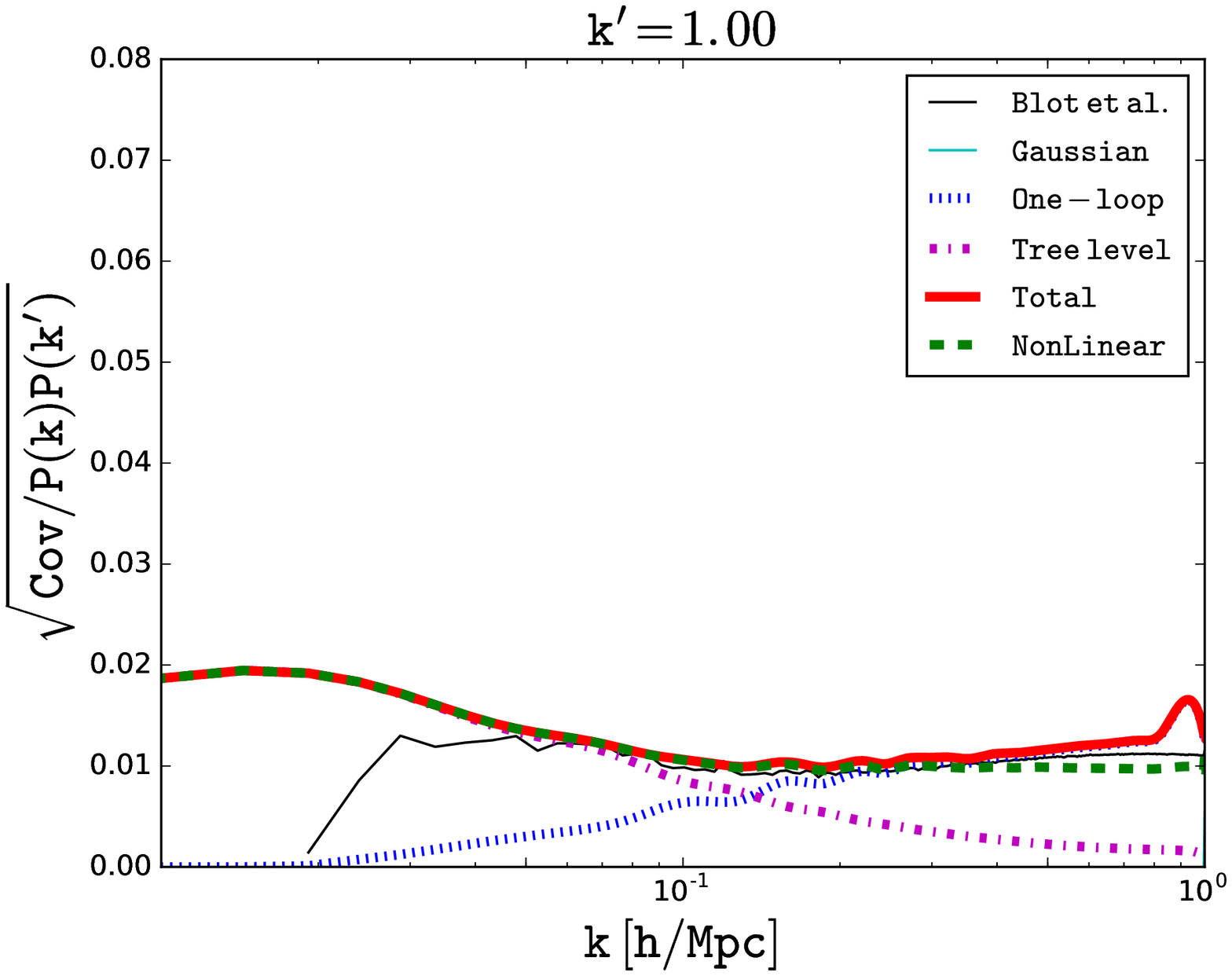}
    \caption{Comparing analytic model with B15 covariance using 1-loop exact functional derivatives at redshift 0.5.}
    \label{fig:blot_1loop05}
\end{figure}

\begin{figure}
    \centering
    \includegraphics[width=0.47\textwidth]{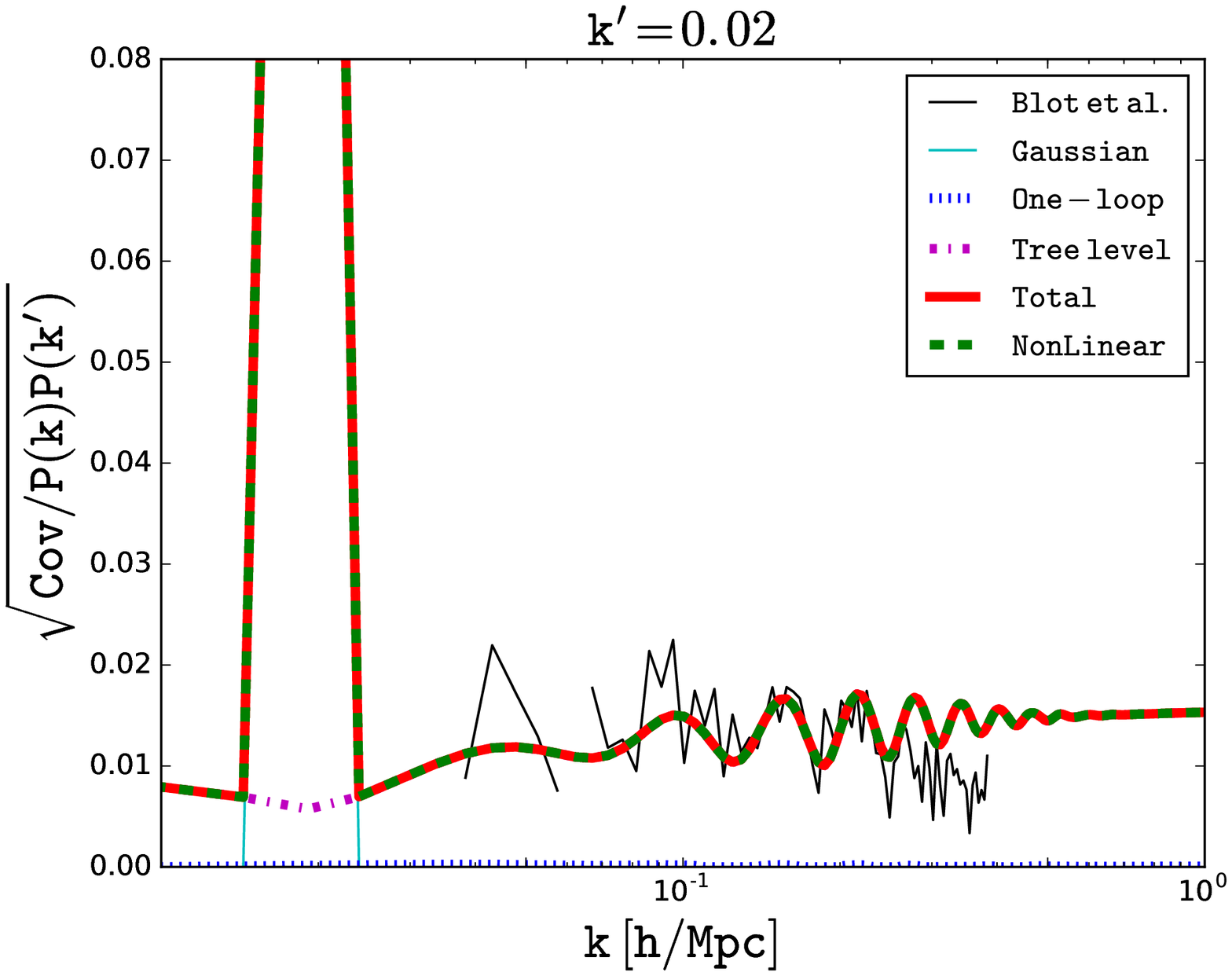}
    \includegraphics[width=0.47\textwidth]{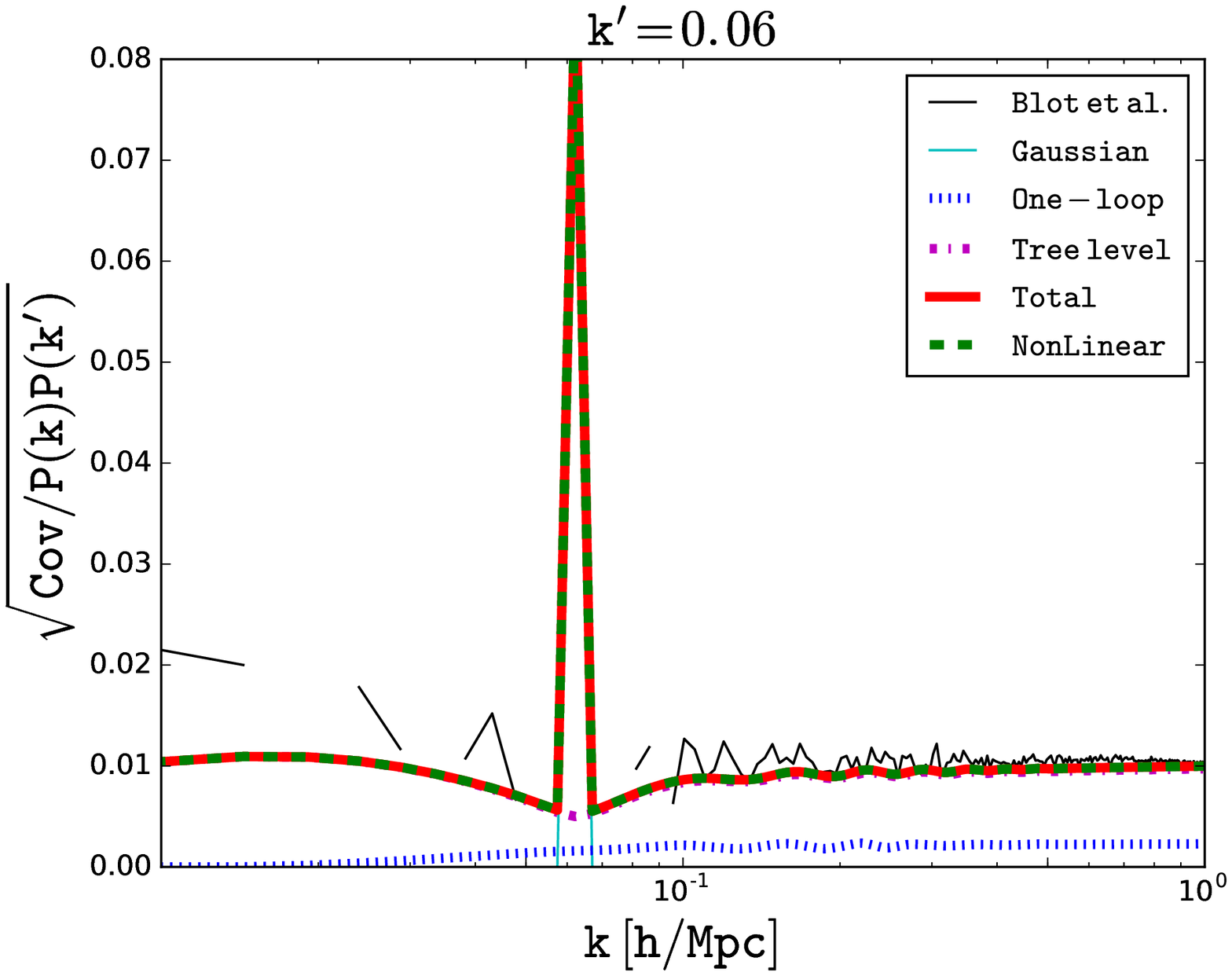}
    \includegraphics[width=0.47\textwidth]{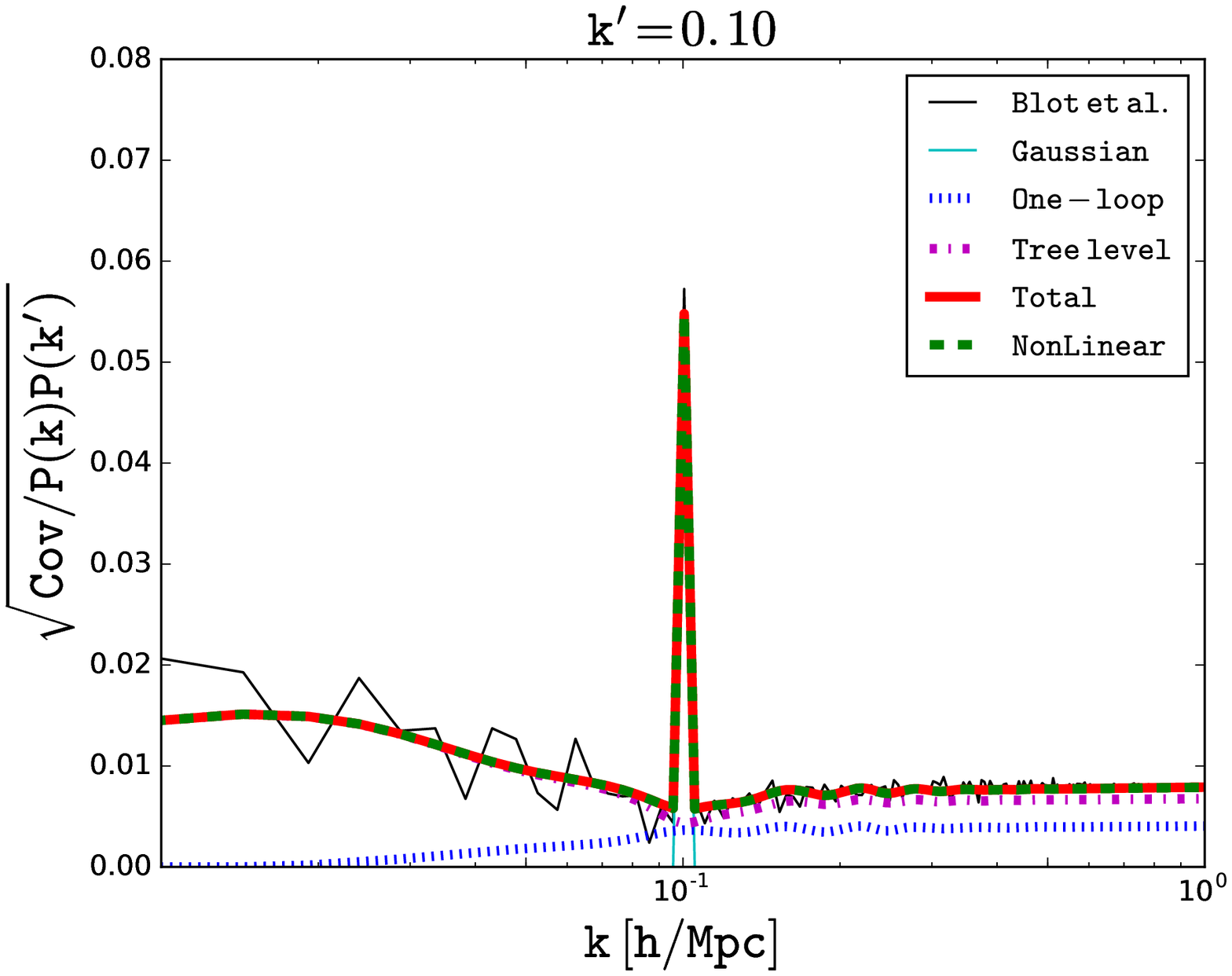}
    \includegraphics[width=0.47\textwidth]{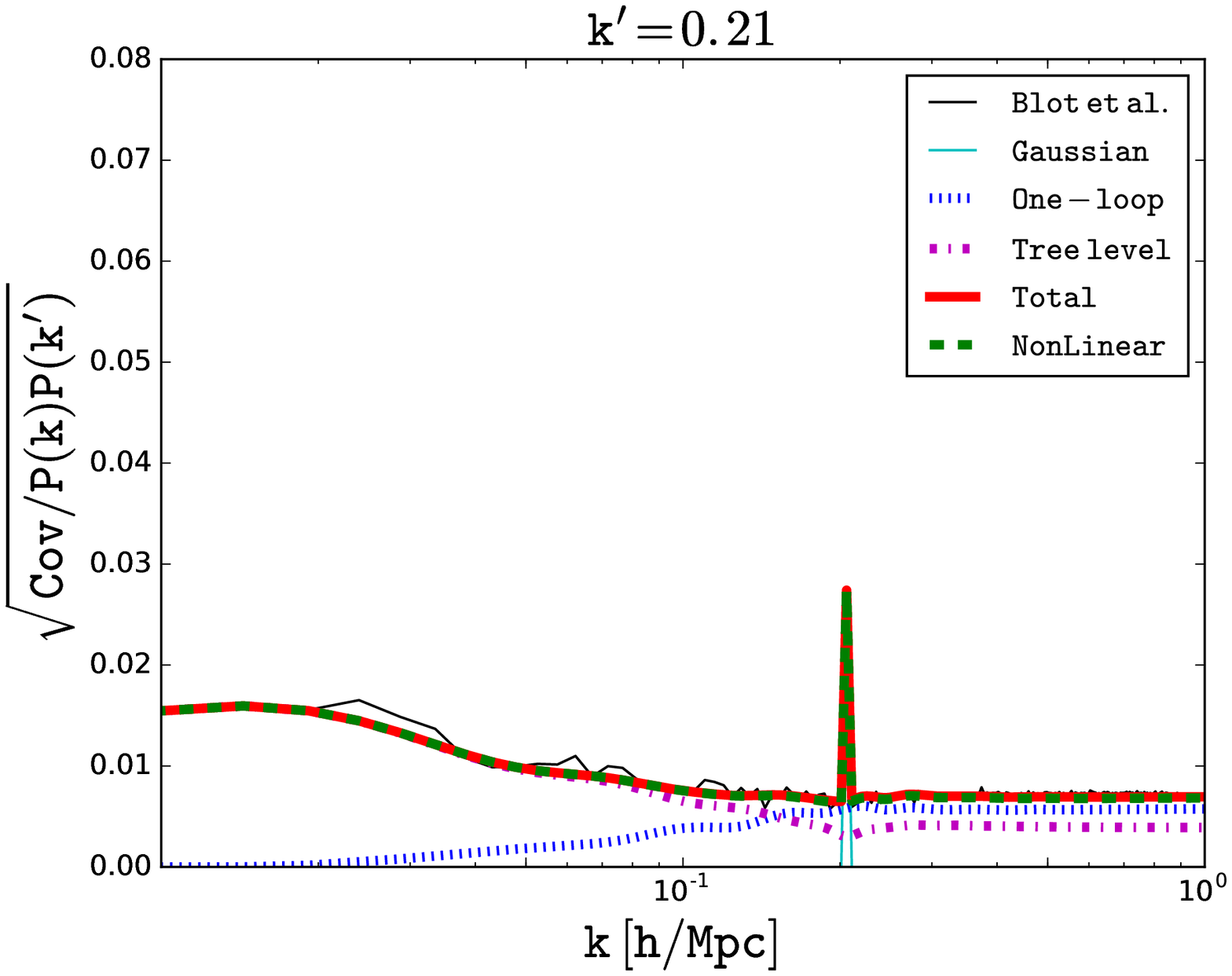}
    \includegraphics[width=0.47\textwidth]{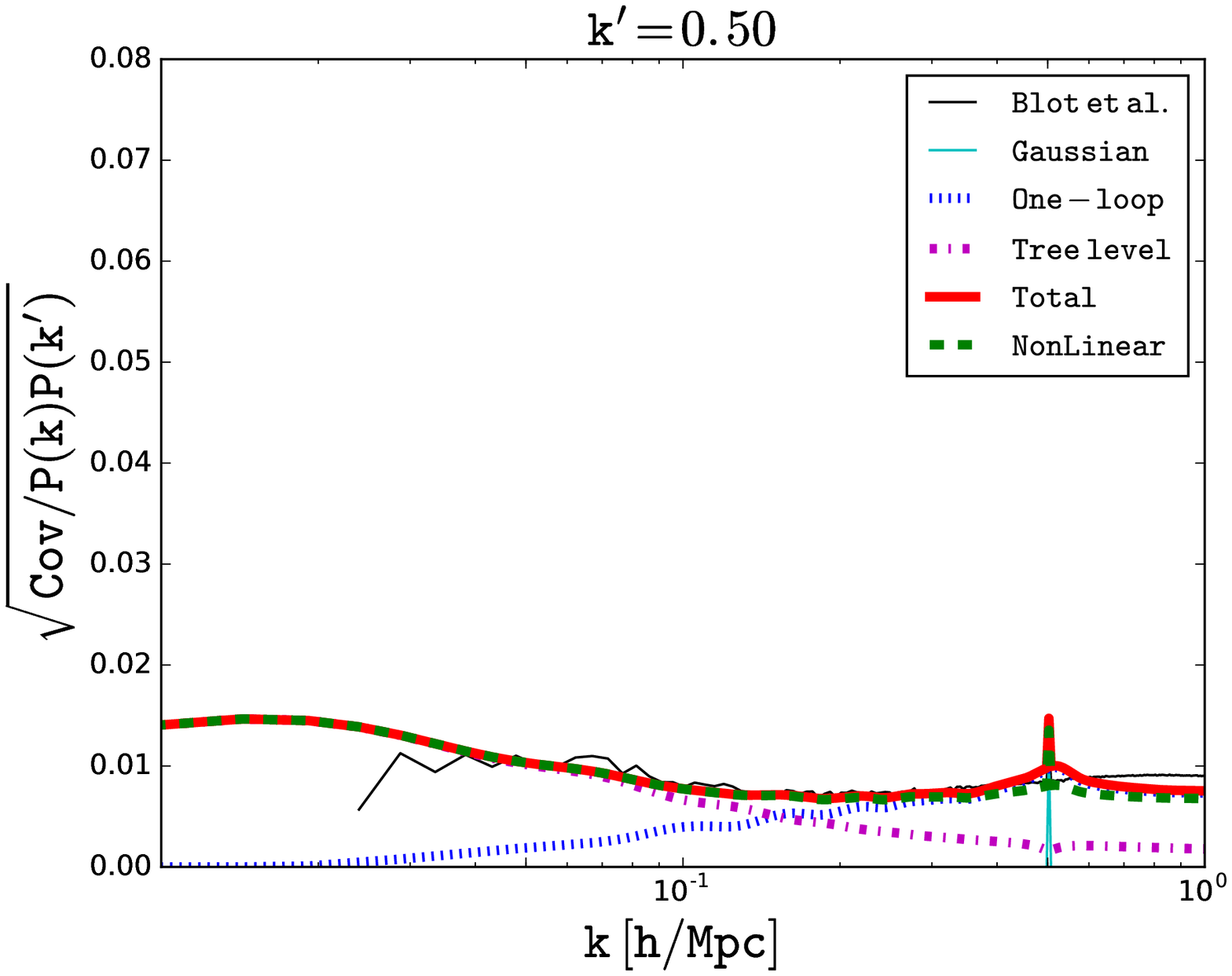}
    \includegraphics[width=0.47\textwidth]{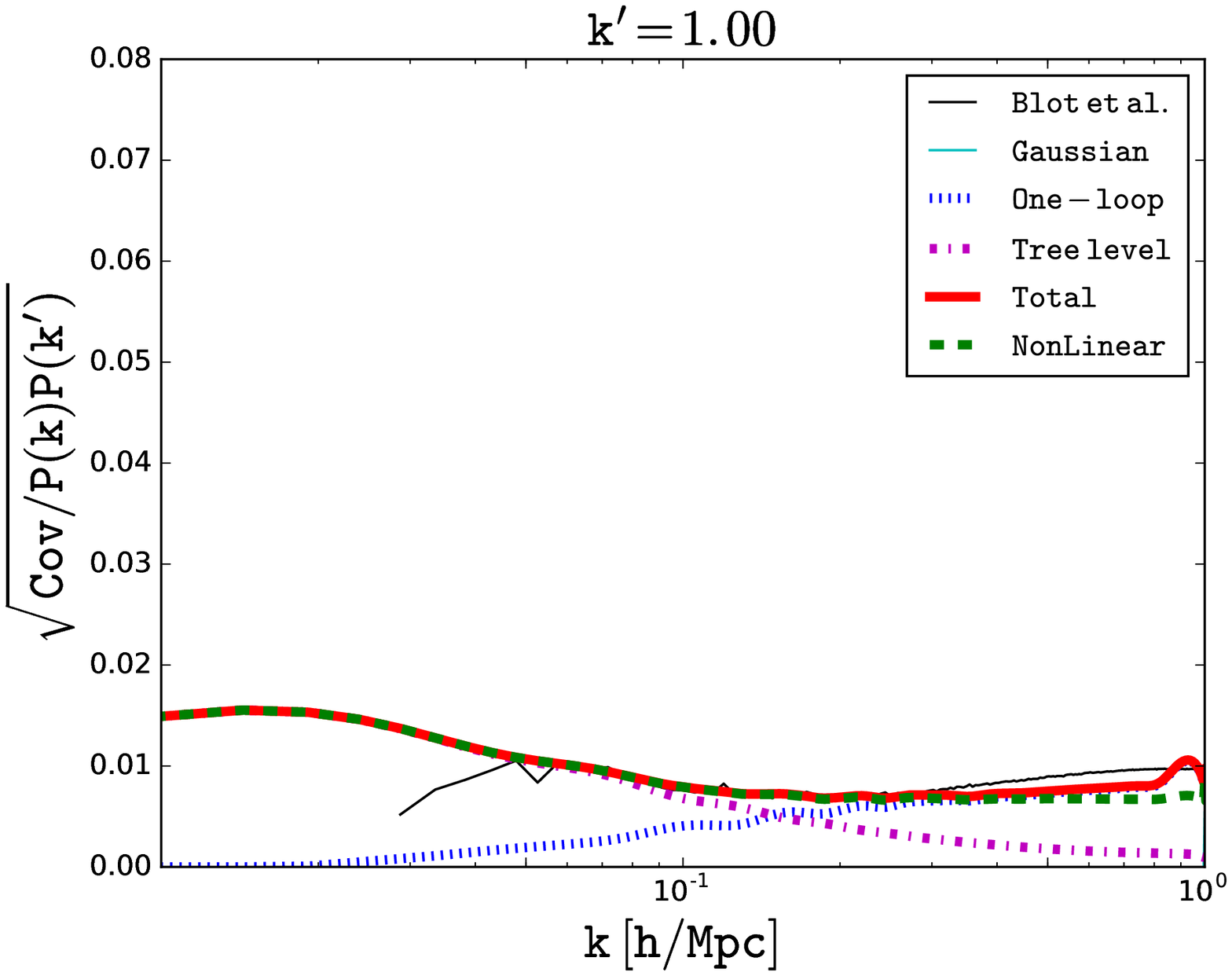}
    \caption{Comparing analytic model with B15 covariance using 1-loop exact functional derivatives at redshift 1.0.}
    \label{fig:blot_1loop10}
\end{figure}

\begin{figure}
    \centering
    \includegraphics[width=0.47\textwidth]{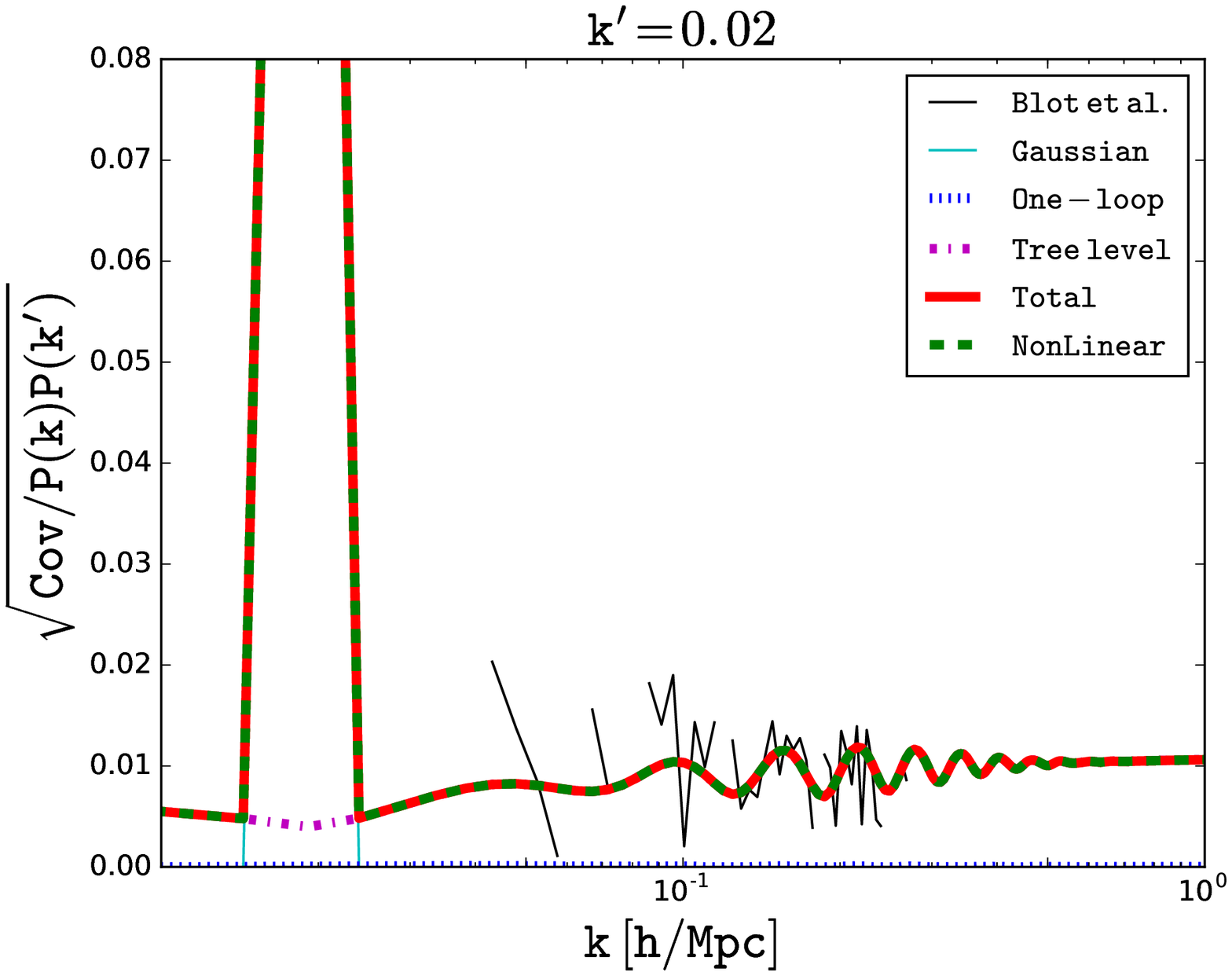}
    \includegraphics[width=0.47\textwidth]{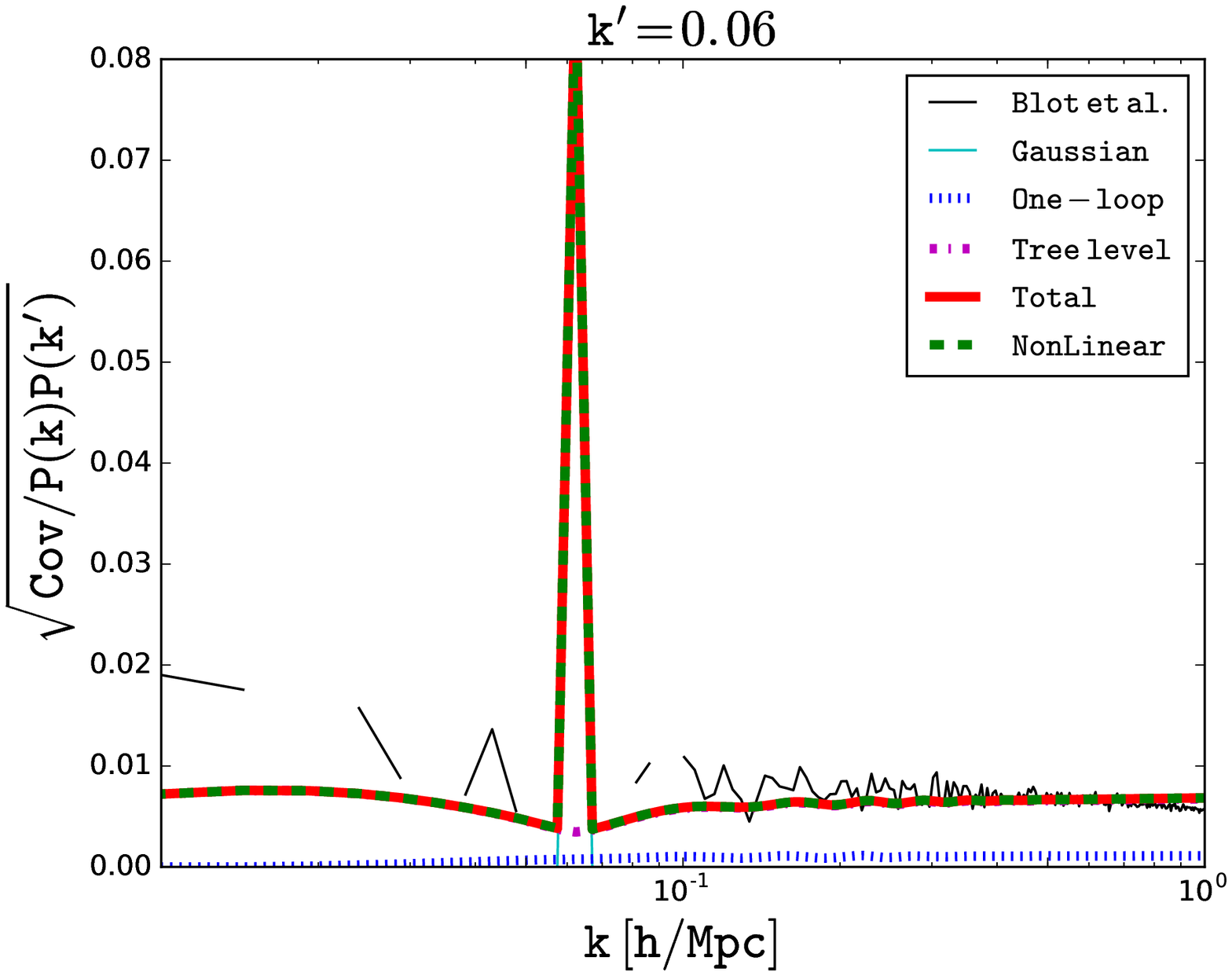}
    \includegraphics[width=0.47\textwidth]{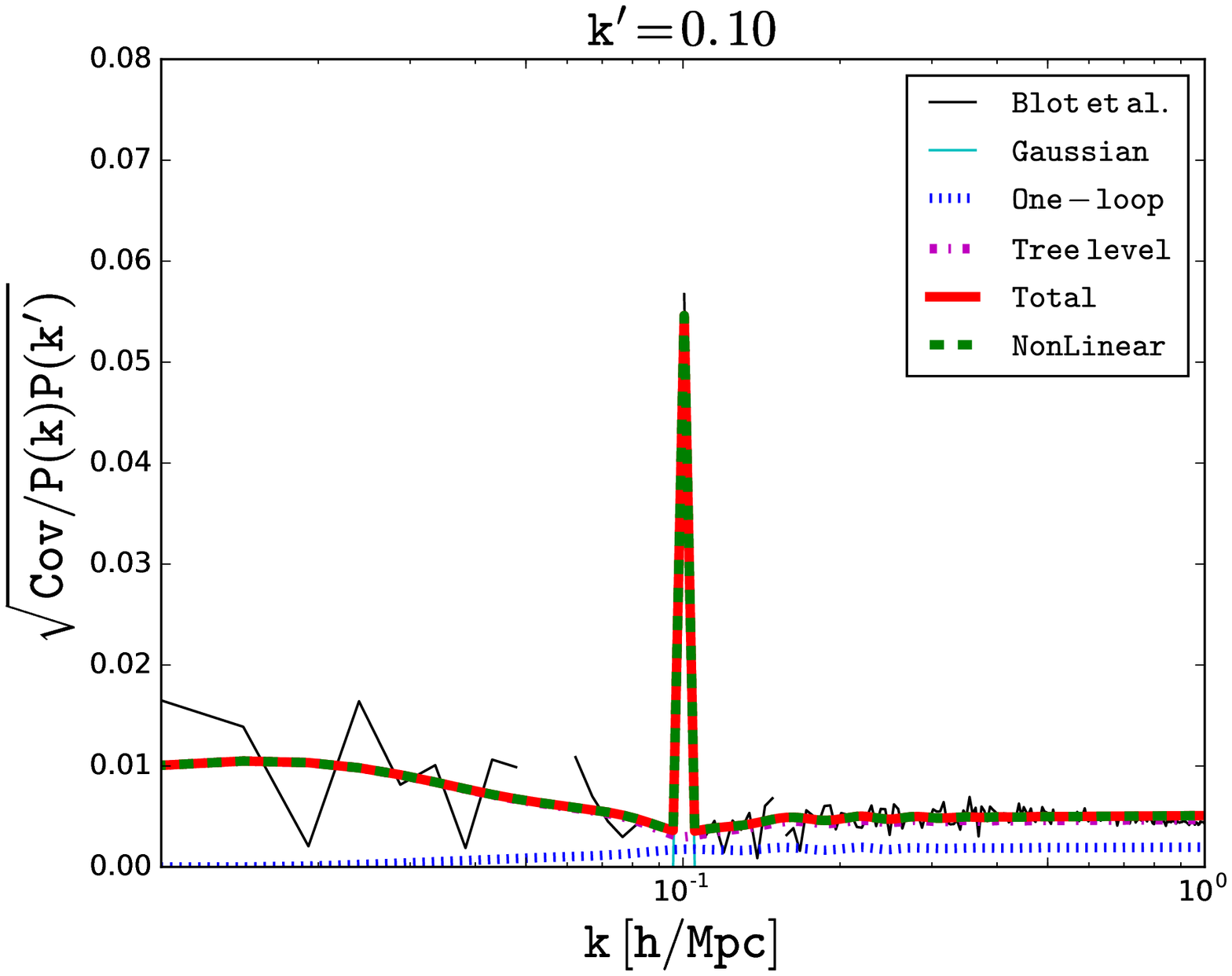}
    \includegraphics[width=0.47\textwidth]{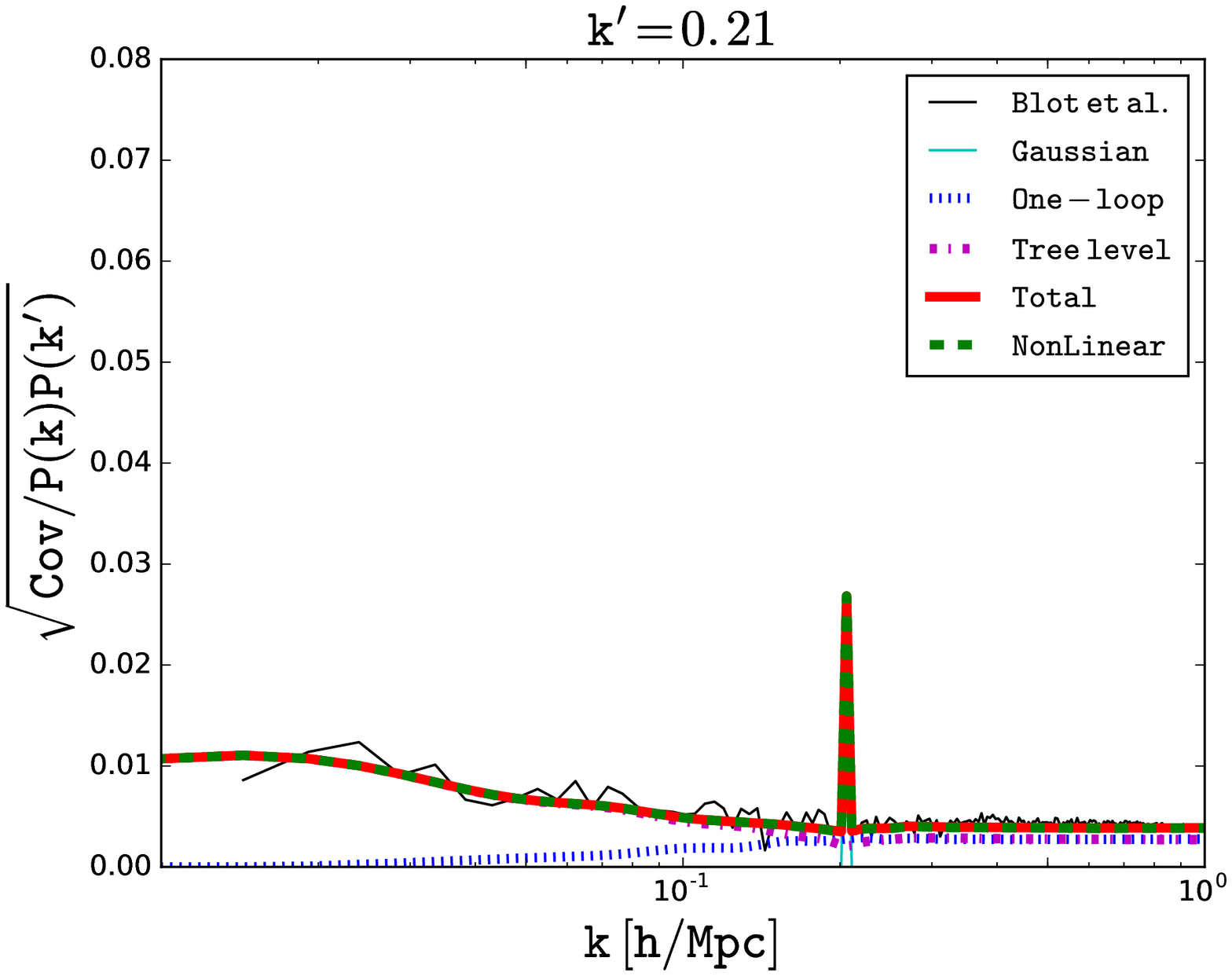}
    \includegraphics[width=0.47\textwidth]{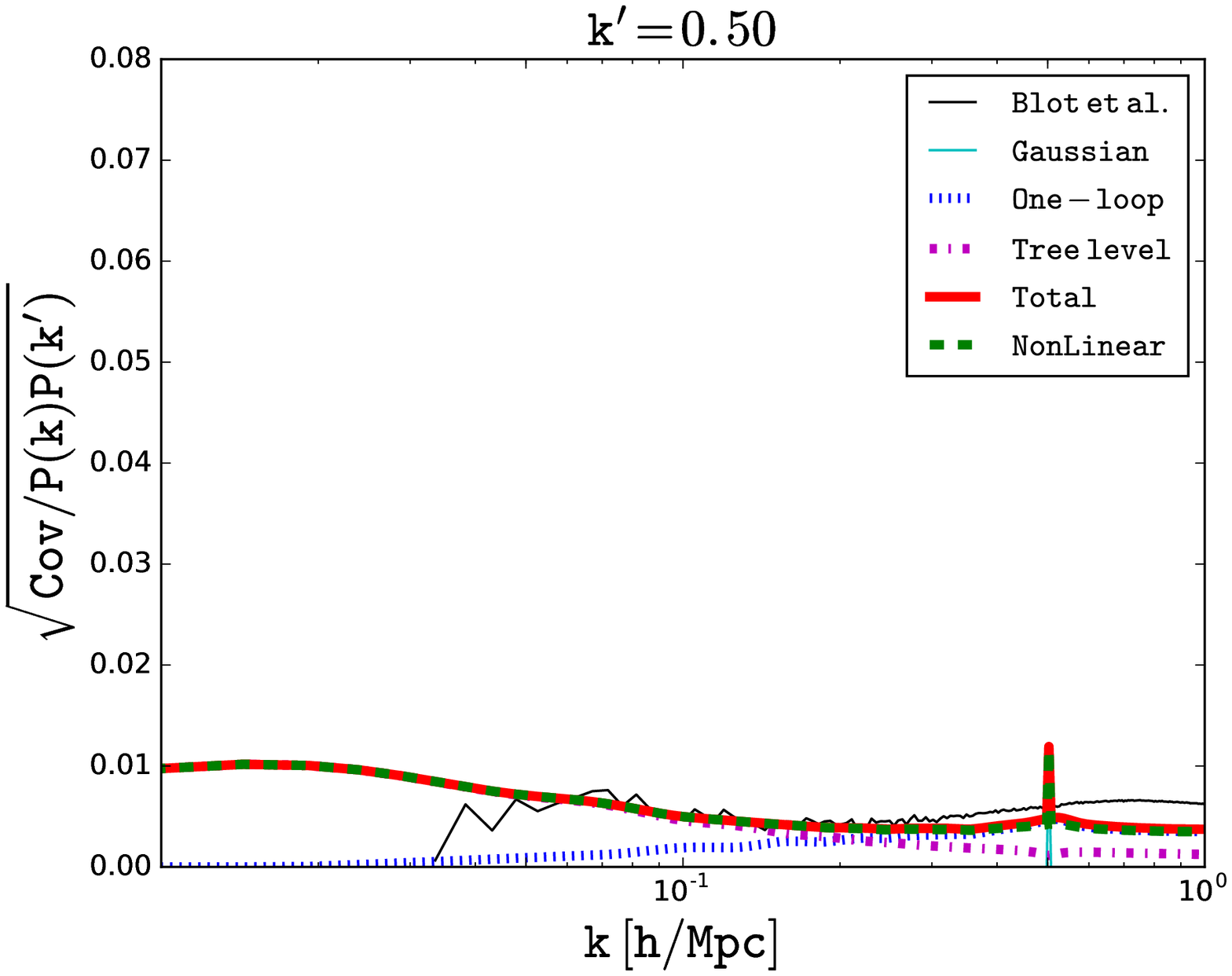}
    \includegraphics[width=0.47\textwidth]{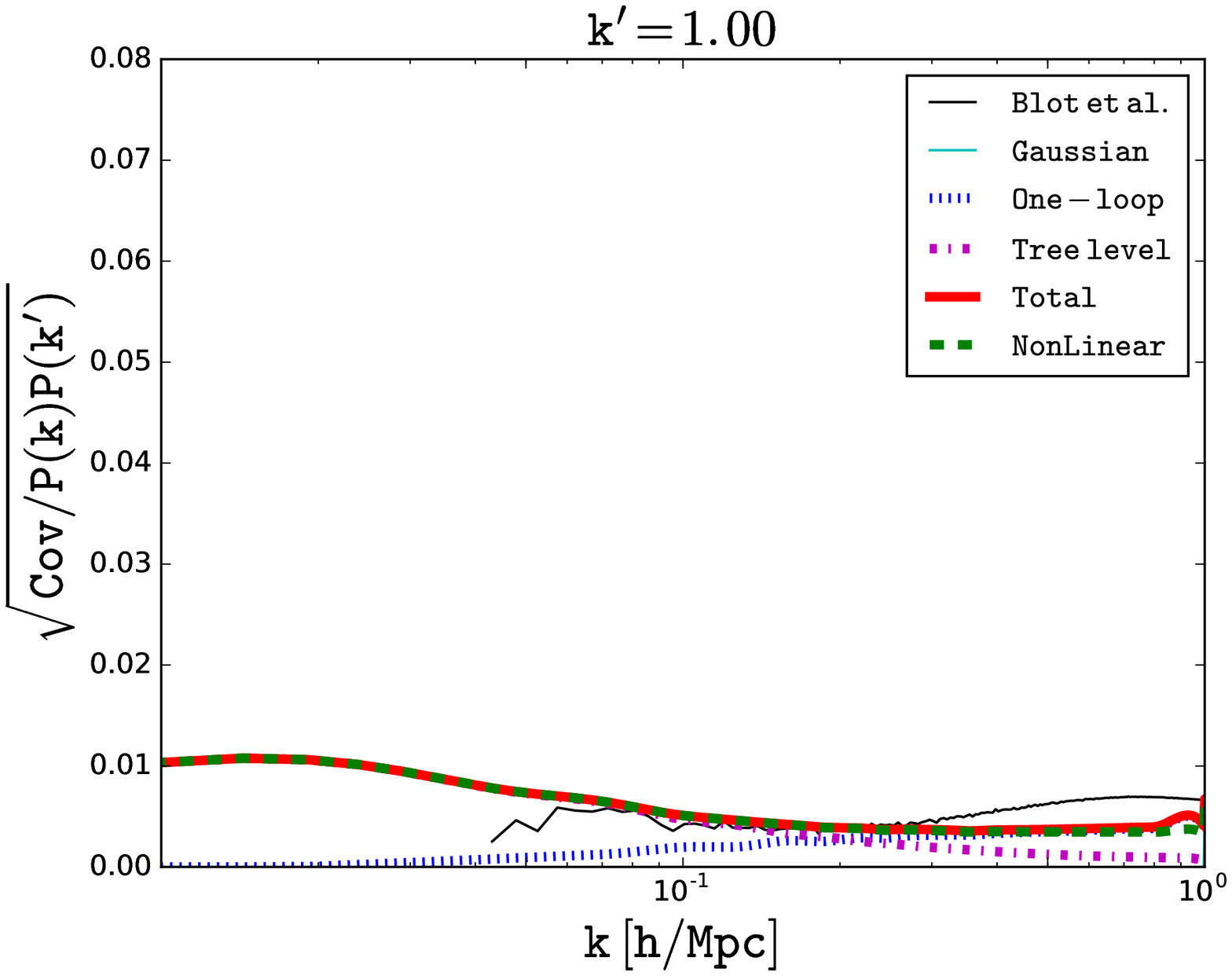}
    \caption{Comparing analytic model with B15 covariance using 1-loop exact functional derivatives at redshift 2.0.}
    \label{fig:blot_1loop20}
\end{figure}

Figure \ref{fig:blot_1loop00}, \ref{fig:blot_1loop05}, \ref{fig:blot_1loop10}, \ref{fig:blot_1loop20} shows the comparison of our model for the covariace matrix to B15 covariance matrix at redshifts 0.0, 0.5, 1.0, 2.0 respectively. Each panel shows various contributions (1-loop, tree-level, Gaussian) to the covariance, a full model and the corresponding simulation output.

\begin{figure}
    \centering
    \includegraphics[width=0.47\textwidth]{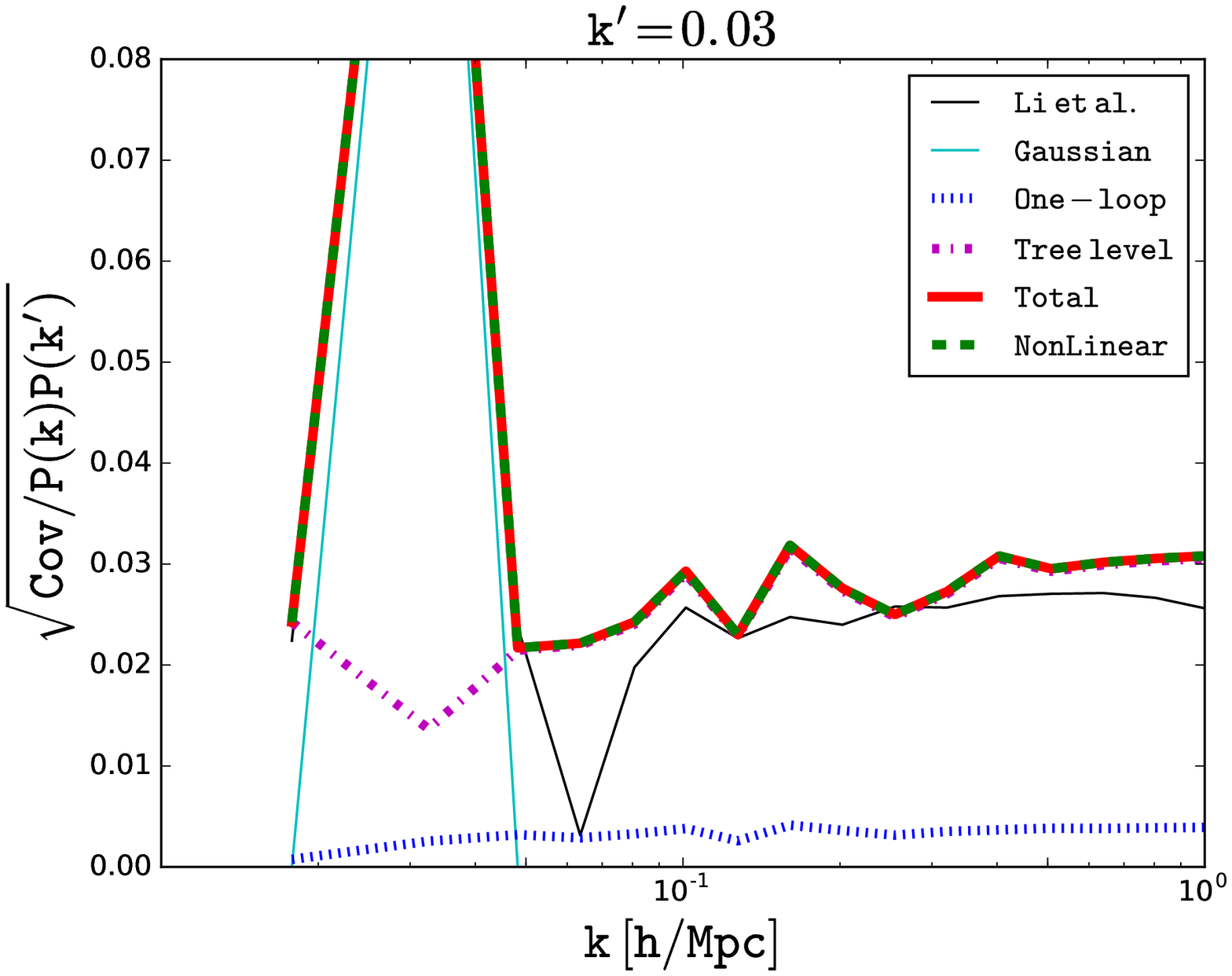}
    \includegraphics[width=0.47\textwidth]{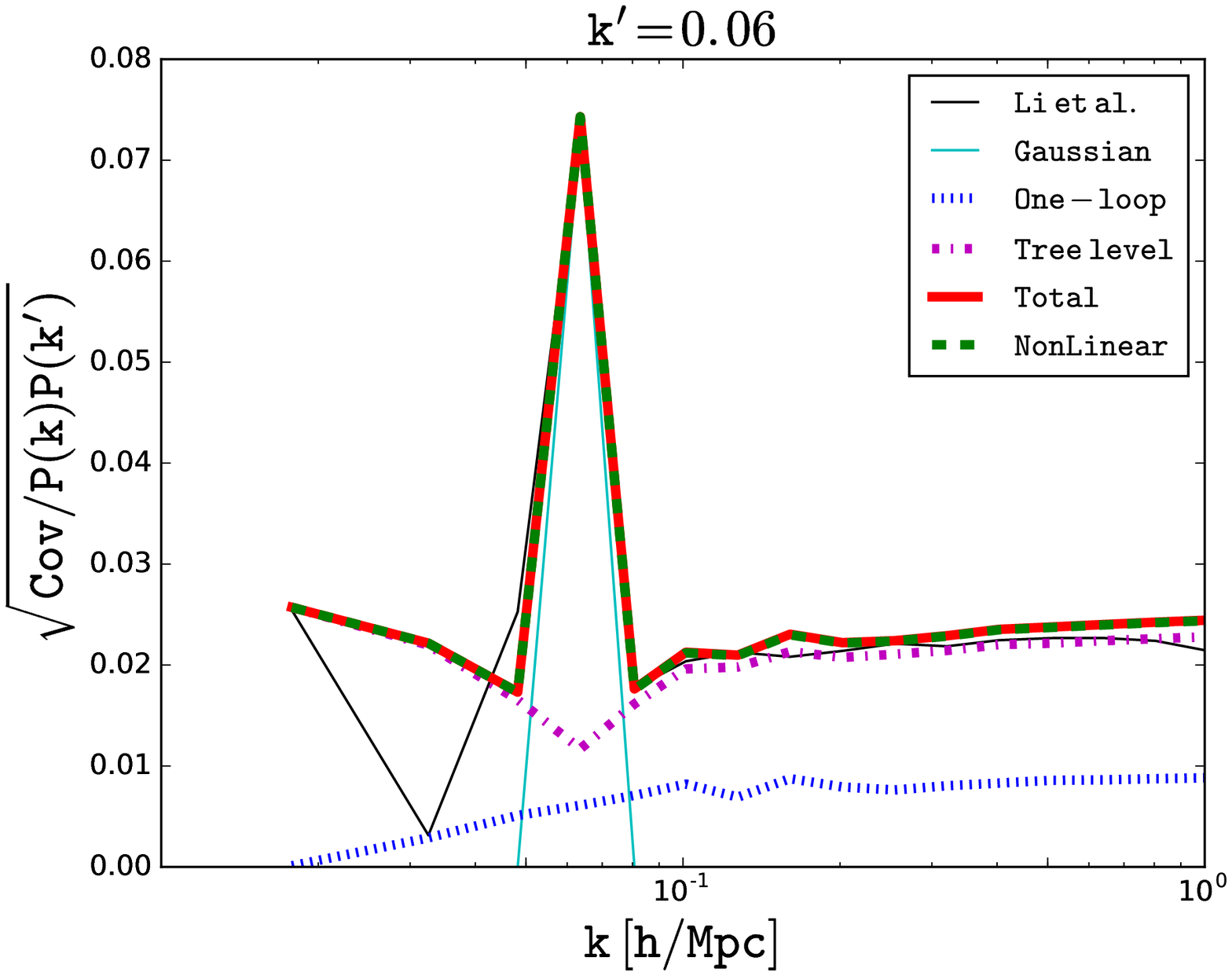}
    \includegraphics[width=0.47\textwidth]{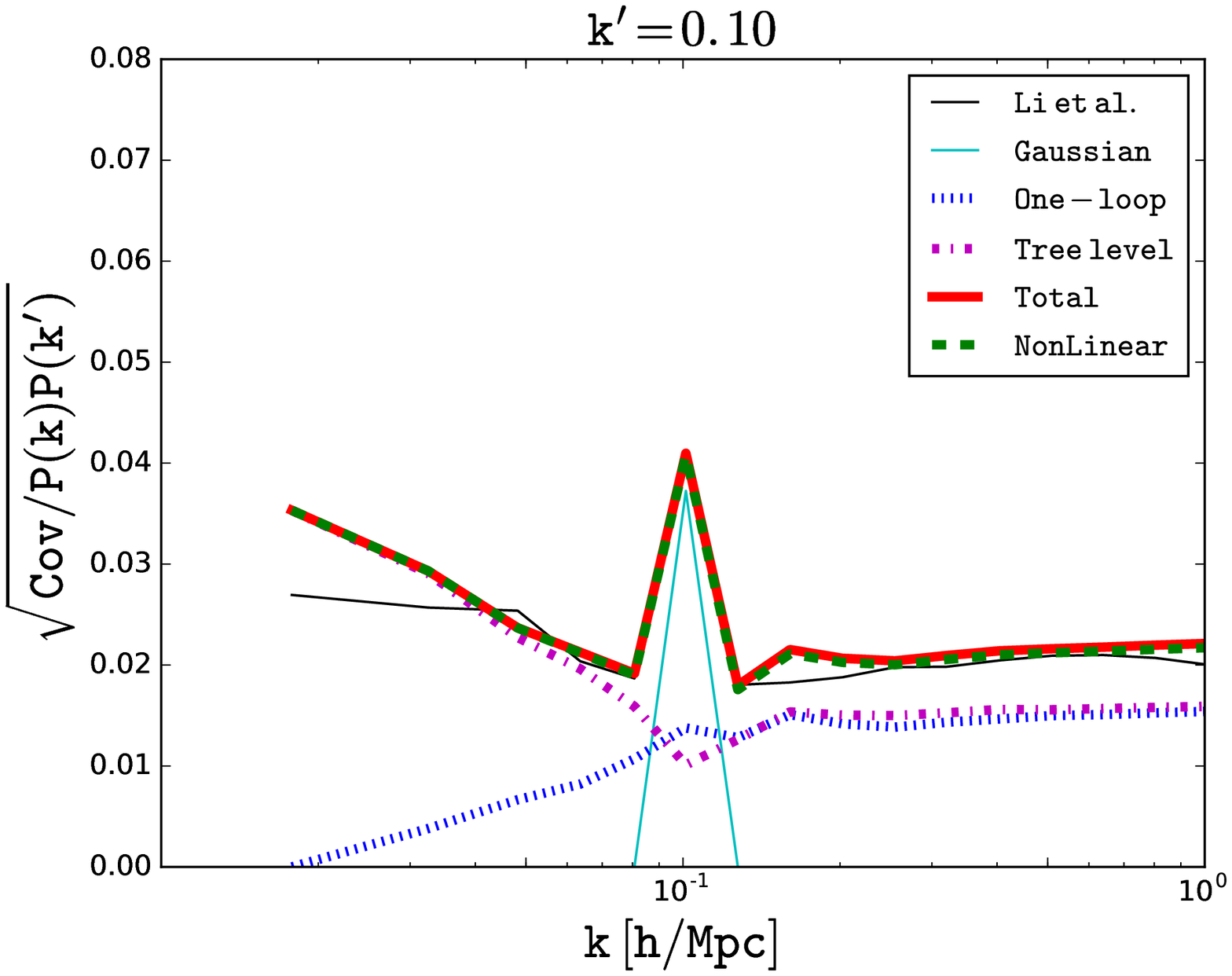}
    \includegraphics[width=0.47\textwidth]{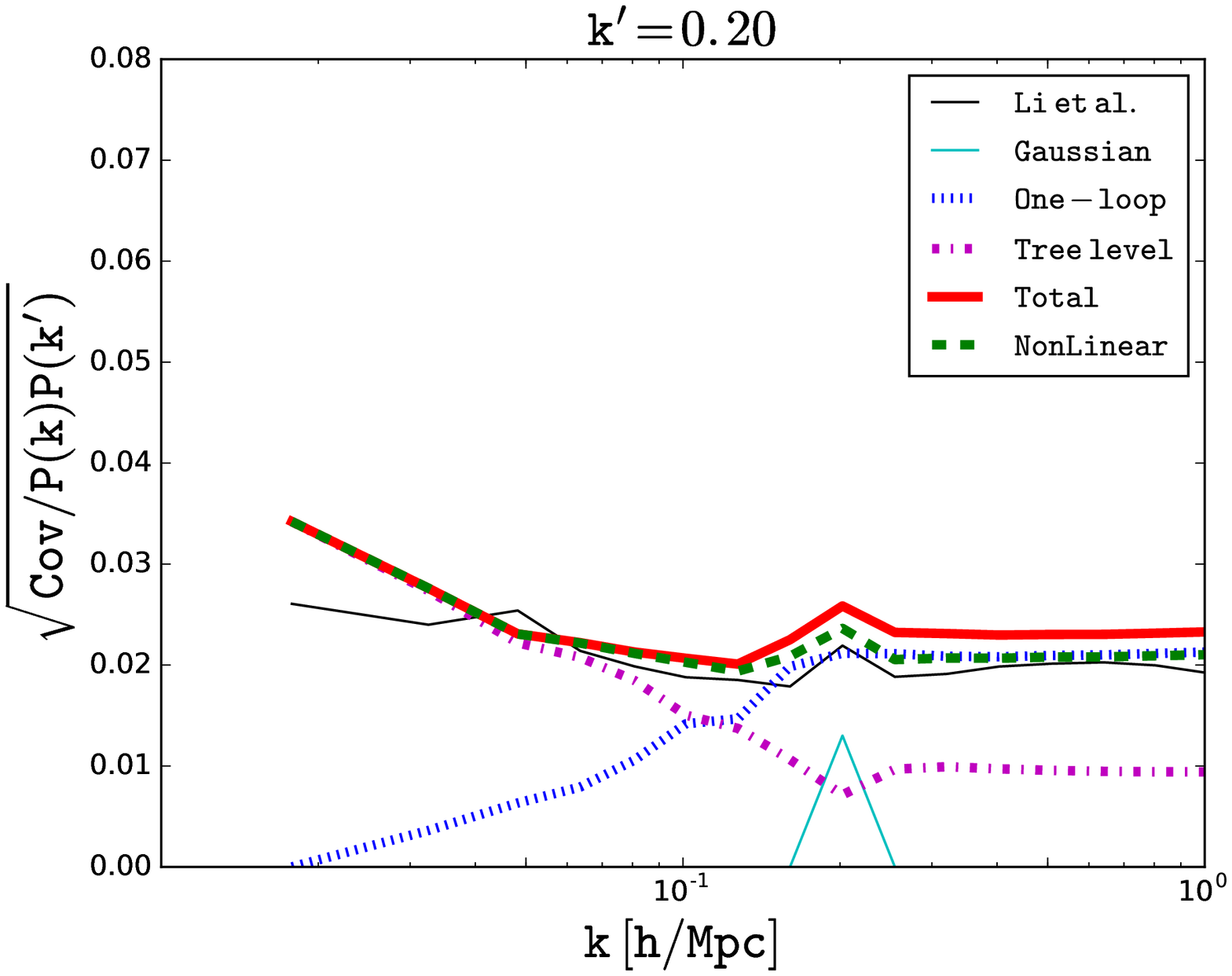}
    \includegraphics[width=0.47\textwidth]{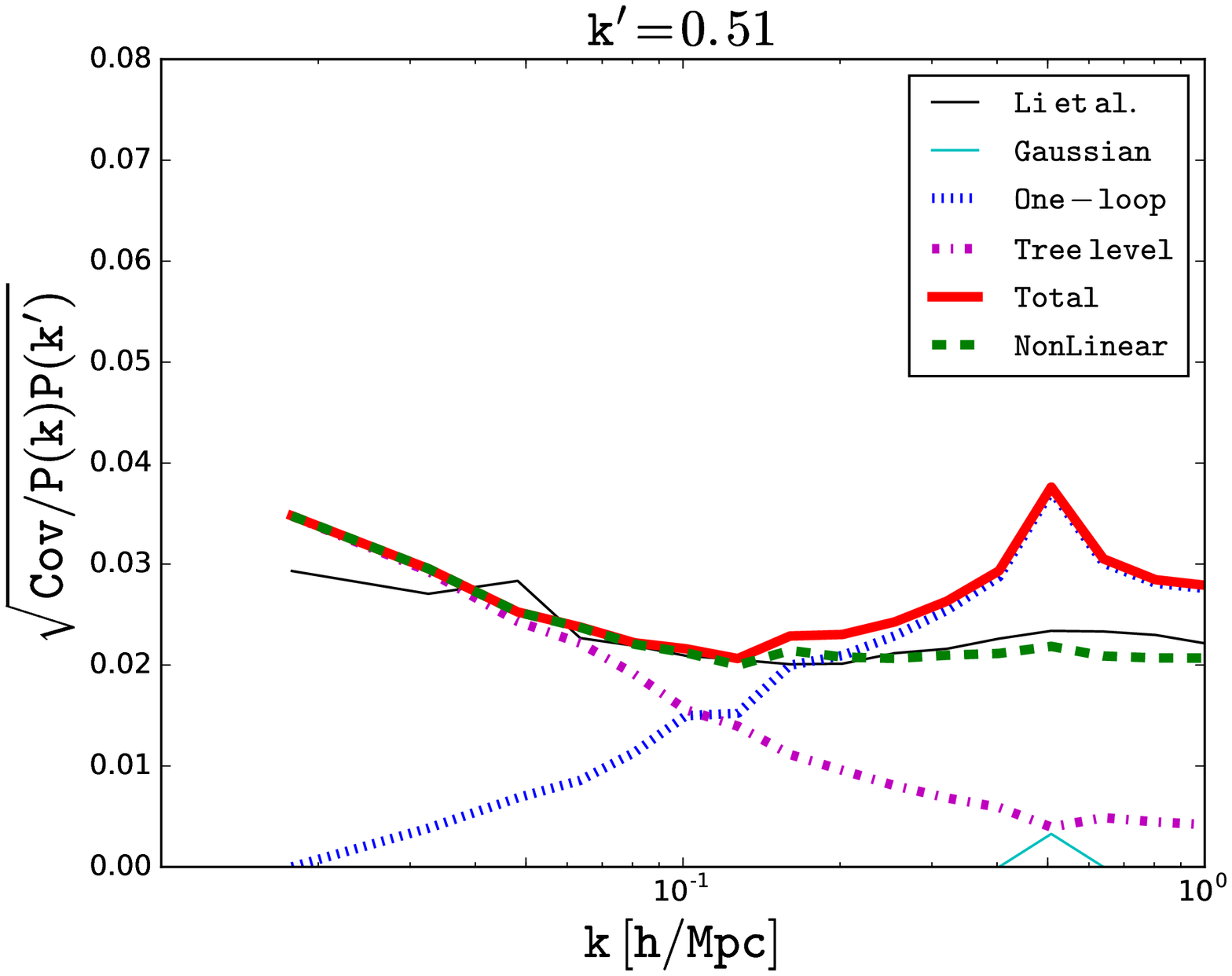}
    \includegraphics[width=0.47\textwidth]{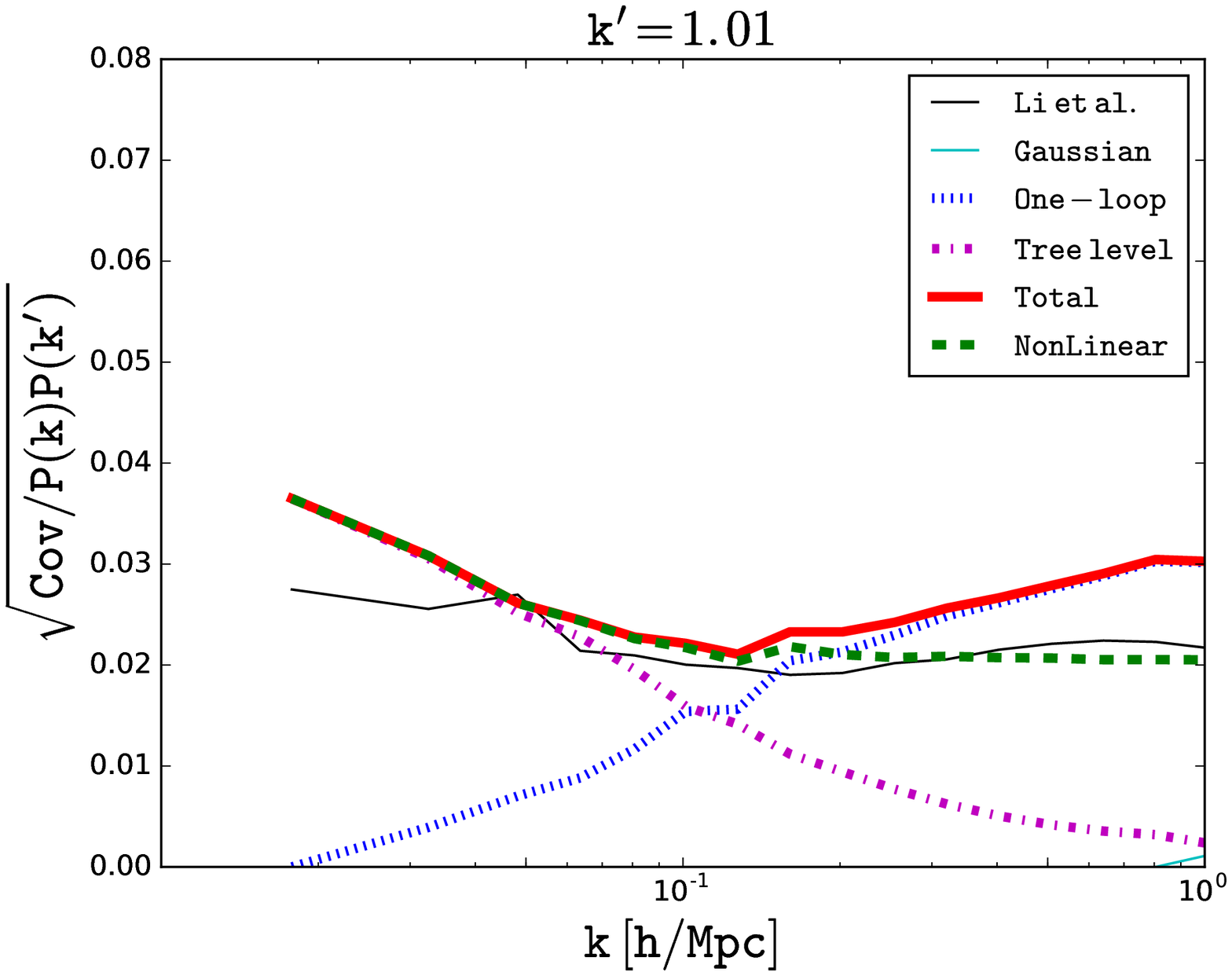}
    \caption{Comparing analytic model with L14 covariance using 1-loop exact functional derivatives at redshift 0.0.}
    \label{fig:li_1loop}
\end{figure}

At redshift zero, there is a good agreement in quasi-linear regime up to $k\sim 0.2 h^{-1}{\rm Mpc}$, but as we go to the non-linear regime, the 1-loop term becomes large and overestimates the covariance. We have explored the nonlinear model of equation \ref{eqn:cov1loopd} with one free parameter and found that with $q_{nl}=0.2$ we find a good agreement even for higher $k$. A similar trend can also be seen for a comparison with L14 covariance at redshift zero in figure \ref{fig:li_1loop}.

As we go higher in redshift, the 1-loop term scales inversely with the eighth power of the growth factor whereas, both  tree level terms scale inversely with the sixth power of the growthfactor. Therefore, the 1-loop contribution to the covariance decreases more rapidly than other contributions. Figure \ref{fig:blot_1loop05} also shows the nonlinear model of equation \ref{eqn:cov1loopd} with $q_{nl}=0.6$. In addition, we expect the correction to $S$ from nonlinear scales to become less important, because $q_{nl}$ increases. As a result, we find a much better agreement between our model and the simulations up to $k\sim 0.8 h^{-1}{\rm Mpc}$.

\begin{figure}
    \centering
    \includegraphics[width=0.47\textwidth]{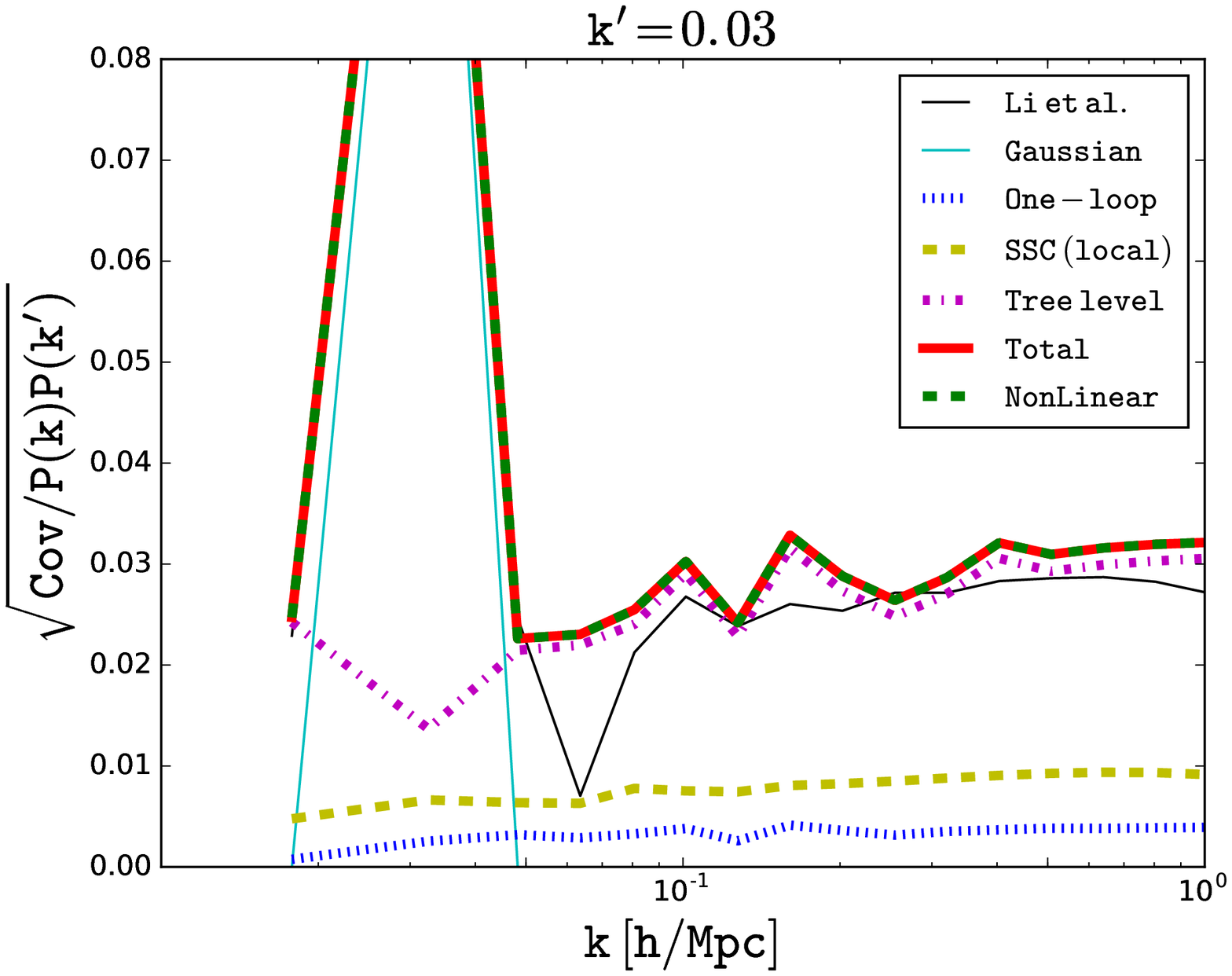}
    \includegraphics[width=0.47\textwidth]{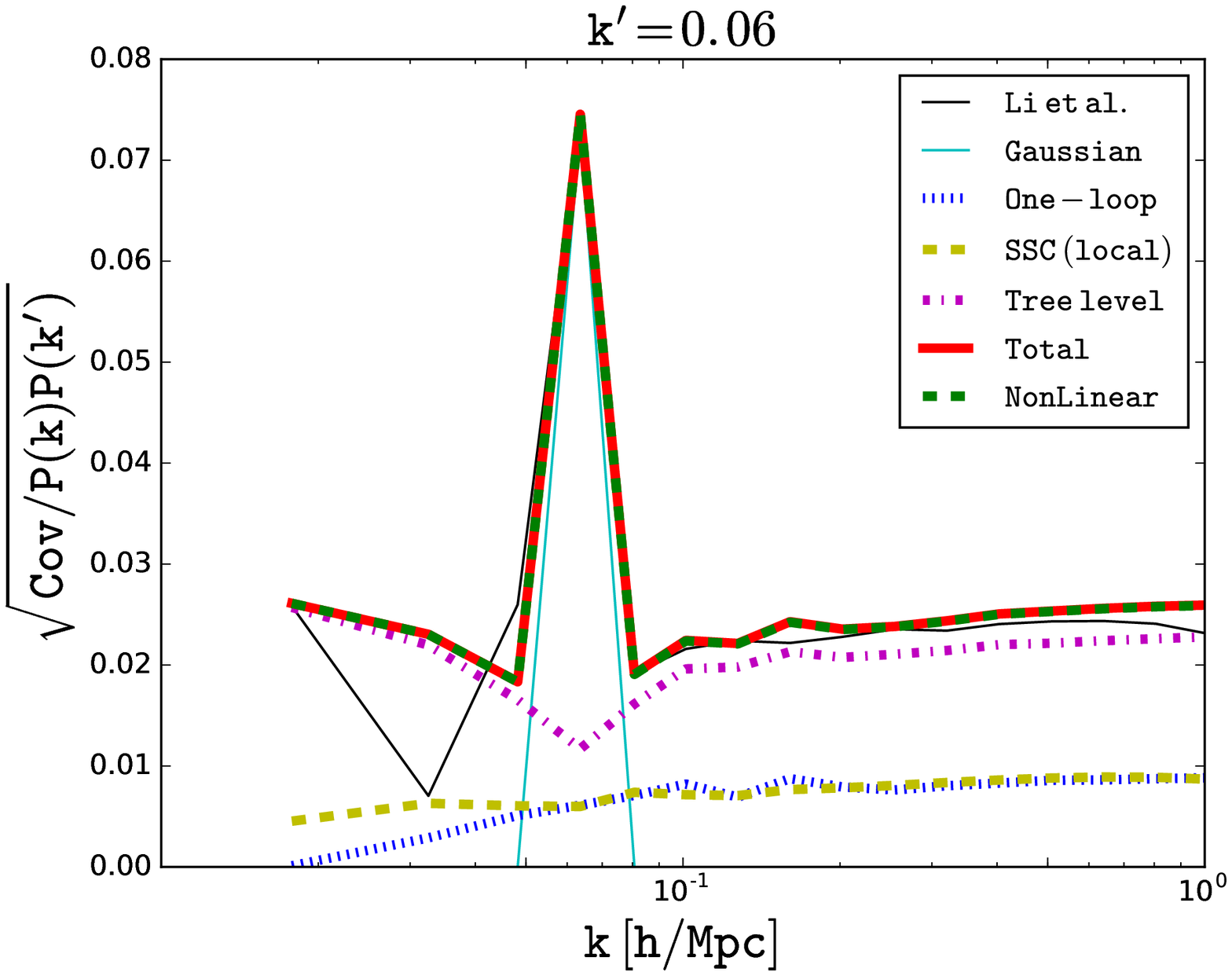}
    \includegraphics[width=0.47\textwidth]{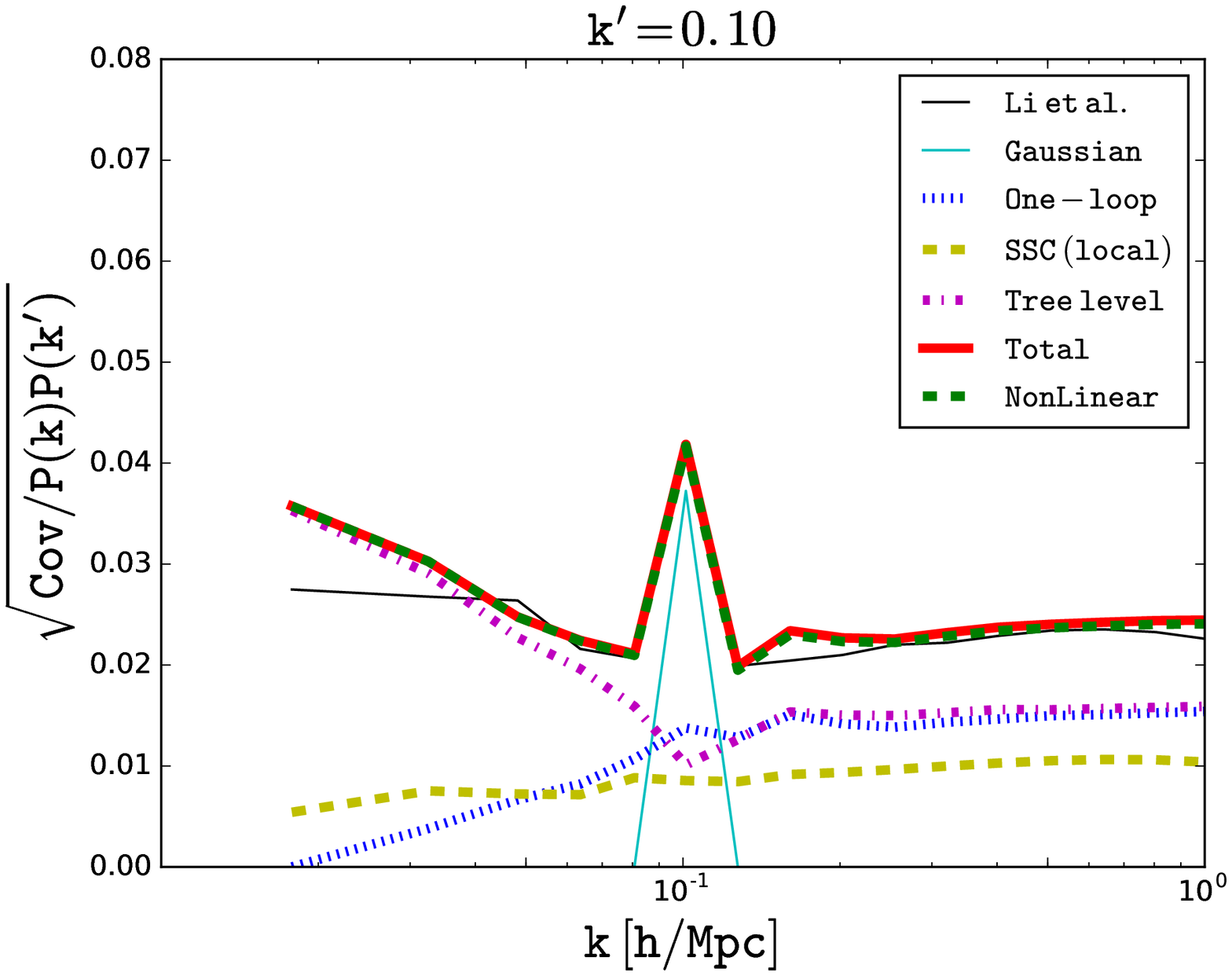}
    \includegraphics[width=0.47\textwidth]{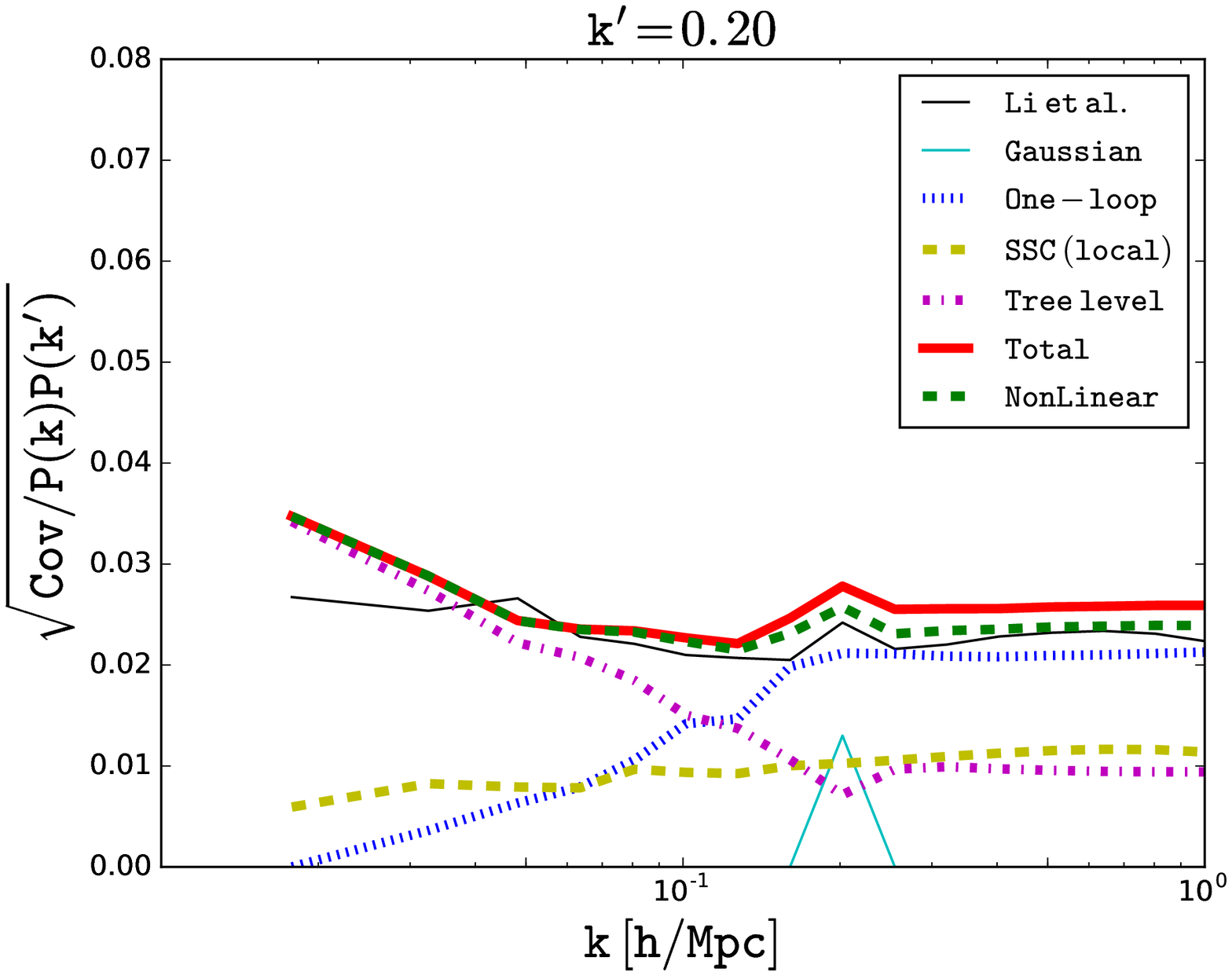}
    \includegraphics[width=0.47\textwidth]{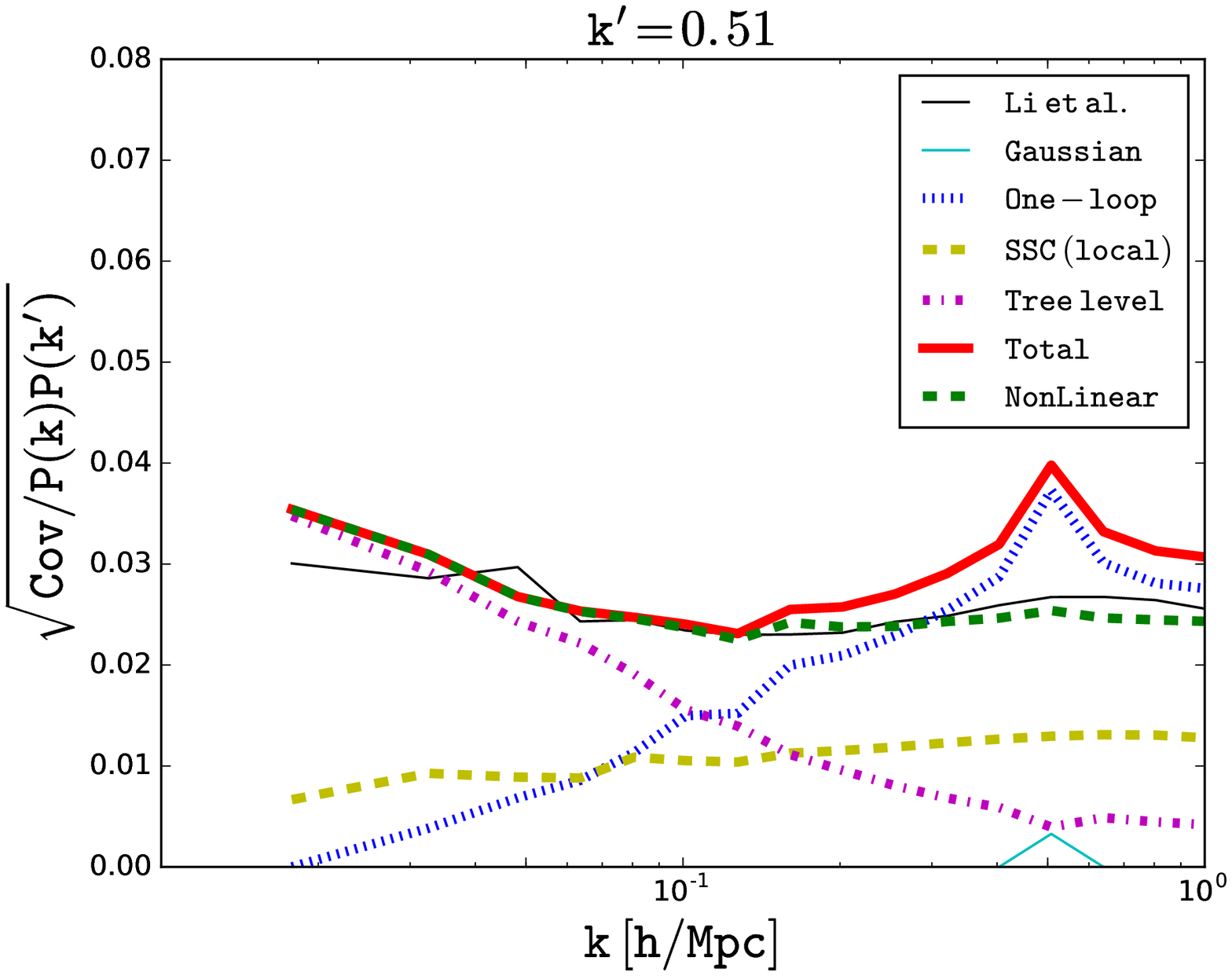}
    \includegraphics[width=0.47\textwidth]{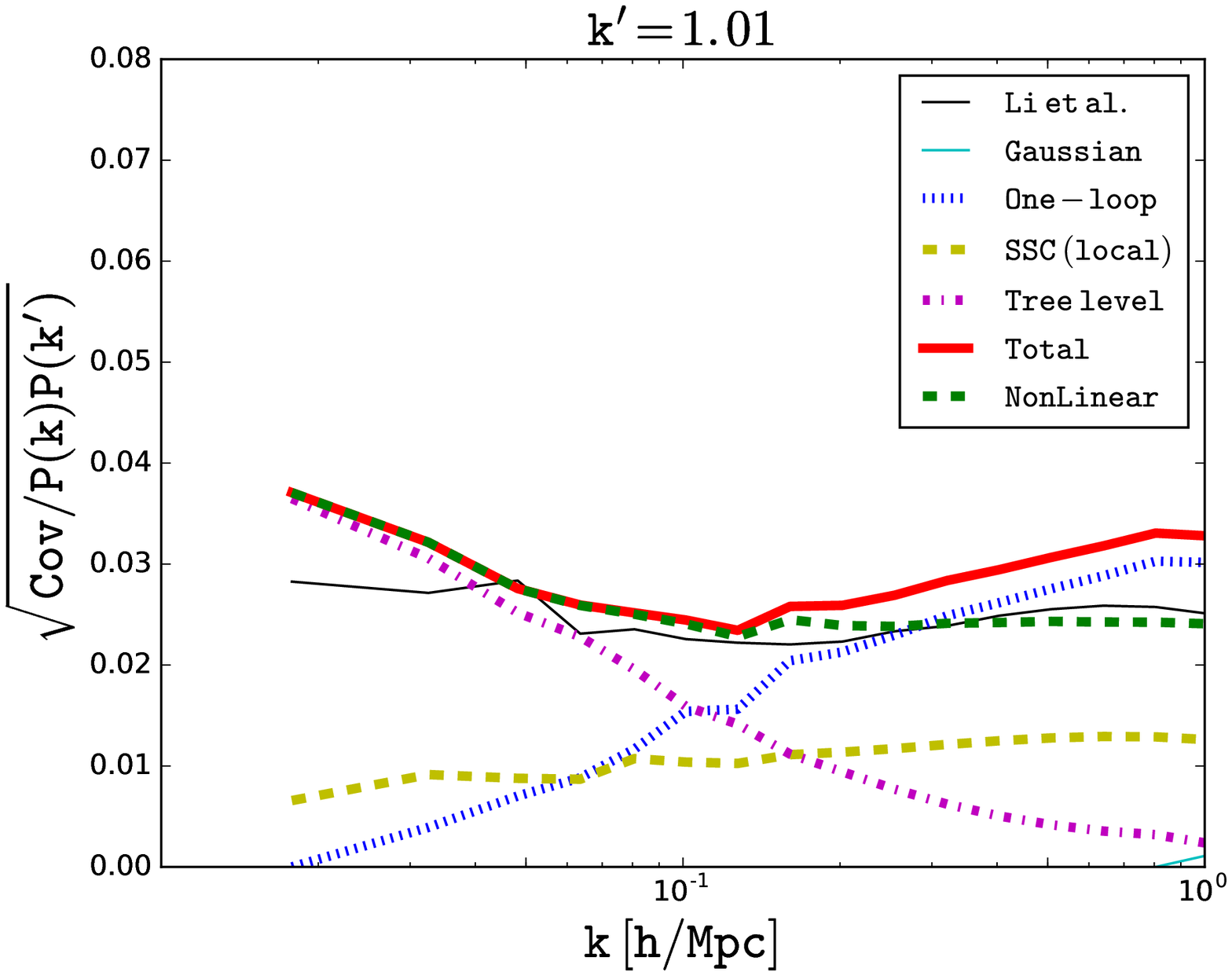}
    \caption{Comparing analytic model with L14 covariance using 1-loop exact functional derivatives at redshift 0.0 with local SSC term.}
    \label{fig:li_1loopSSClocal}
\end{figure}

\begin{figure}
    \centering
    \includegraphics[width=0.47\textwidth]{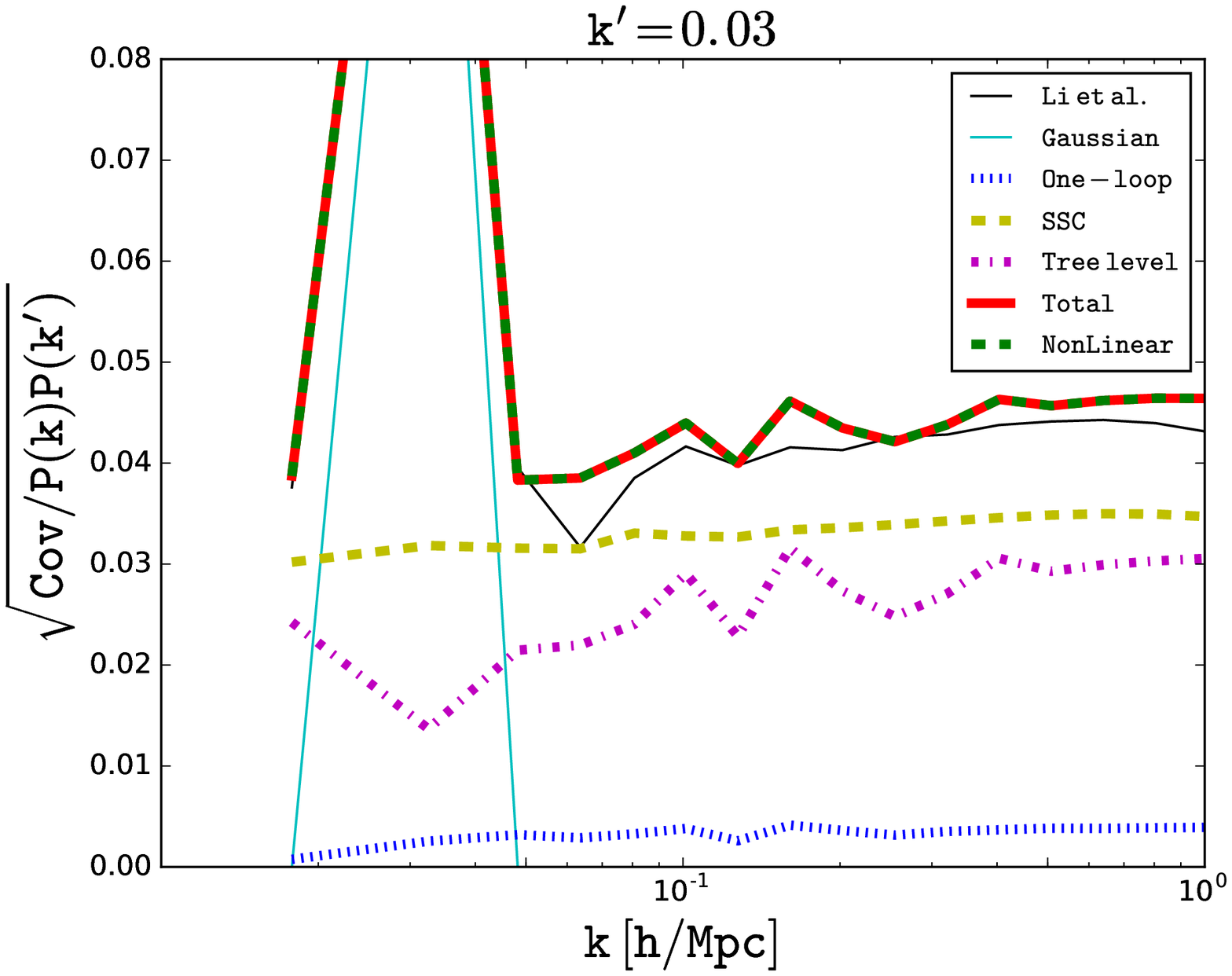}
    \includegraphics[width=0.47\textwidth]{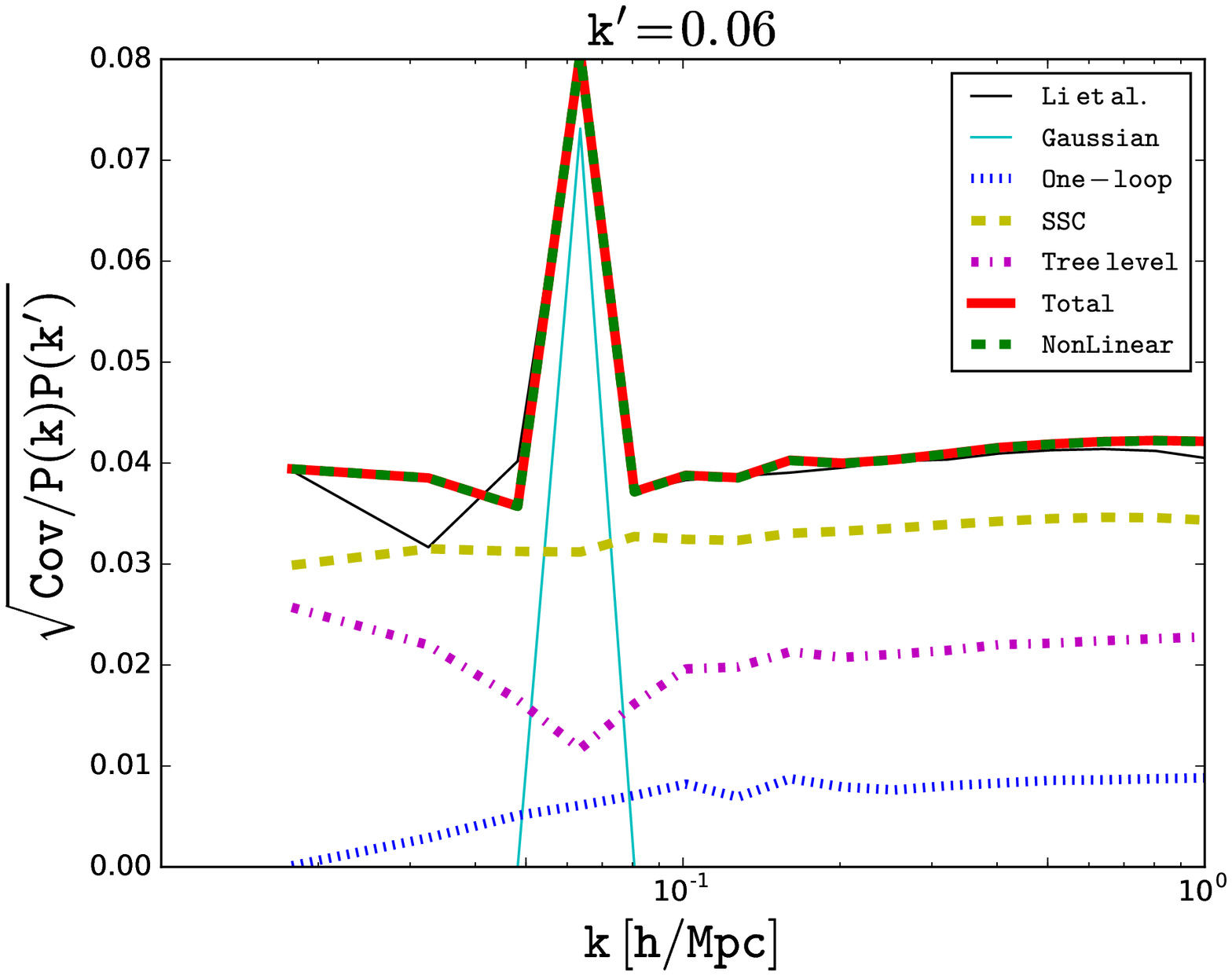}
    \includegraphics[width=0.47\textwidth]{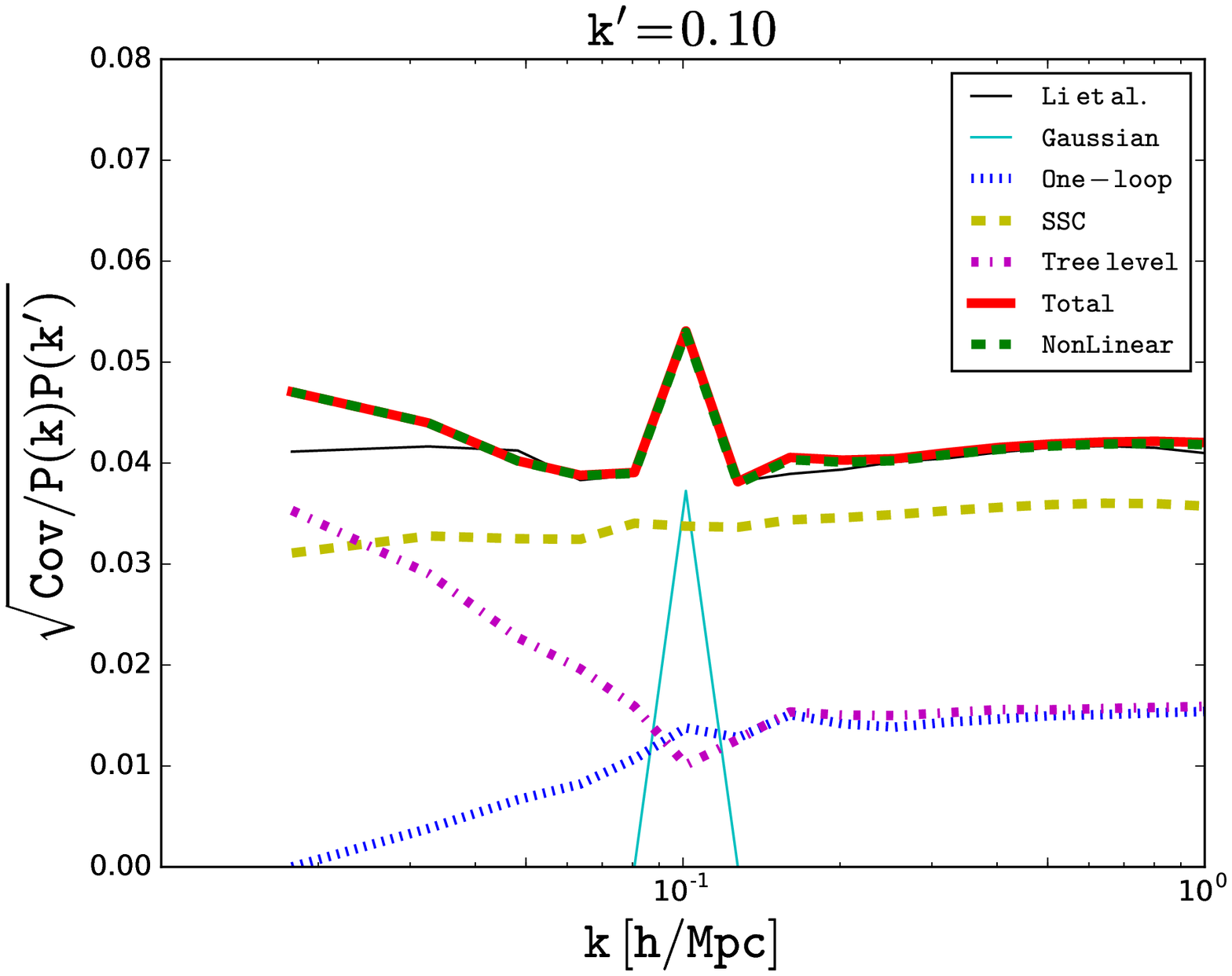}
    \includegraphics[width=0.47\textwidth]{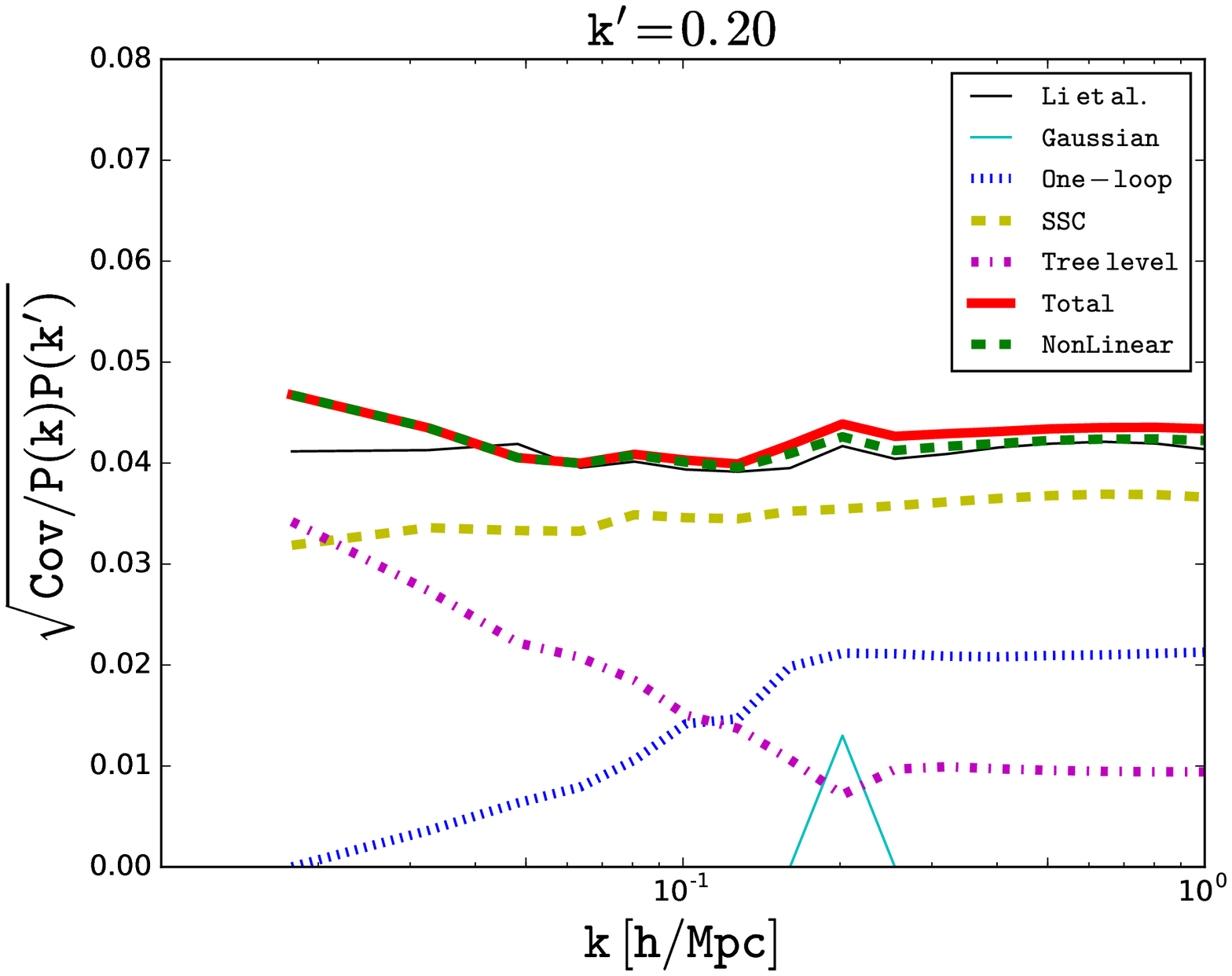}
    \includegraphics[width=0.47\textwidth]{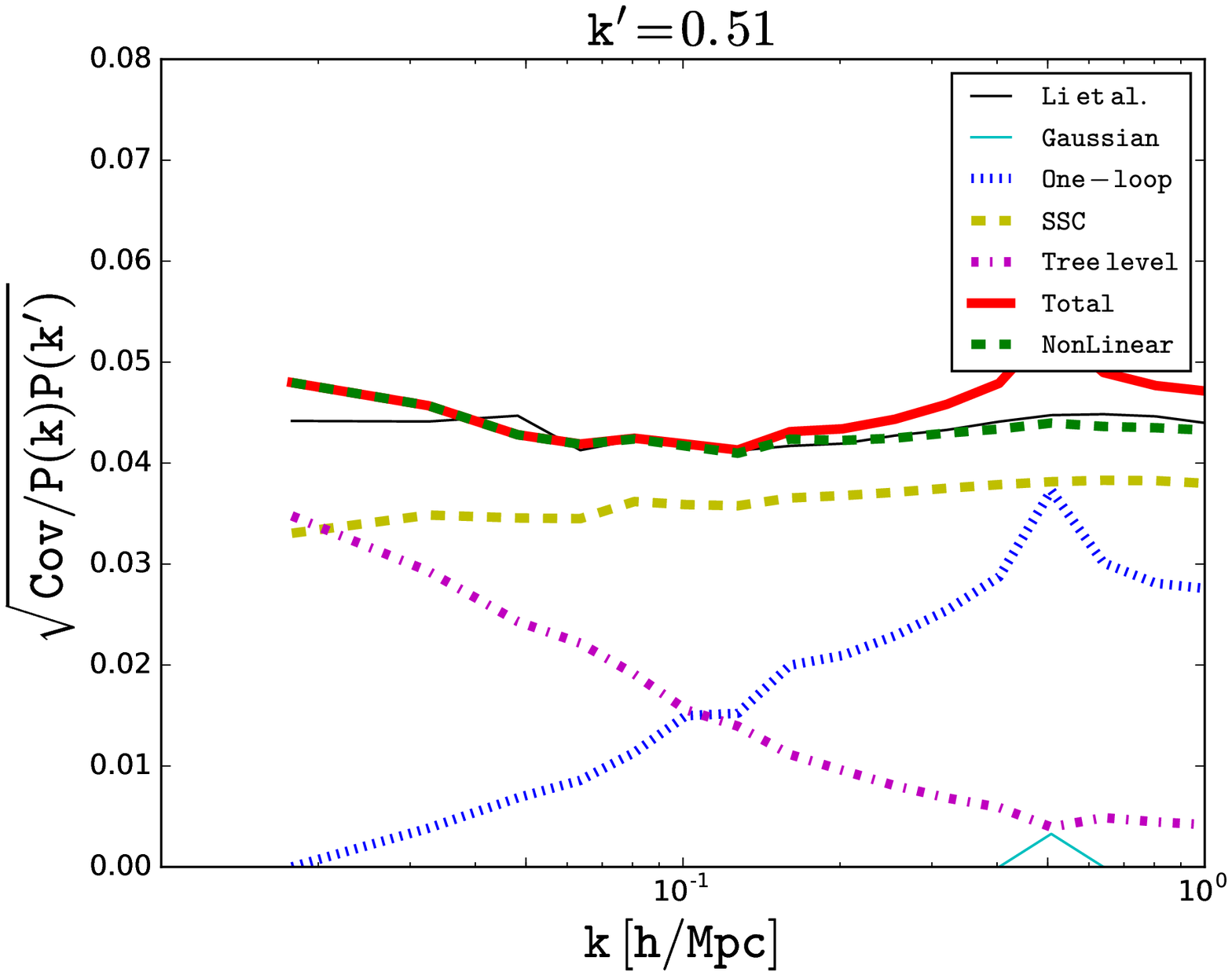}
    \includegraphics[width=0.47\textwidth]{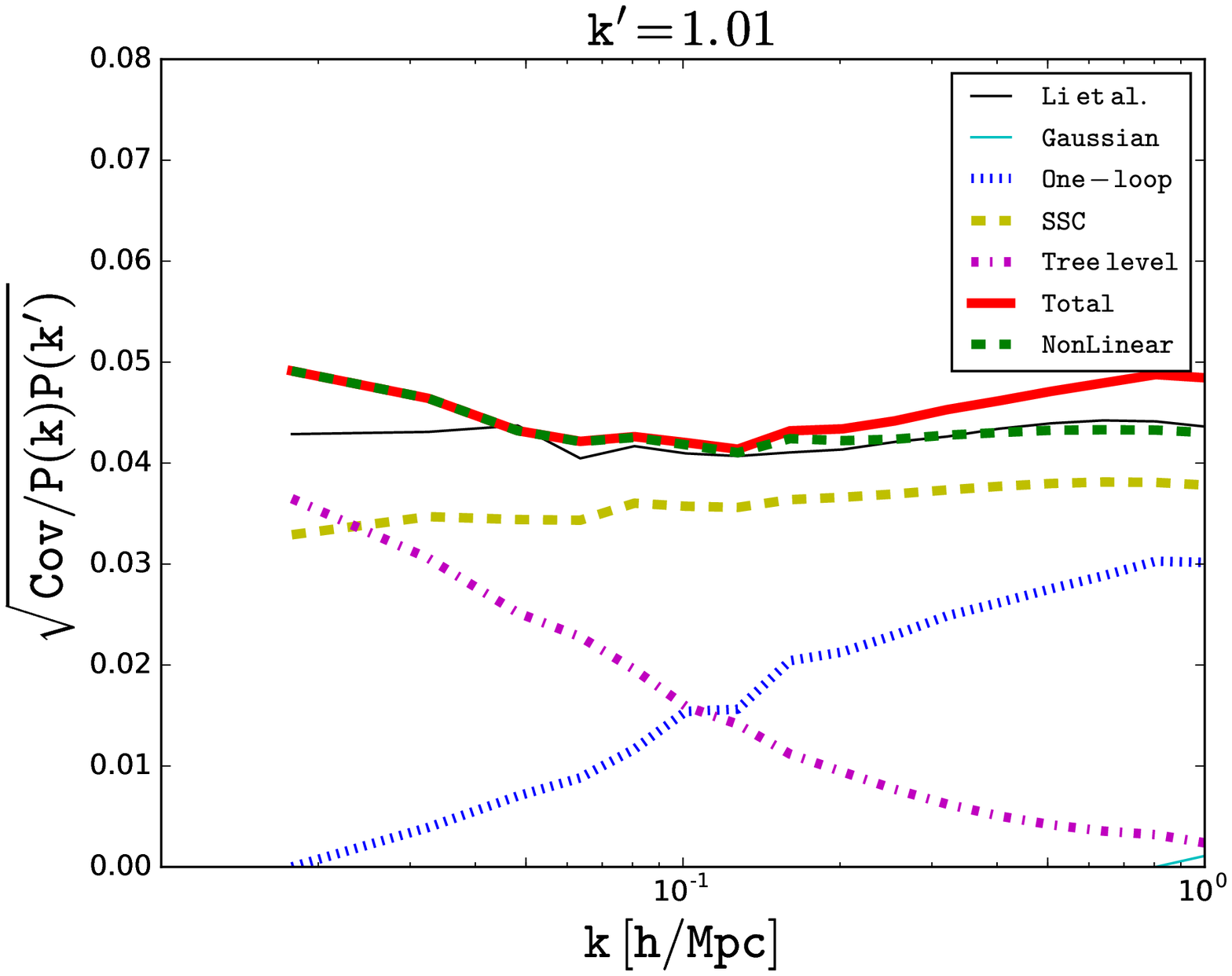}
    \caption{Comparing analytic model with L14 covariance using 1-loop exact functional derivatives at redshift 0.0 with global SSC term.}
    \label{fig:li_1loopSSC}
\end{figure}

In figure \ref{fig:li_1loopSSClocal} and \ref{fig:li_1loopSSC}, we show the comparison of the covariance of L14 dataset, with local and global SSC term respectively added to the calculation. For large scales, nearly quasi-linear scales, global SSC term dominates, and the 1-loop contribution is sub-dominant. Therefore, the agreement with the simulations is better than without SSC contribution even without additional nonlinear correction. For higher redshifts, the 1-loop term is suppressed relative to SSC, and we expect an even better agreement with simulations.

\section{Eigenmode decomposition of the covariance matrix}\label{sec:decomposition}

\begin{figure}
    \centering
    \includegraphics[width=0.48\textwidth]{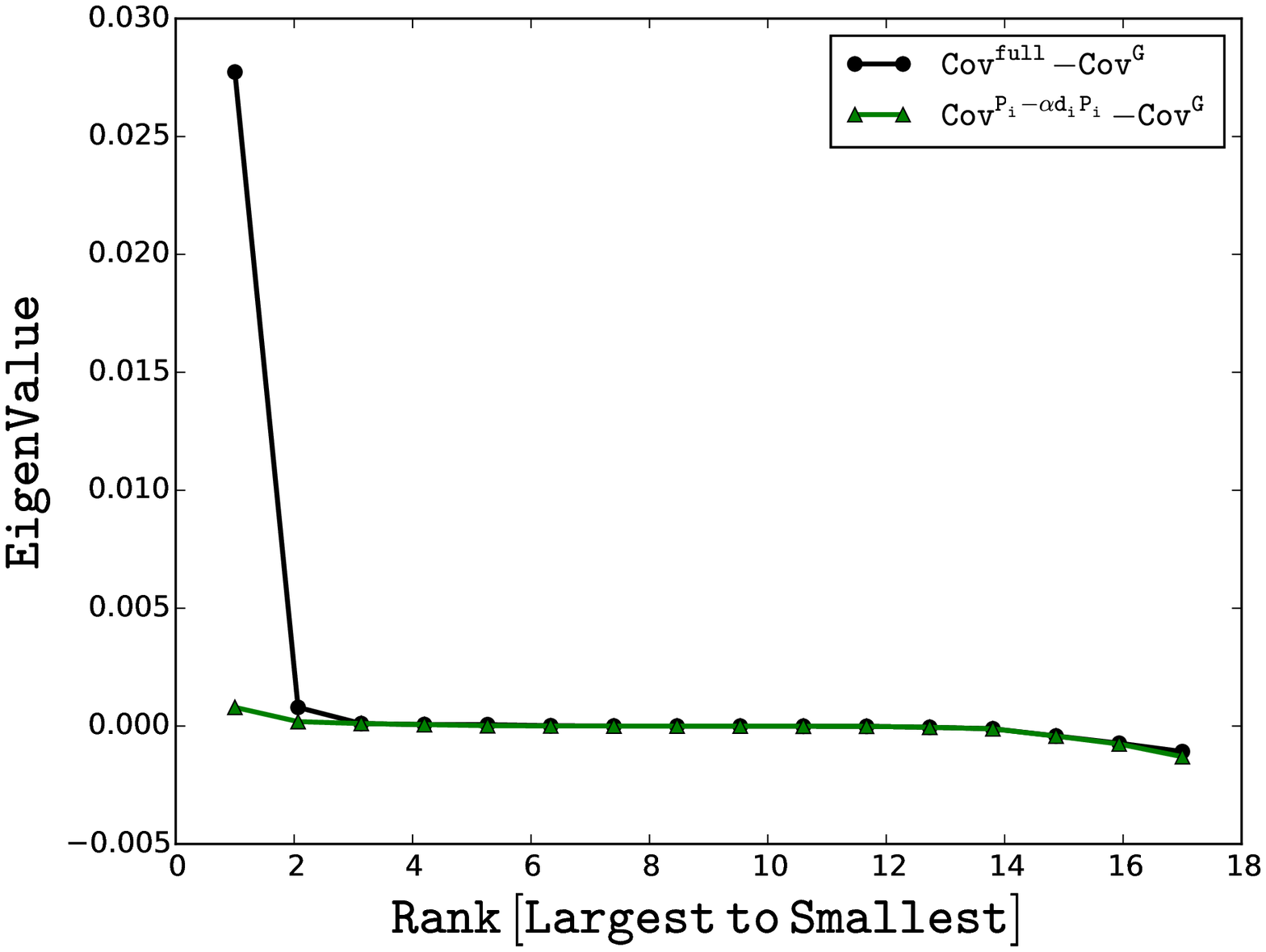}
    \includegraphics[width=0.48\textwidth]{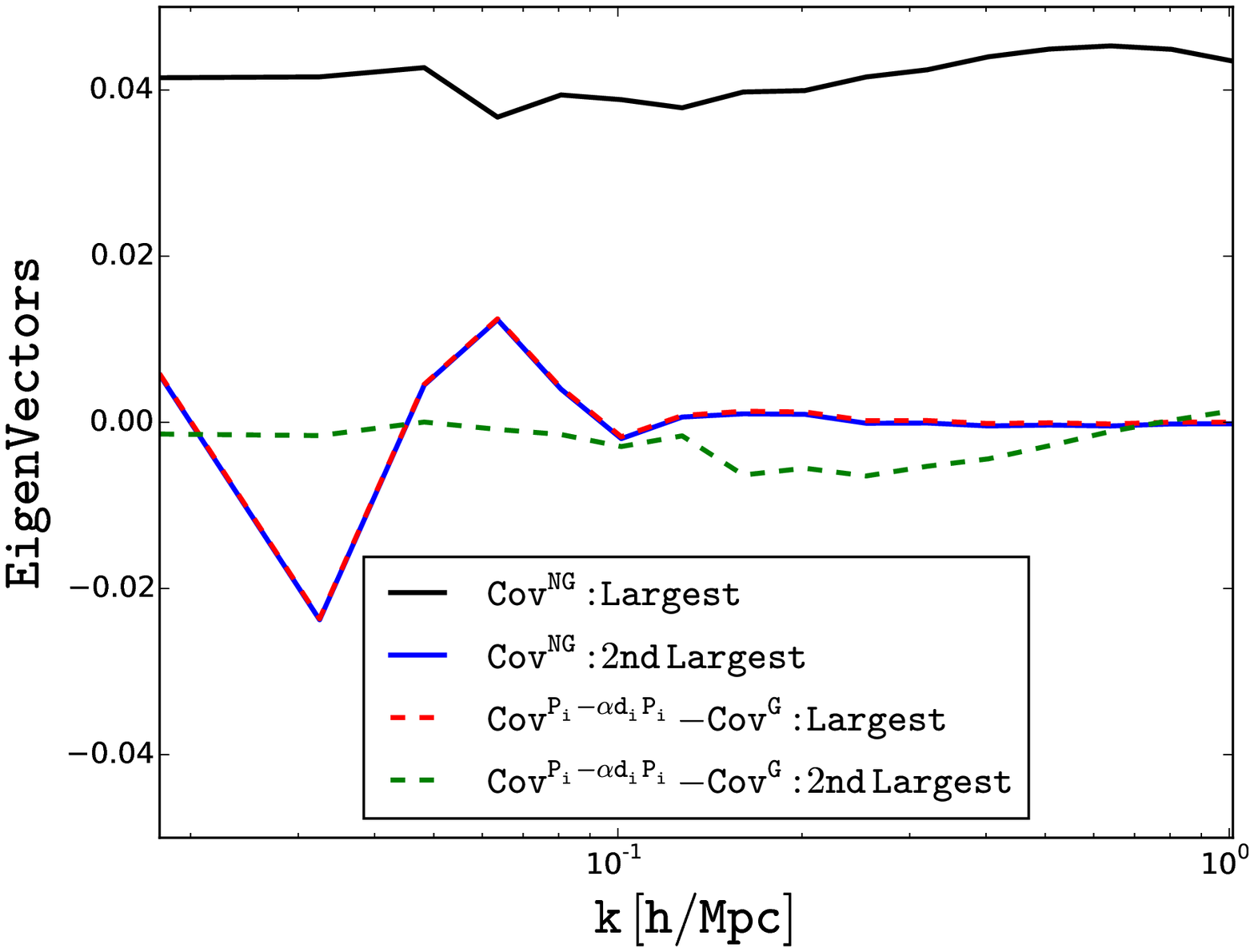}
    \caption{Showing Eigen-values and Eigen-vectors (only two largest) for the covariance matrix (with SSC) and the reduced covariance matrix
            of L14 dataset.}
    \label{fig:eigenvalues}
\end{figure}

The SSC part of the covariance is a rank 1 tensor (equation \ref{eqn:ssc}), which means that it has a single non-zero eigenvalue. The 1-loop term expression in equation \ref{eqn:cov1loop} also suggests a similar form. While the tree level trispectrum does not  decompose like that, it is mostly important at low $k$ where the Gaussian contribution dominates. It is thus interesting to ask about the eigenvalue structure of the entire covariance matrix \citep{2012MNRAS.423.2288H}. Here we use a principle component analysis (PCA) on the non-Gaussian part of the covariance. We start by calculating the covariance matrix of global SSC version of L14 (total 3584 power spectra) data set using equation \ref{eqn:covdefinition}. We perform a principal component analysis of the matrix $C_{ij}$, such that

\begin{equation}
    C_{ij} = \dfrac{\mathbf{Cov}_{ij}^{\rm full} - \mathbf{Cov}_{ij}^{\rm Gauss}}{P(k_i)P(k_j)}
\end{equation}
\\
The eigenvalues of the matrix $C_{ij}$ are shown in the left panel of figure \ref{fig:eigenvalues}. Clearly one eigenvalue is much larger than all of the others. Therefore, the full matrix $C_{ij}$ can be well approximated using a single principal component $d_1 = \sqrt{\lambda_1}v_1$, where $\lambda_1$ is the largest eigenvalue and $v_1$ is the corresponding eigenvector, such  that $C_{ij}=d_id_j^{\dag}$. The right panel of figure \ref{fig:eigenvalues} shows the first two principal components ($d_1$ and  $d_2 = \sqrt{\lambda_2}v_2$).

\begin{figure}
    \centering
    \includegraphics[width=0.48\textwidth]{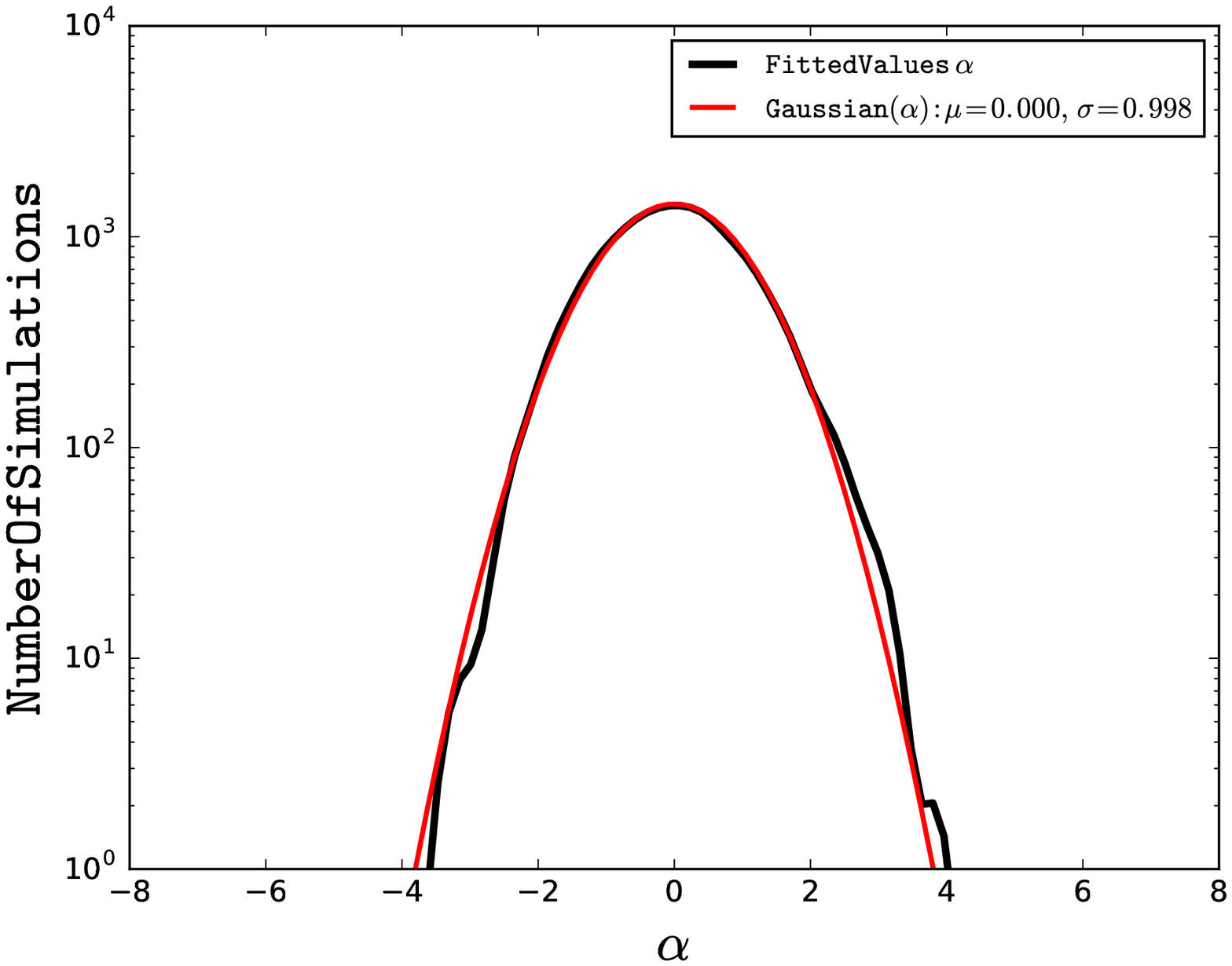}
    \caption{Distribution of $\alpha$ as in equation \ref{pkd}.}
    \label{fig:alpha}
\end{figure}

In order to show the dominance of the first component explicitly, we perform another exercise by removing the first eigenvector from the underlying dataset. To do so, we subtract a contribution from each of the power spectra, such that the new power spectra ($P^{\prime}_i(k)$) are,

\begin{equation}
    P^{\prime}(k_i) = P(k_i) - \alpha d_1(k_i) \langle P(k_i)\rangle.
\label{pkd}
\end{equation}
\\
Here $\alpha$ is the best fit coefficient when fitting $P_i(k)$ to $(1+\alpha d_1(k))\langle P(k) \rangle$. Figure \ref{fig:alpha} shows the distribution of $\alpha$, which is very Gaussian. This is not surprising, since  it is dominated by the SSC term variance on the scale of the survey box, which is Gaussian distributed.  We computed the covariance matrix of the new reduced dataset $P^{\prime}(k_i)$. The corresponding eigenvalues can be seen in the left panel of figure \ref{fig:eigenvalues}. It can be noticed that the largest eigenvalue of the reduced covariance matrix is very close to the second largest eigenvalue of the original covariance. Similar trend can also be seen in the eigenvectors in the right panel of figure \ref{fig:eigenvalues}.

\begin{figure}
    \centering
    \includegraphics[width=0.48\textwidth]{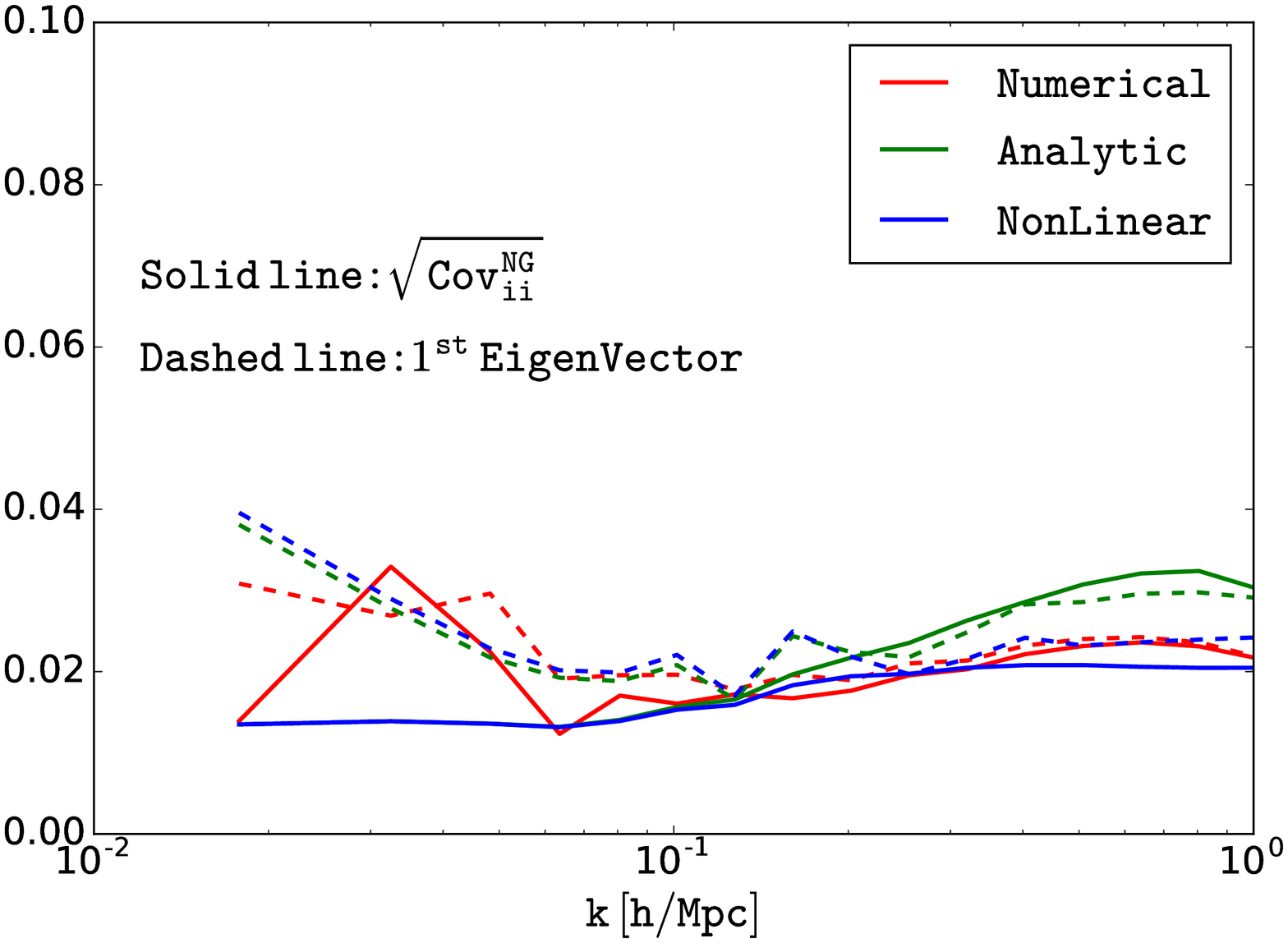}
    \includegraphics[width=0.48\textwidth]{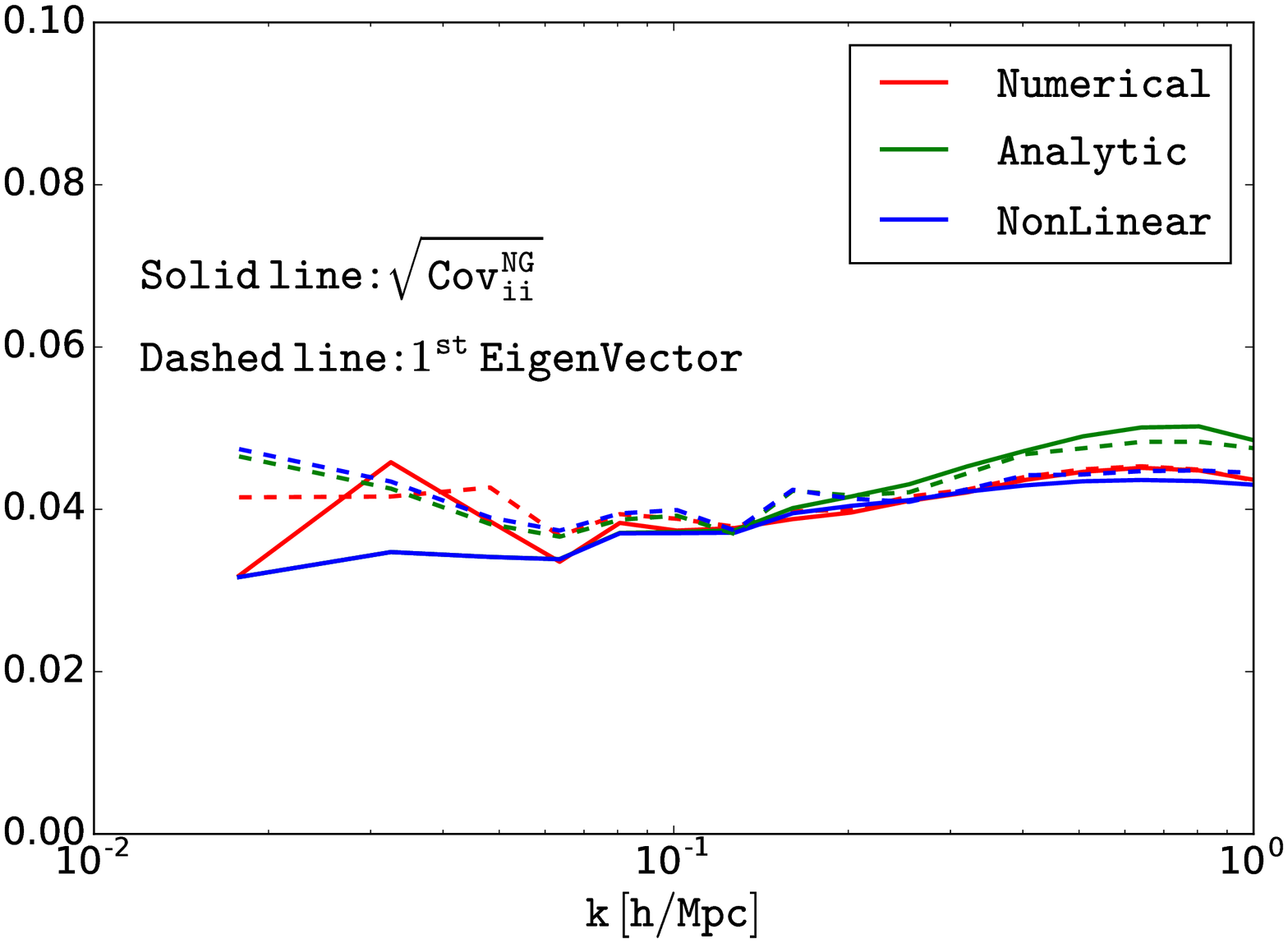}
    \caption{The first eigenvector $d_1$ (in dashed lines) and the diagonal component of the covariance matrix
    for analytic (1-loop and damped 1-loop, which we call nonlinear) versus numerical approach, at $z=0$. The left panel shows the covariance without SSC term, the right panel with global SSC term. We see that analytic model makes about 20\% error at $ k \sim 1 h {\rm Mpc^{-1}}$ without SSC, and about 10\% error with SSC. The nonlinear model reduces this error to a few percent only in 
the SSC case, in agreement with previous figures.}
    \label{fig:NumVsAnl}
\end{figure}

\begin{figure}
    \centering
    \includegraphics[width=0.47\textwidth]{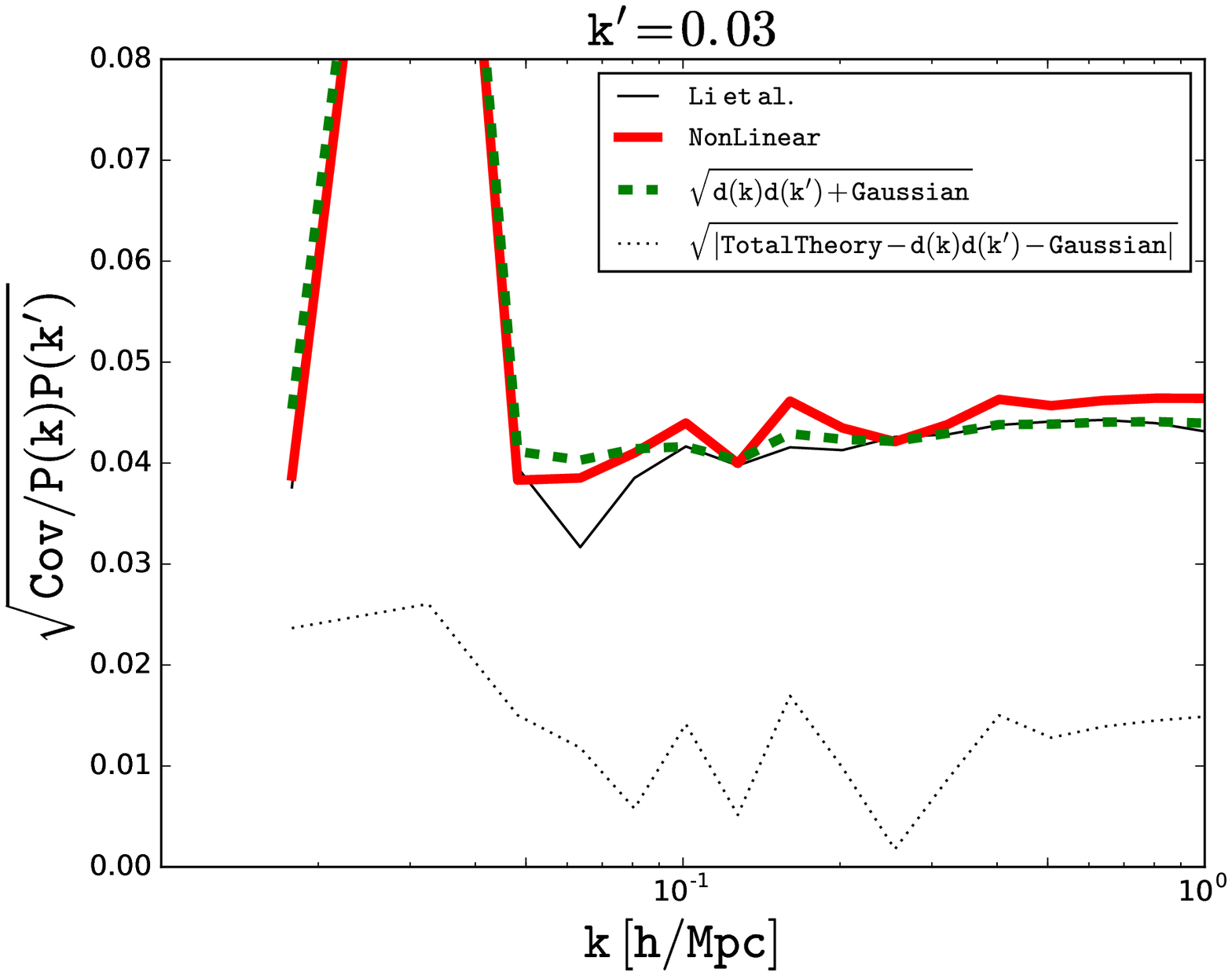}
    \includegraphics[width=0.47\textwidth]{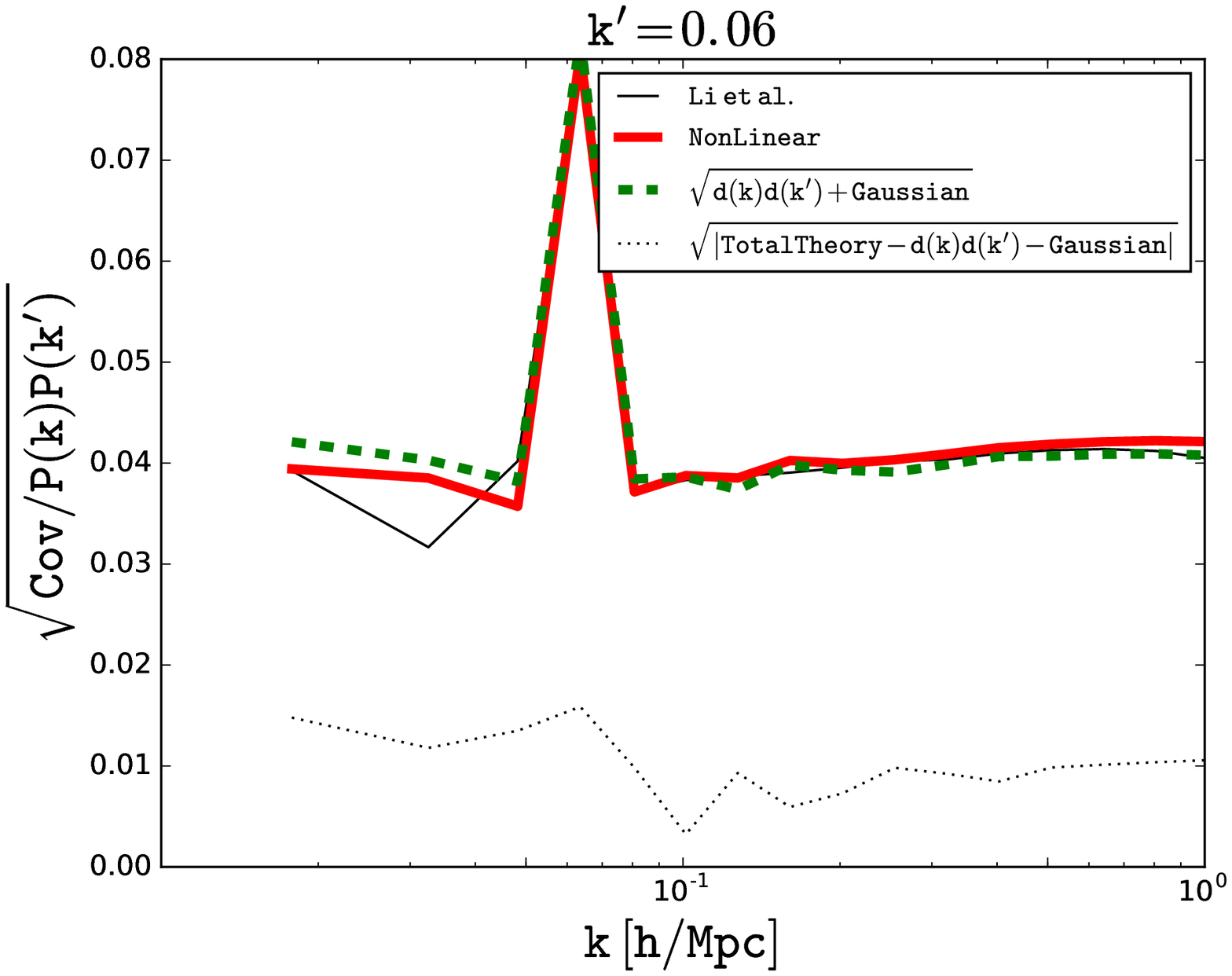}
    \includegraphics[width=0.47\textwidth]{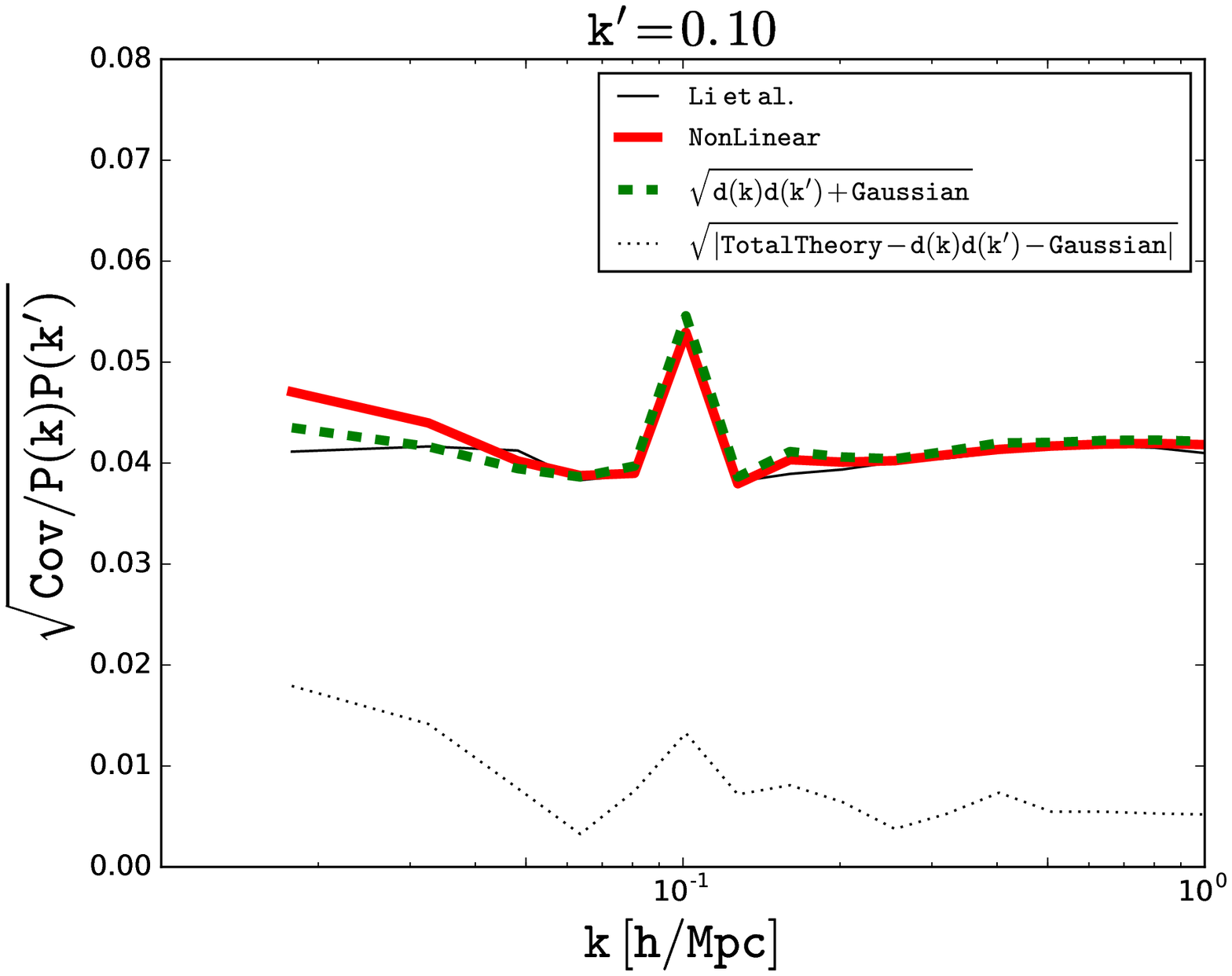}
    \includegraphics[width=0.47\textwidth]{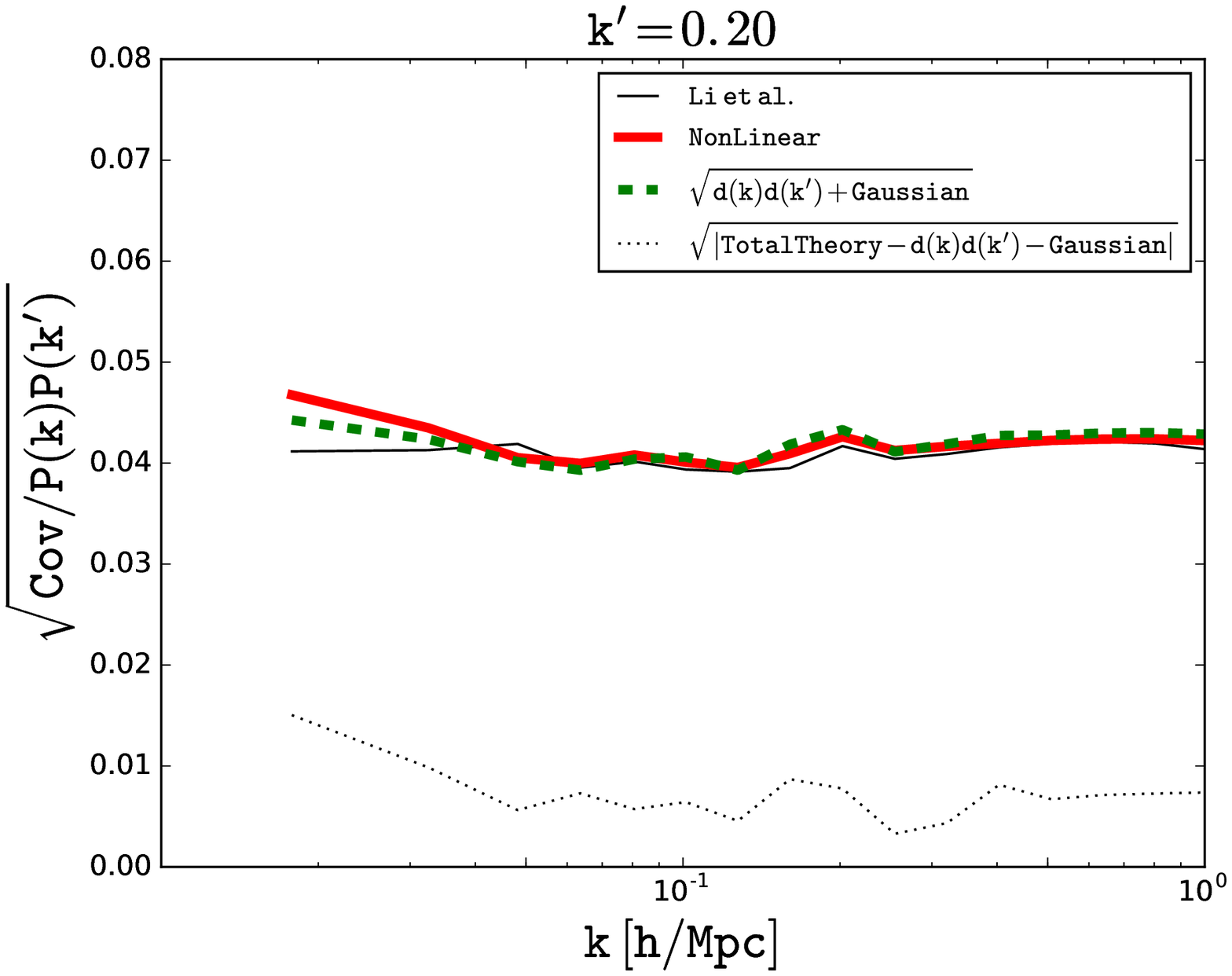}
    \includegraphics[width=0.47\textwidth]{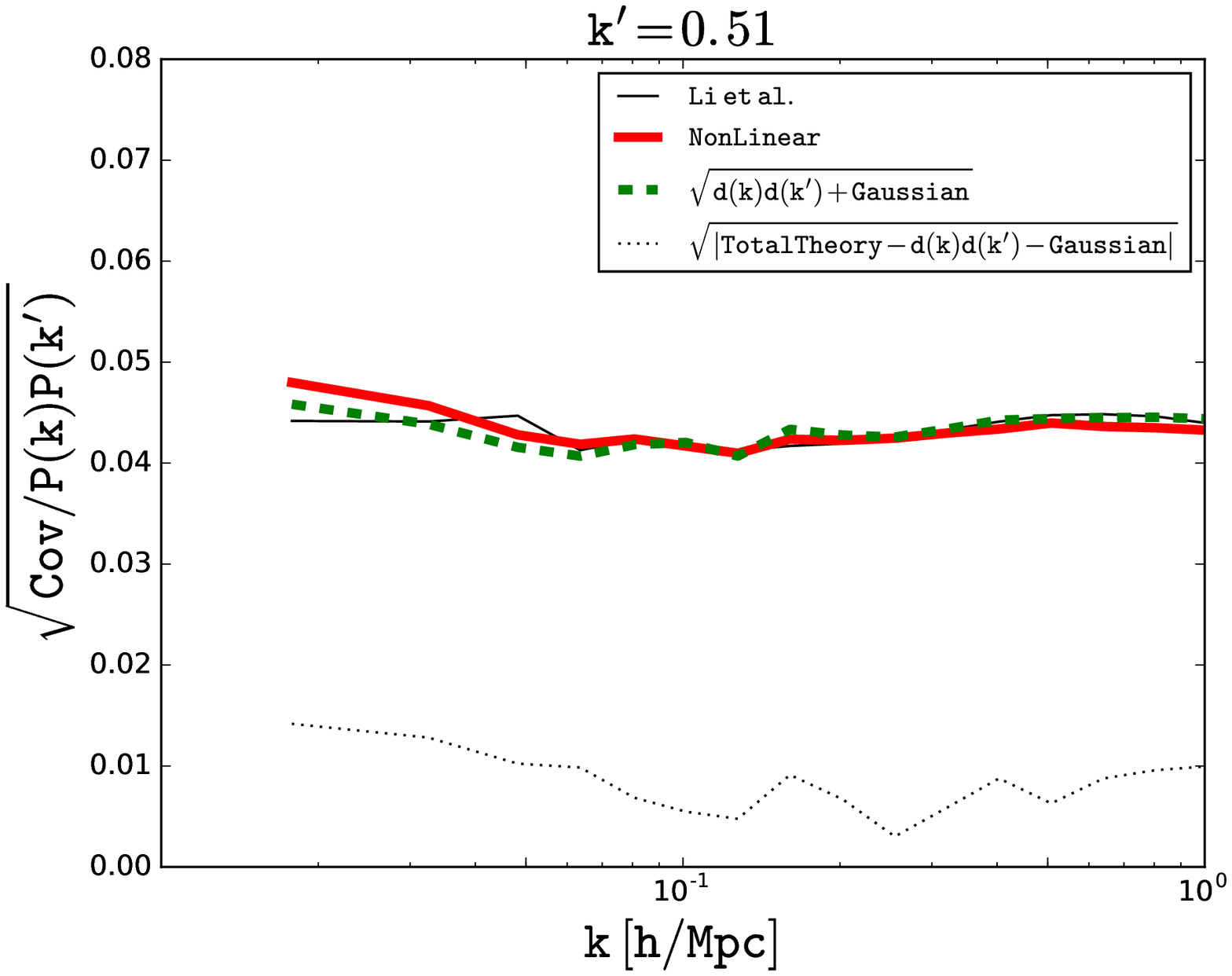}
    \includegraphics[width=0.47\textwidth]{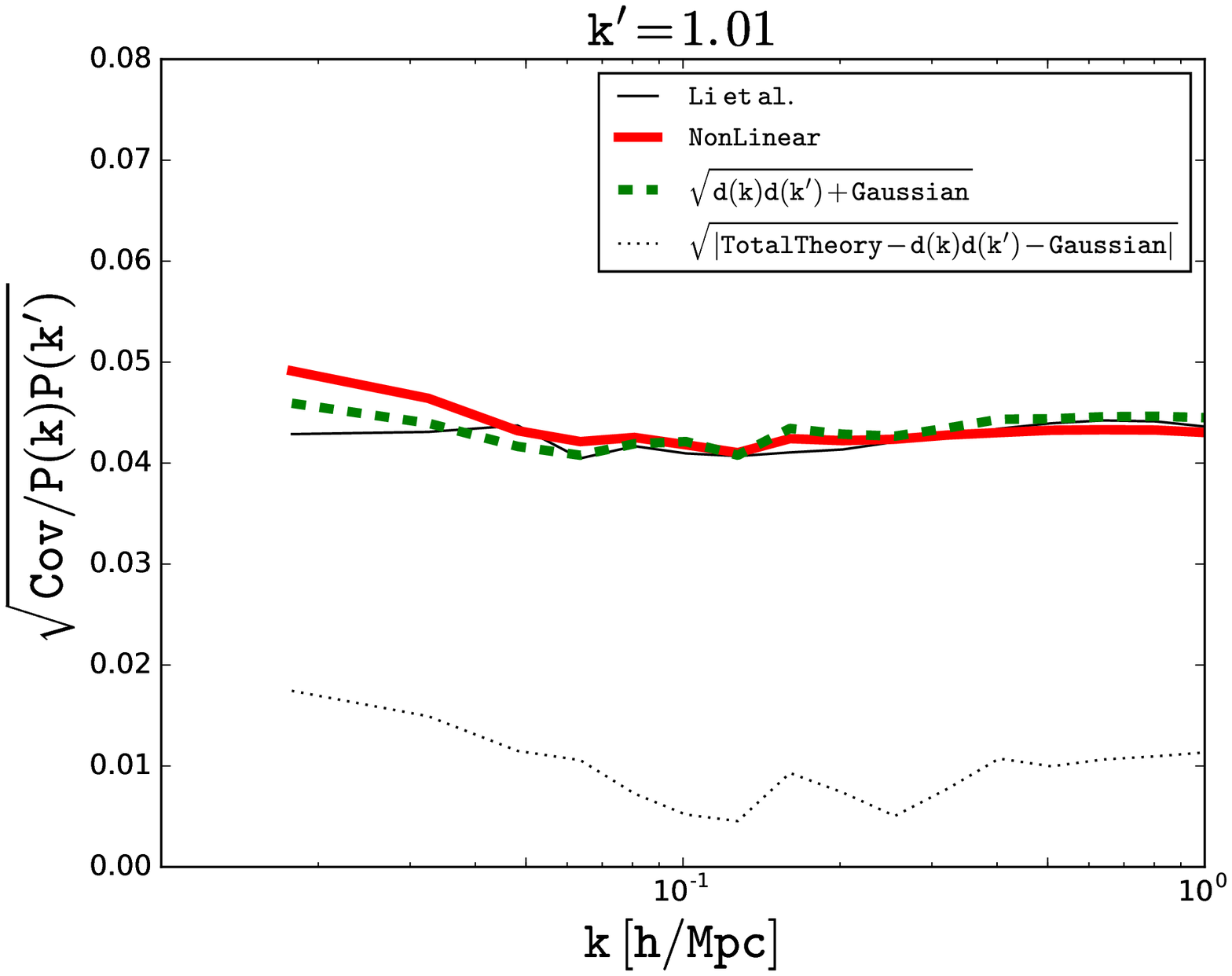}
    \caption{Comparing analytic model with L14 covariance (global SSC case).  
We compare the single eigenvector model (dashed green) and total 1-loop predictions 
(solid thick red) against simulations (thin black). The residual covariance after subtracting the single eigenvector component
is shown with dotted line.}
    \label{fig:li_1loopSSCPCA}
\end{figure}

We also performed the diagonal decomposition of the analytic matrix, i.e., using the covariance from our analytic model. Figure \ref{fig:NumVsAnl} shows the comparison of the analytic and numerical covariance matrix decomposition, particularly for the first eigenvector and the diagonal part of the covariance. The left panel shows the comparison without the SSC term whereas the right panel shows the comparison with global SSC term. As expected, the agreement improves for the full covariance matrix with SSC term. Figure \ref{fig:li_1loopSSCPCA}, shows a comparison of the single eigenvector model of the full covariance matrix with the L14 simulations, as well as the diagonal.
We see that analytic model makes about 20$\%$ error at $ k \sim 1 h {\rm Mpc^{-1}}$
without SSC, and about 10$\%$ error with SSC. The nonlinear model where high $q$ modes are damped 
reduces this error to a few percent only in
the SSC case, in agreement with previous figures.

\begin{figure}
    \centering
    \includegraphics[width=0.6\textwidth]{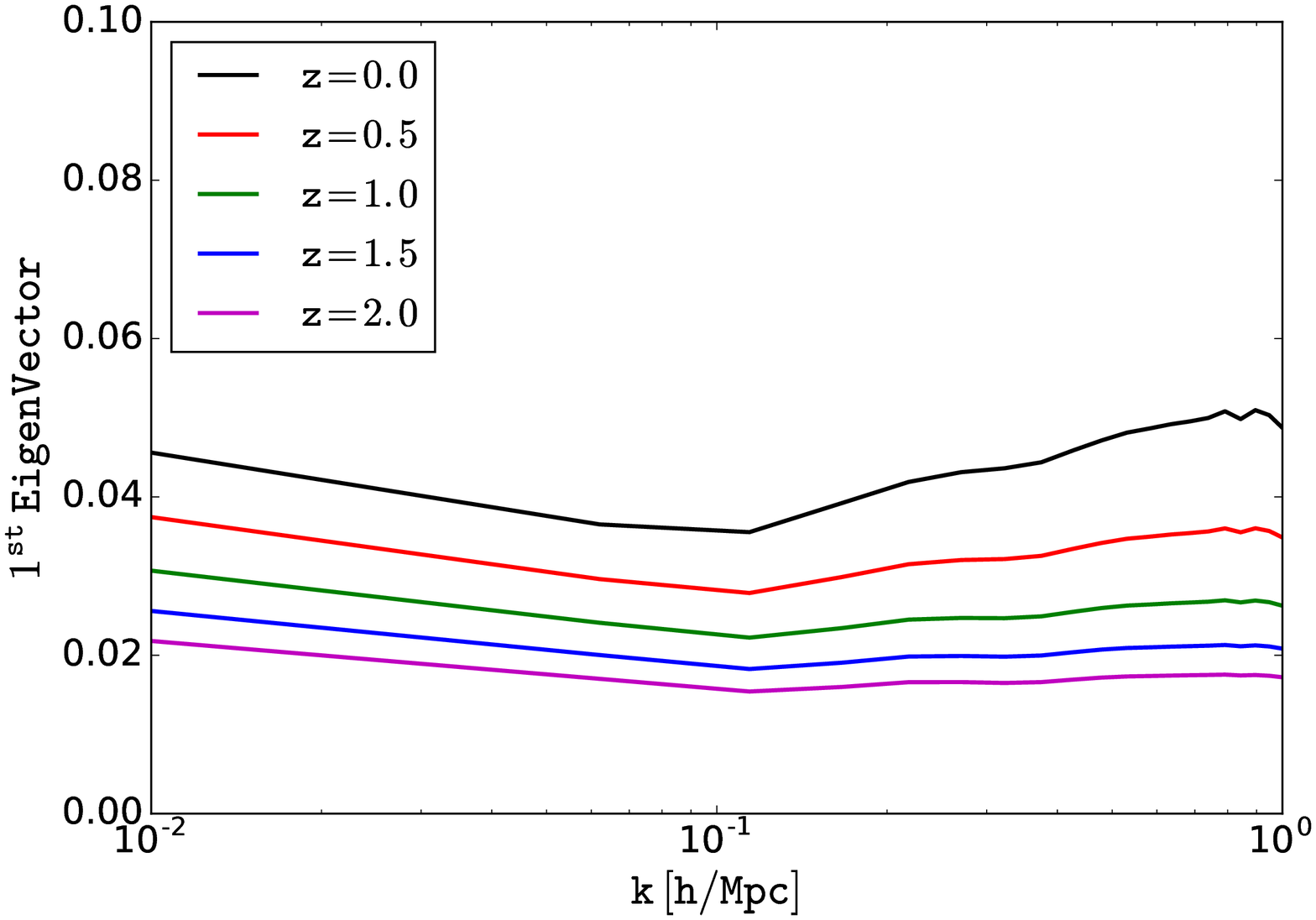}
    \caption{Redshift evolution of the first eigenvector of the full covariance matrix using L14 data including
    SSC covariance.}
    \label{fig:d1z}
\end{figure}

Having the decomposition of the analytic form of the covariance matrix gives us a handle to perform the redshift evolution of the principle component without explicitly computing the covariance at each redshift. Figure \ref{fig:d1z} shows the first eigenvector at five different redshifts (with global SSC).

\subsection{Degeneracies with cosmological parameters}\label{sec:fisher}

The analysis above suggests that instead of using a full covariance matrix, one can use the Gaussian covariance, and replace the non-Gaussian part  with a fictitious external parameter $\alpha$, whose response is given by the first eigenvector $d_1(k)P(k)$, as in equation \ref{pkd}. The parameter $\alpha$ has zero expected mean, with $\langle \alpha^2 \rangle$ determined by the covariance matrix calculations above. However, one can also try to determine it from the data. In this section, we explore the degeneracies between the first eigenvector of the full covariance matrix with SSC contribution to the response of various cosmological parameters to the matter power spectrum. A related analysis with just SSC has been performed in \cite{2014PhRvD..90j3530L}. We use the eigenvector of the full covariance matrix of the L14 dataset.

\begin{figure}
    \centering
    \includegraphics[width=0.48\textwidth]{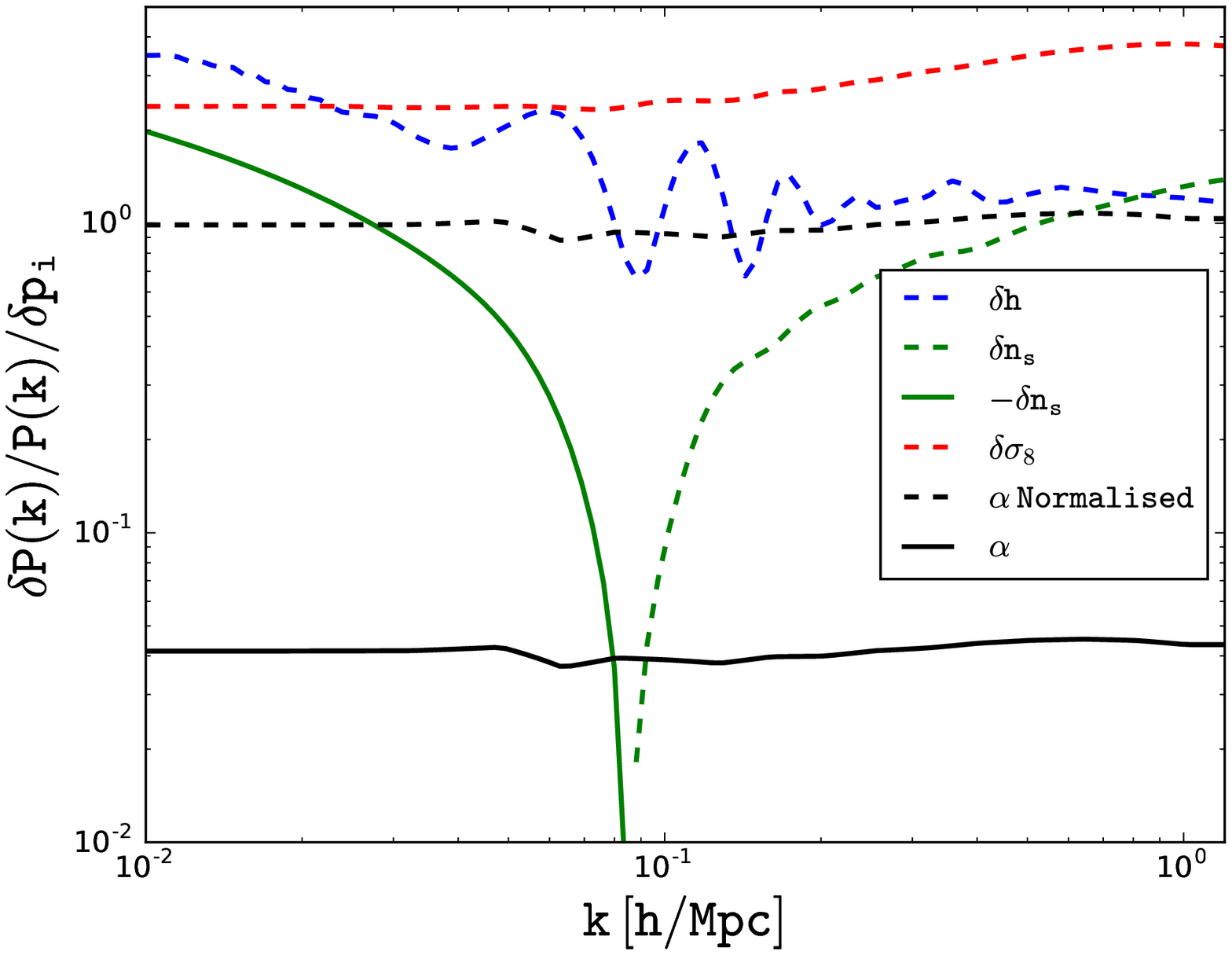}
    \includegraphics[width=0.48\textwidth]{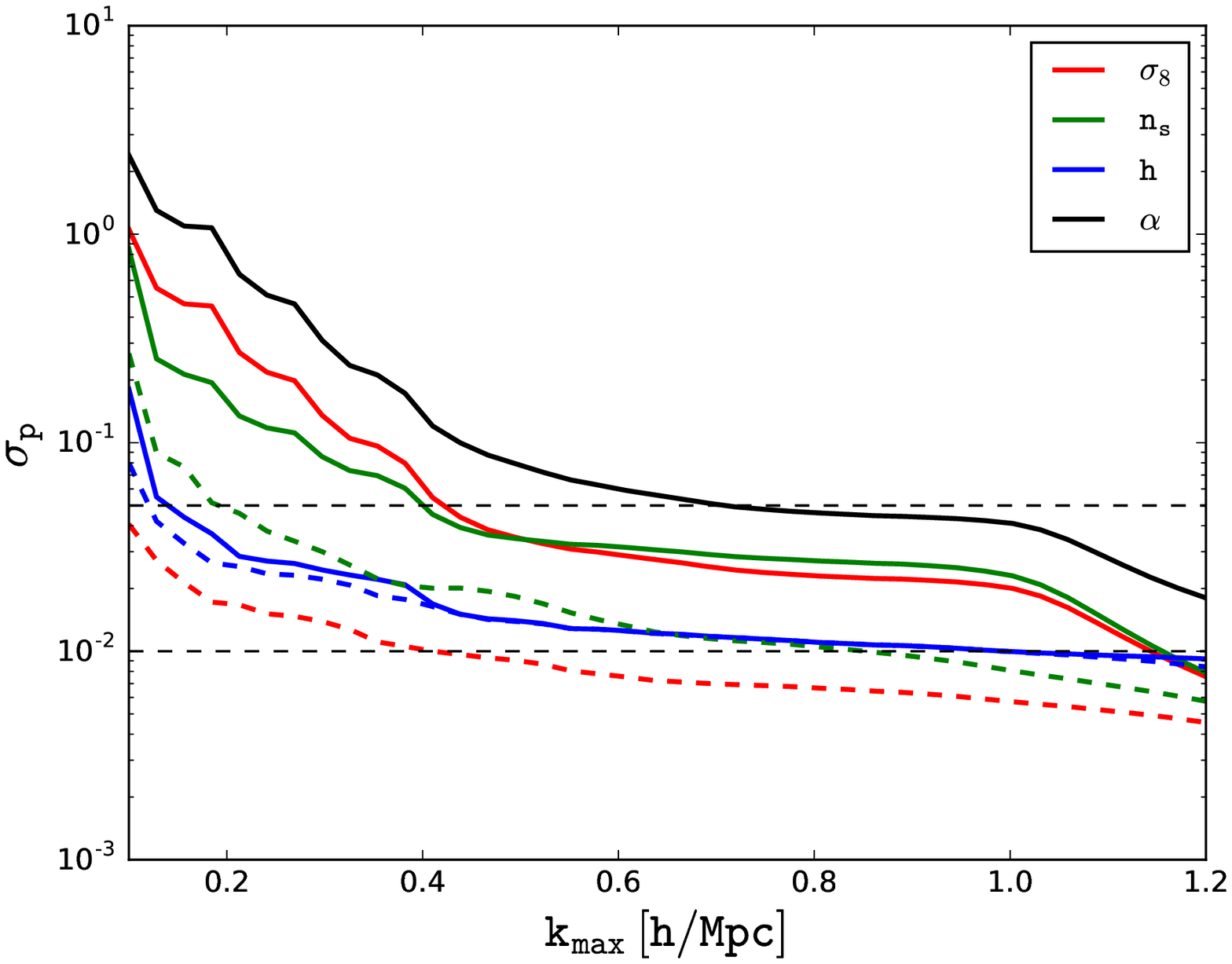}
    \caption{Left: the response of the cosmological parameters and the covariance parameter to the matter power spectrum.
            Right: showing the forecast errors on each parameter for fixed $\alpha$ (in dashed lines) and marginalizing over
            $\alpha$ (in solid lines).}
    \label{fig:fisher}
\end{figure}

The left panel of Figure \ref{fig:fisher} shows the comparison of this response to the cosmological parameters responses. At low $k$ there is a degeneracy between the $d_1(k)$ and the amplitude of the fluctuations ($\sigma_8$) to the matter power spectrum. In this regime, the covariance matrix calculation of the first eigenvalue gives a prior, shown as the horizontal dashed line on the right-hand side. However, this degeneracy is broken in the highly non-linear regime ($k>0.3 h {\rm Mpc}^{-1}$), suggesting that the data may determine the covariance matrix amplitude if one uses high $k$ information.

To explore the effects of this degeneracy on cosmological parameter estimation, we performed a Fisher matrix analysis for the two cases: (i) when $\alpha$ is fixed, and hence, only the response of the cosmological parameters plays a role, (ii) while marginalizing over $\alpha$, incorporating the response of the $\alpha$. The full fisher matrix can be written as,

\begin{equation}
    F_{\mu \nu} = \dfrac{\partial P(k_i)}{\partial p_{\mu}} C^{\rm Gauss}_{ij} \dfrac{\partial P(k_j)}{\partial p_{\nu}}
\end{equation}
\\
where, $p$ is a set containing cosmological parameters and $\alpha$. We used three cosmological parameters for this analysis - amplitude of the fluctuations $\sigma_8$, spectral index $n_s$ and Hubble constant in the units 100 km/s/Mpc $h$. Therefore,

\begin{equation}
    p \equiv \{\sigma_8, n_s, h, \alpha \}
\end{equation}
\\
For this particular exercise we used only the Gaussian part of the covariance matrix. We carried out this exercise for different $k_{\rm max}$ (from 0.1 to 1.2), and computed the expected errors on each parameter.

The right panel of figure \ref{fig:fisher} shows the expected errors on the parameters. The solid lines represent the expected errors on the parameters when marginalized over $\alpha$, and dashed lines when $\alpha$ kept fixed. Due to the strong degeneracy between $\alpha$ and $\sigma_8$, the marginalization errors are large for small $k$. However, as we increase $k_{max}$ this degeneracy is broken, giving improved constraints on all cosmological parameters including $\alpha$.


\subsection{Convergence of the covariance matrix and the required volume of simulations}

As we argued the non-Gaussian covariance is dominated by a single eigenmode, whose $k$ dependence is fixed, and only its amplitude $\alpha$ varies from a realization to a realization. By the central limit theorem, we expect the distribution of $\alpha$'s to be close to a Gaussian. This is shown in figure \ref{fig:alpha}, normalized to its variance. We see that the distribution is indeed very close to a Gaussian.

In the context of this model we can ask how many simulations are needed to fully determine the covariance matrix. To determine the $k$ dependence of the eigenvector $d_1(k)$ we just need 2 simulations: the difference in $P(k_i)$ between the two simulations will be proportional to $d_1(k_i)P(k_i)$. The proportionality factor is not relevant if we use the method described in the previous subsection, where $k$ dependence is used to determine the amplitude and remove the non-Gaussian part of the covariance matrix. 

If we wish to determine the full covariance matrix including its normalization we also need to determine $\langle \alpha^2 \rangle$. Since $\alpha$ has a Gaussian distribution we can obtain a fractional error of $(2/N)^{1/2}$ on its variance with $N$ simulations. While Gaussian distribution is the best case scenario, it is still a relatively slow convergence: if one wishes to determine covariance matrix to 10\% one needs 200 simulations, while 1\% requires 20000 simulations.

However, we can also exploit the scalings of individual terms of covariance matrix with volume. The Gaussian term is analytic and the SSC term can be well determined by a response to a constant curvature (separate universe methods), and the value of $\sigma_b^2$ on the  scale of the survey given by equation \ref{SSCsig}. Note that this term does not scale simply as $1/V$, it also  depends on the power spectrum $P(q)$. 

We are thus left with the modes inside the volume of the survey. Equation \ref{ngcon} states that the covariance is  given by the trispectrum divided by the volume $V$.  One can compute the trispectrum using a small volume simulations and scale by the survey volume.  But one must include all of the trispectrum contributions.  The mode contribution to the simplified version of the  1-loop integral is shown in figure \ref{fig:p2k2}. Modes with $q<0.02 h {\rm Mpc^{-1}}$ contribute about 1\% of the total  integrand. If we are willing to tolerate 1\% error then the required box size to get all the contributing  modes is $V \sim (300 h^{-1}{\rm Mpc})^3$.  In general, we, therefore, expect that the minimal volume per simulation is roughly of this size. 

If the desired error of the covariance is 10\% then we can run 200 simulations of $V \sim (300h^{-1}{\rm Mpc})^3$, i.e. a total volume of $V \sim 5(h^{-1} {\rm Gpc})^3$. If the desired error is 1\% then the required volume is $V \sim 500( h^{-1} {\rm Gpc})^3$.  To reduce the disconnected part of the covariance matrix, which is effectively noise for the connected part, one can simply take  broader bins of $k$. As shown in figure \ref{fig:trispectrum} the covariance is expected to be very insensitive to the binning.  Note that the effective volume of future surveys is of order tens to hundreds of $(h^{-1} {\rm Gpc})^3$, so the required volumes are  of order a single survey volume. But the simulations still need to be done on periodic boxes, to eliminate the SSC term, which would  otherwise dominate the covariance of sub-boxes (although see below for an alternative method where this is not needed).  SSC term is added separately to the total covariance matrix, using the correct  value of $\sigma_b^2$ from equation \ref{SSCsig}.

\begin{figure}
    \centering
    \includegraphics[width=0.8\textwidth]{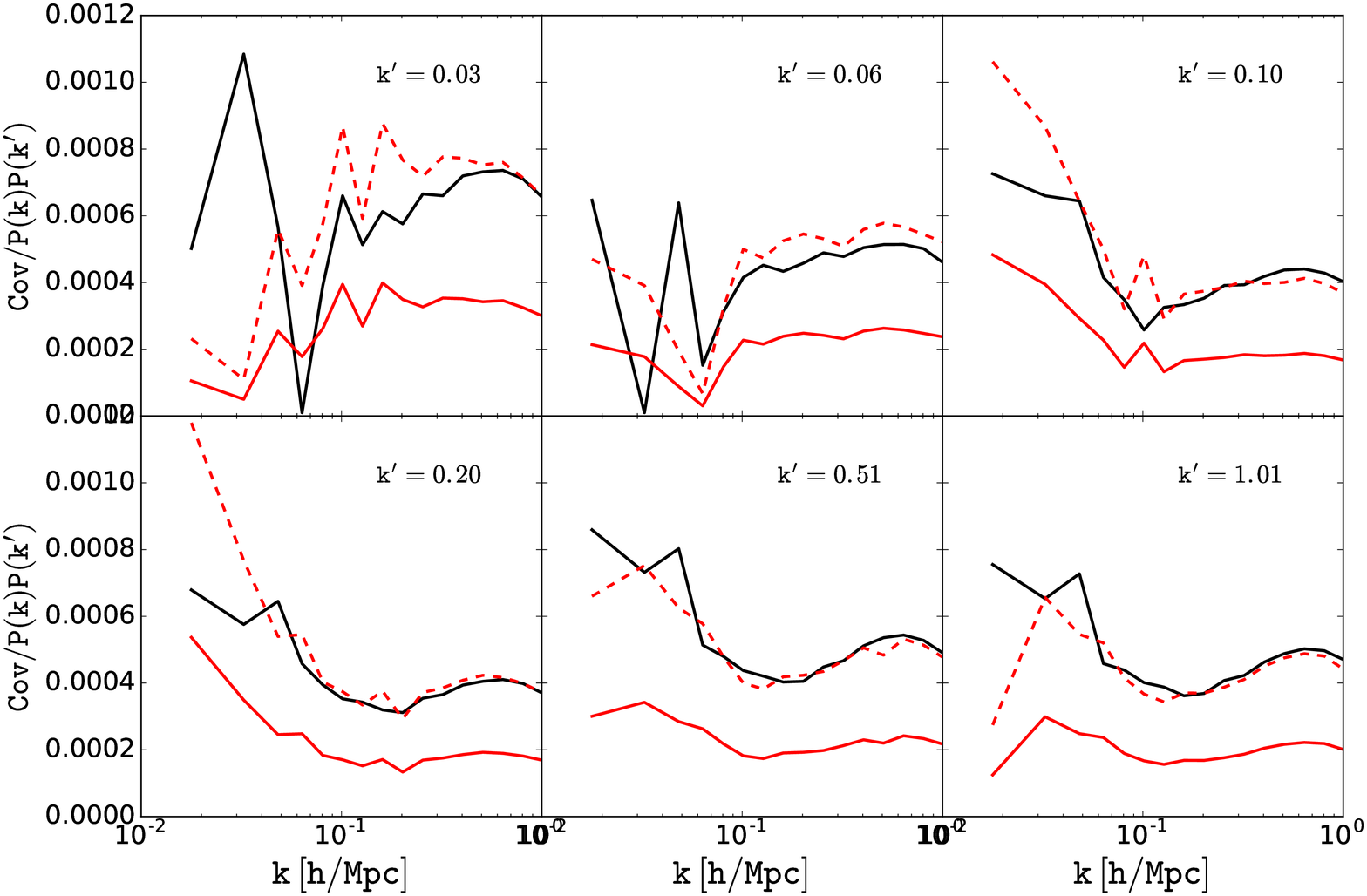}
    \caption{Comparing L14 covariance (solid black) to that of B15 (red dashed), where B15 covariance has been renormalized 
to the same volume as L14, i.e., by a factor $(650/500)^{3}$. Unrescaled B15 is shown as red solid.}
    \label{fig:compare_li_blot}
\end{figure}

To test this we compare the two covariance matrices at our disposal, which have been simulated at different volumes.  We rescale them by the volume ratio, which is $(650 h^{-1}{\rm Mpc})^3$ for B15 and $(500 h^{-1}{\rm Mpc})^3$ for L14, a factor of 2.2. We see in figure \ref{fig:compare_li_blot} that this gives a very good agreement between the two.  There are residual differences at the level of 10\%. At low $k$ these are probably  due to noise fluctuations. It is unclear if the remaining differences are due to the differences in the cosmological model, or due to some differences in the simulations.

Finally, we mention that one may be able to get the covariance matrix simply from the survey data itself.  As we argued above, the volumes of typical future surveys are tens of $(h^{-1} {\rm Gpc})^3$, which is not too different from the volume  needed to reliably simulate covariance matrix. One can divide a survey into many sub-volumes of a typical size of $(300 h^{-1}{\rm Mpc})^3$  and measure the power spectrum in each sub-volume. To reduce the fluctuations for the disconnected part one should use relatively broad $k$ bins, and remove the disconnected part (including any possible window function effects that induce mixing between the bins).  The main remaining issue is that dividing an observed volume into  sub-volumes also induces SSC term on the sub-volume (same is  true for jackknife or bootstrap methods that use a similar subdivision). This requires thus to estimate and remove the SSC term from the  sub-volumes. This can be done  by correlating the spatially dependent sub-volume power spectrum with the mean density on the sub-volume, which is a form of a squeezed limit  bispectrum \citep{2014JCAP...05..048C}. Once the mean SSC response is determined one can multiply it with the sub-volume mean density and subtract  it out of the sub-volume power spectrum, thereby removing SSC term from the sub-volumes. From the remaining power spectra, one  can now compute the covariance matrix whose connected term will be dominated by the modes inside the sub-volume, and which is expected  to scale as $1/V_{\rm sub-volume}$. Once the connected part of the covariance matrix is determined one can scale it as $V_{\rm sub-volume}/V$ to get the  corresponding covariance of the whole survey from modes inside the sub-volume. To this one needs to add SSC term by computing $\sigma_b^2$ from equation  \ref{SSCsig} using the window function defined by the whole survey, multiplied by the response as derived from the squeezed limit  bispectrum. Finally, one adds the disconnected part of the covariance matrix, including the survey window function effects. 


\subsection{Connecting perturbation theory to the halo model}

The halo model has been a very successful model for nonlinear clustering of the dark matter power spectrum (see e.g. \cite{2002PhR...372....1C} for a review). It decomposes the nonlinear power spectrum into the correlations between  halos, the so-called two-halo term, and correlations within the halos, the so-called one-halo term. The two-halo term has been originally modeled as a linear power spectrum smoothed on the virial radius scale  of the halos. This does not capture effects like the smoothing of baryonic acoustic peaks. A  simple generalization has been presented in \cite{2015PhRvD..91l3516S}, where the two-halo term is instead replaced by a Zeldovich approximation (ZA) power spectrum. The difference between the full power spectrum and ZA is defined to be the one-halo term. This difference has to vanish for $k \rightarrow 0$, meaning that  the one-halo term needs to be compensated and is not a constant, which happens at very low $k$.  The one-halo term on large scales is given by Poisson fluctuations at zero lag, and can be calculated as the second moment of mass integrated over the halo mass function. 

An alternative description pursued here is to use perturbation theory. PT has been  applied to the power spectrum and at 1-loop order, it is expected to do well only at very low $k$,  while at higher $k$ it requires modest EFT corrections, which make 1-loop sufficiently accurate  up to $ k \sim 0.2 h {\rm Mpc^{-1}}$  \citep{2016JCAP...05..027F,2016JCAP...03..007B,2016JCAP...03..057V}. In \cite{2015PhRvD..91l3516S} it was argued that we can define a regime where both  halo model and 1-loop PT (with EFT) are equally valid so that one can calibrate the one-halo term  on PT, and then use the halo model to extend it to higher $k$. This allows a smooth continuation from PT  to a high $k$ regime where the halo model is valid.  When compared to the halo model predictions in terms of the second moment integrated over the halo mass function, this requires a halo mass definition that is somewhat higher than  the standard virial mass definition $M_{200}$. 

A similar discussion applies to the covariance matrix. In the halo model the covariance at high $k$  arises from  Poisson fluctuations.
A particularly simple form has been  developed in \cite{2014MNRAS.445.3382M}, where the connected term has been derived as 

\begin{equation}
\mathbf{Cov}_{ij}^{\rm connected}=P(k_i)P(k_j)V^{-1}\delta_{A_0}^2.
\end{equation} 

In the halo model, $\delta_{A_0}^2$ can be related to the fourth moment of mass integrated over the halo mass  function. This calculation somewhat overpredicts the true value when compared to simulations  \citep{2014MNRAS.445.3382M}, but it is reasonably close. 

One can see that the form of the halo model prediction is identical to PT calculation in the limit where there is a  single eigenvector that does not depend on $k$, which is approximately valid, as seen in figure \ref{fig:NumVsAnl}.  In PT the variance  is given by the value of $S$ in  equation \ref{S}. This is replaced by $\delta_{A_0}^2$ in the halo model.    If we insist that the two descriptions agree with each other in the regime of overlap ($k \sim 0.2 h {\rm Mpc^{-1}}$) then this determines the value of $\delta_{A_0}^2$ from PT.  
Hence the fourth moment  of the halo mass function is determined by equation \ref{S}, just as 
the second moment of halo mass function is  determined by 1-loop PT \citep{2015PhRvD..91l3516S}.  
This is a consistency that restricts the form of the halo mass  function.  Note that the agreements are not perfect: while for the power spectrum the standard  one-halo term calculation under-predicts PT (unless the halo mass is increased),  it slightly over-predicts it for the covariance, but these differences could also be caused by  uncertainties in the halo mass function and halo mass definitions.  
This picture allows a convenient connection of PT to the halo model on  small scales. For covariance, we expect the small scale clustering to be dominated  by one-halo term from lower mass halos, which are less rare, and hence we  expect ${\rm Cov}_{ij}^{\rm connected}/P(k_i)P(k_j)$ to decrease as we go to higher $k$. 

Ultimately, this agreement is just a justification for the halo model  being applicable to both the power spectrum and its covariance. While the one halo term 
has to be the dominant term at high $k$ in the power 
spectrum, essentially
by definition of the halo mass function,
for higher order moments the the halo model 
needs to be justified case by case. We have seen that it works well for covariance. For variance of 
covariance the halo model prediction is given by the eighth moment of the mass integrated over the 
halo mass function, and predicts very large relative 
fluctuations, of order 30\% for a volume of $1({\rm Gpc/h})^3$ 
\citep{2014MNRAS.445.3382M}.  The corresponding 1-loop PT calculation is given by $112\pi^3\int P(q)^4d^3q / [\int P(q)^2d^3q]^2/V$
and it gives two orders of magnitude lower variance of covariance, suggesting Poisson fluctuations are 
not very important, as also explicitly verified by simulations in figure \ref{fig:alpha} (note that this figure 
includes SSC term). This suggests that the halo model cannot be reliably 
applied to the calculations of the eigth moment.


\section{Conclusions}\label{sec:discussion}

In this paper, we developed a perturbative model for the covariance matrix of the power spectrum, going up to 1-loop in PT, but without including all the 1-loop terms (see \cite{2015arXiv151207630B} for the full 1-loop calculation). In addition, we go beyond 1-loop by  including the nonlinear damping of response functions \citep{2014arXiv1411.2970N}, which improves our  results at higher $k$. Overall, our approach predicts the results from simulations to about ten percent accuracy in the quasi-linear regime. The largest contribution to the covariance is beat coupling or super-sample covariance (SSC), which arises due to the coupling of the modes to the ones which are larger than the survey scale and which has been extensively  studied previously \citep{2014PhRvD..89h3519L,2014PhRvD..90j3530L}. This effect comes from the modes outside the survey and cannot be captured by the  jackknife or bootstrap methods, which subdivide the full volume into many smaller volumes. In fact, these methods overestimate the covariance terms because they generate SSC from the smaller regions. We propose an alternative  approach which removes this problem. On large scales, the second largest contribution comes from tree-level terms from modes inside the survey, while on smaller scales 1-loop terms dominate. The dominant 1-loop term, and the only one we include, comes from the sampling variance fluctuations of large-scale modes within the survey volume, which induce a coherent response by small-scale modes which we model using 1-loop power spectrum.  When we allow for a damped nonlinear response to modes that are highly nonlinear 
\citep{2014arXiv1411.2970N}, we obtain some additional improvements, especially at lower redshifts. Damping does not affect results at higher redshifts because all of the modes that have significant sampling fluctuations are linear.

We explore the structure of the covariance matrix and find that its non-Gaussian part is strongly dominated by a single eigenmode. This suggests that the non-Gaussian response has always the same shape, only its amplitude varies. We analyze the probability distribution of this amplitude and find that it is close to a Gaussian. Thus, the convergence rate of covariance matrix simulations scales as a Gaussian, with $(2/N)^{1/2}$ giving the relative error after $N$ simulations. One possible alternative approach is to ignore the non-Gaussian covariance in the analysis and include instead the eigenvector response as a fictitious external parameter in the analysis. This parameter can in principle be determined from the data itself, but it is quite degenerate with other cosmological parameters.

While the analysis given in this paper provides several important insights into the nature of the covariance matrix of the two-point correlators, our results cannot be directly applied to the data. For weak lensing observations, which are measuring the dark matter correlations, one needs to perform the projection along the line of sight. This leads to a decorrelation of the non-Gaussian covariance because different $l$ correspond to different effective redshift, and hence different effective volume. This differs from our analysis where all modes are maximally correlated because they are coupled to the same long wavelength modes within the given volume. So far this has  only been addressed in the context of the halo or SSC model \citep{2014PhRvD..90l3523S,2016arXiv160105779K}.

The 3-d analysis performed here is more likely to be of immediate application to the galaxy clustering, but this would require adding biasing and redshift space distortions to the model, and it is unclear whether   successful analytic models can be built. Nevertheless, the decomposition of the covariance matrix into the  three components, disconnected, connected but generated from modes outside the survey, and connected and  generated from modes inside  the survey, allows one in principle to build the full covariance matrix from relatively small simulation volumes, and possibly even from the data itself, by subdividing into sub-volumes and determining each of the three components  separately. It is worth pursuing this further to determine the optimal approach that delivers the highest  accuracy with this technique. 

\section*{Acknowledgment}

We thank Matias Zaldarriaga for useful discussions.
U.S. acknowledges support from NASA grant NNX15AL17G.
Z.V. is supported in part by the U.S. Department of Energy contract to SLAC no. DE-AC02-76SF00515.
I.M. is supported by Fermi Research Alliance, LLC under Contract No. De-AC02-07CH11359 with the United States Department of Energy.


\bibliographystyle{mnrasfile}
\def\apj{ApJ}
\def\apjl{ApJL}
\def\aj{AJ}
\def\mnras{MNRAS}
\def\aap{A\&A}
\def\nat{Nature}
\def\pasj{PASJ}
\def\prd{PRD}
\def\physrep{Physics Reports}
\def\jcap{JCAP}
\bibliography{ms.bib}


\end{document}